\documentclass[12pt,preprint,longabstract]{aastex}
\usepackage{pdflscape}
\usepackage{graphicx}
\usepackage{epstopdf}

\slugcomment{CCCP OB Signatures Paper, \today}
\shorttitle{Carina OB Signatures}
\shortauthors{Gagn\'e et al.}

\begin{document}

\title{Carina OB Stars: X-ray Signatures of Wind Shocks and Magnetic Fields}

\author{Marc Gagn\'e, Garrett Fehon, and Michael R. Savoy}
\affil{Department of Geology and Astronomy, West Chester University, West Chester, PA 19383}
\author{David H. Cohen}
\affil{Department of Physics and Astronomy, Swarthmore College, Swarthmore, PA 19081}
\author{Leisa K. Townsley, Patrick S. Broos, and Matthew S. Povich\altaffilmark{1}}
\affil{Department of Astronomy \& Astrophysics, 525 Davey Laboratory, Pennsylvania State University, University Park, PA 16802}
\author{Michael F. Corcoran}
\affil{CRESST and X-ray Astrophysics Laboratory, NASA/GSFC, Greenbelt, MD 20771}
\author{Nolan R. Walborn}
\affil{Space Telescope Science Institute, Baltimore, MD 21218, USA}
\author{Nancy Remage Evans}
\affil{Smithsonian Astrophysical Observatory, MS 4, 60 Garden St., Cambridge, MA 02138}
\author{Anthony F.J. Moffat}
\affil{D\'epartement de Physique, Universit\'e de Montr\'eal, Succursale Centre-Ville, Montr\'eal, QC, H3C 3J7, Canada}
\author{Ya\"el Naz\'e\altaffilmark{2}}
\affil{GAPHE, D\'epartement AGO, Universit\'e de Li\`ege, All\'ee du 6 Ao\^ut 17, Bat. B5C, B4000-Li\`ege, Belgium and Research Associate FRS-FNRS}
\and
\author{Lida M. Oskinova}
\affil{Institute for Physics and Astronomy, University of Potsdam, 14476 Potsdam, Germany}
\altaffiltext{1}{NSF Astronomy \& Astrophysics Postdoctoral Fellow}
\altaffiltext{2}{Research Associate FRS-FNRS}

\begin{abstract}

The {\it Chandra} Carina Complex contains 200 known O- and B type stars. The {\it Chandra} survey
detected 68 of the 70 O stars and 61 of 127 known B0-B3 stars. We have assembled a publicly available
optical/X-ray database
to identify OB stars that depart from the canonical $L_{\rm X}/L_{\rm bol}$ relation, or
whose average X-ray temperatures exceed $1$~keV.
Among the single O stars with high $kT$ we identify two
candidate magnetically confined wind shock sources: Tr16-22, O8.5~V, and LS~1865, O8.5~V((f)).
The O4 III(fc) star HD 93250 exhibits strong, hard, variable X-rays,
suggesting it may be a massive binary with a period of $>30$~days.
The visual O2~If* binary HD 93129A shows soft 0.6~keV and hard $1.9$~keV emission components,
suggesting embedded wind shocks close to the O2~If* Aa primary, and colliding wind shocks between Aa and Ab.
Of the 11 known O-type spectroscopic binaries, the long orbital-period systems 
HD~93343, HD 93403 and QZ Car have higher shock temperatures than short-period systems such as
HD~93205 and FO~15.
Although the X-rays from most B stars may be produced in the coronae of unseen, low-mass
pre--main sequence companions, a dozen B stars with high $L_{\rm X}$ cannot be explained
by a distribution of unseen companions. One of these, SS73~24 in the Treasure Chest cluster,
is a new candidate Herbig Be star.

\end{abstract}


\section{Introduction}

The {\it Chandra} Carina Complex Project (CCCP) \citep{townsley11a}
survey area contains over 200 massive stars: the luminous blue variable (LBV) $\eta$~Car,
the Wolf Rayet (WR) stars WR 22, WR 24 and WR 25, 70 known O stars, and
127 B0-B3 stars with determined spectral types and photometry consistent with a distance of
2.3 kpc to the Carina cluster Trumpler 16 (Tr 16) \citep{smith06}.
In the next section we describe the database of OB stars in more detail. Briefly, the massive star
population in Carina can be divided into groups based on spectral type, luminosity class and binarity:
LBV/WR stars (4), early-O binaries (6), early-O single dwarfs and giants (10), late-O binaries (9), late-O dwarfs
and giants (44), single O-type supergiants (1), B0-B3 stars (127), and B5-B7 stars (3).

In addition to the OB stars with measured spectral types considered here and by \citet{naze11},
this volume contains four other articles which discuss the massive-star population in Carina:
\citet{povich11a} present a list of 94 candidate OB stars, selected on basis on their X-ray emission
and infrared spectral energy distributions (SED); their spectral types have not been determined.
\citet{evans11} identify candidate late-B stars in Tr~16 based on their UBV photometry.
\citet{townsley11a} analyze the CCCP spectra of the three WR stars and
\citet{parkin11} present a detailed X-ray spectral/temporal analysis of the
double spectroscopic binary QZ Car, O9.7~I~+~O8~III.
In this paper, we examine the X-ray, optical, and infrared characteristics of 
the 200 OB stars with determined spectral types and use X-ray spectra and
light curves of this large sample of well-studied OB stars to better understand the physical
mechanisms that produce X-rays in massive stellar systems.

The ubiquitous X-ray emission from O and early B stars is generally thought to 
arise in shocks embedded in the powerful radiation-driven winds of these 
stars. However, there are exceptions to this paradigm, with a subset of 
early-type stars showing harder and stronger X-ray emission than can be 
explained by embedded wind shocks (EWS) alone.
We point to three recent X-ray surveys that have examined these questions in some detail:
the {\it XMM-Newton} survey of NGC~6231 in the Sco~OB1 association \citep{sana06a,sana06b},
the {\it Chandra} survey of Tr~16 \citep{evans03,evans04},
and the {\it Chandra} Orion Ultradeep Project \citep{stelzer05}.

The Orion Nebula Cluster contains only two O stars, and both have unusual
X-ray emission: $\theta^1$~Ori C, O7~Vp \citep{gagne05a}, and 
$\theta^2$~Ori A, O9.5~SB \citep{feigelson02,schulz06}. Of the early-B
stars, \citet{stelzer05} identify two groups: strong-wind sources (earlier than B4)
that give rise to X-rays in wind shocks (relevant to this study), and
weak-wind sources (later than B4) whose X-rays, when detected, may be
produced by late-type pre-main-sequence (PMS) companions \citep[see][]{evans11}.

In their study of Tr~16 and part of Trumpler~14 (Tr~14), \citet{evans03,evans04}
analyzed the ACIS-I spectra of some of the prominent O stars in Carina,
identifying HD~93250, O4~III(fc), and Tr16-244, O3/4~I, as highly unusual,
and proposing a new colliding-wind binary system: Tr16-22, O8.5~V. Aside from these
anomalously active stars, \citet{evans03} find that the canonical
$L_{\rm X}  \approx 10^{-7} L_{\rm bol}$ relationship is obeyed.

In their comprehensive study of the OB stars in NGC~6231,
\citet{sana06b} confirm the $L_{\rm X}/L_{\rm bol}$ relationship in the soft
{\it XMM-Newton} bands below 2.5~keV, but show a breakdown in the correlation in the
2.5-10 keV hard band. They also confirm the prominent kink in
$L_{\rm X}/L_{\rm bol}$ below $\log L_{\rm bol} = 38$ first noted by
\citet{berghoefer97} \citep[see also][]{naze11}.

In this paper, we
focus mainly on three emission mechanisms, in addition
to EWS, to understand the X-rays observed from O and B stars in the CCCP:
colliding wind shocks (CWS), magnetically confined wind shocks (MCWS),
and coronal emission from unseen pre-main--sequence (PMS) companions,
though we discuss other mechanisms in \S1.5.
In particular, we focus on those stars
that show high $L_{\rm X}/L_{\rm bol}$, hard X-ray spectra, or show notable time variability.

\subsection{Embedded Wind Shocks}

The default mechanism for X-ray production on O and early B stars is 
embedded wind shocks (EWS), generally assumed to be associated with the 
Doppler deshadowing instability intrinsic to line driving 
\citep{lucy80,lucy82,owocki88,feldmeier97}. The instability 
predicts that wind streams having different velocities interact, shock 
heating a modest fraction of the wind beyond a few tenths of a stellar radius.
Several stellar radii above the photosphere, the 
shocked portion of the wind is traveling at speeds 
approaching the wind terminal velocity of thousands of km~s$^{-1}$.
\citep{dessart05a,dessart05b}.

But the relative velocities of the interacting wind streams are generally only a 
few hundred km~s$^{-1}$, generating shock temperatures of just a few million K. This 
shock-heated plasma radiates a line-dominated soft X-ray spectrum. This 
soft X-ray emission generally shows little variability, indicating that 
there are numerous separate shock-heated regions of the wind at any given time.
Wind clumping is a universal phenomenon in hot-star winds 
\citep{lepine99, lepine00, moffat08} and \citet{feldmeier03} showed that 
the wind clumping affects the emergent X-ray line profiles.

The soft X-ray emission is attenuated by the colder, X-ray absorbing 
wind in which it is embedded. This wind attenuation affects the shapes 
of individual line profiles, as is evident in high-resolution X-ray 
spectroscopic studies of nearby O stars, which also confirm Doppler 
broadening commensurate with the wind terminal velocity, lending strong 
support to the EWS mechanism 
\citep{kahn01,cassinelli01,zhekov07,walborn09,cohen10}.

The wind  attenuation can harden the overall X-ray spectrum for early O stars with 
very high mass-loss rates. The attenuation of the X-rays from embedded 
wind shocks also affects the emergent X-ray luminosity.  Empirically, a 
scaling of emergent X-ray luminosity with bolometric luminosity, 
$L_{\rm X}  \approx 10^{-7} L_{\rm bol}$, has been long known 
\citep{pallavicini81}, and is largely confirmed in the broad OB population
in the CCCP \citep{naze11}.
No strong theoretical expectation for this scaling has been found, 
although wind attenuation can explain it in principle \citep{owocki99}.

For Carina O stars with X-ray emission dominated by the EWS mechanism, 
we should expect to see soft X-ray emission (characterized by $kT < 1$ 
keV) at a level corresponding to $L_{\rm X} \approx 10^{-7} L_{\rm bol}$, and 
with little variability.  For B stars, significantly lower relative X-ray 
luminosities are seen, with values as low as $L_{\rm X} \approx 10^{-9} 
L_{\rm bol}$ by spectral subtype B2~V \citep{cohen97}.

While embedded wind shocks presumably exist in all early-type stars with 
strong radiation-driven winds, detailed X-ray studies of nearby O and 
early B stars show that a fraction of them have properties that are at 
odds with the EWS scenario. These usually involve higher X-ray 
luminosities and harder X-ray spectra than is predicted to arise in 
embedded wind shocks alone, and often involve X-ray time variability on 
orbital or rotational timescales. When high-resolution X-ray spectra 
are available, these often show narrower X-ray emission lines, as well. 
Corroborating evidence from other wavelengths generally shows evidence 
for either wind-wind collisions in binary systems or magnetic 
confinement of the wind. In the next four subsections, we summarize some of these
theoretical and observational studies.

\subsection{Colliding Wind Shocks}

In binary systems where both stars have strong winds, shock heating in 
the wind interaction zone between the two stars can lead to stronger X-ray 
emission than that seen in the EWS scenario 
\citep{stevens92,pittard09}. The X-ray properties of CWS sources 
depend on the wind and orbital parameters of the two stars, and they 
vary with orbital phase in the case of systems with eccentric orbits.
\citet{antokhin04} predict that X-ray luminosity scales as the reciprocal of the binary separation.
In many O+O, WR+O, and LBV+O-type systems in which both components have high mass-loss rates,
CWS X-rays tend to dominate those from embedded wind shocks,
with X-ray flux increasing for more equal-momentum winds and intermediate separations.

For example, the O+O binaries in the {\it XMM-Newton} survey of hot stars \citep{naze09}
have, on average, $\sim3$ times higher $L_{\rm X}/L_{\rm bol}$ than single O stars.
We note that most single and binary O stars in the {\it XMM} survey
show relatively low-temperature shocks. In fact, the only O+O binaries in the {\it XMM}
surveys with hard X-ray spectra are the long-period Carina binaries HD~93403, O5~III(fc) + O7~V \citep{naze09},
and HD~93343, O8~V + O7-8.5~V \citep{antokhin08}. Long-period, WR+O and LBV+O binaries
like WR 140 and $\eta$~Car also show very high-temperature shocks \citep{pollock05,corcoran01}.

\citet{pittard09} model the colliding wind emission from pairs of O6 V + O6 V stars with varying orbital periods;
their simulations predict peak temperatures of 3-4 MK and 10-20 MK for models
{\sc cwb1} (3-d period) and {\sc cwb2} (10-d period), respectively. We will examine these issues in more detail
in \S 6.3.

It can sometimes be difficult to confirm the binary nature of systems 
with X-ray properties indicative of CWS emission. Spectroscopic or photometric orbital variations
are difficult to measure when the orbital inclination is low, 
although persistent observing \citep{hoffmeister08} and complementary 
approaches \citep{nelan04} reveal long-suspected companions in CWS 
systems. After many years of monitoring, \citet{gamen08}
derived a 207.7-d orbital period for the Carina WN6h colliding-wind binary WR~25.
Observations of non-thermal radio emission often provide clues 
about the presence of wind-wind interactions in binaries, even when 
binary orbital parameters cannot be measured
\citep{dougherty00,debecker07}. Ideally, X-ray monitoring
over multiple orbital periods is needed to show the orbital modulation
expected from colliding wind systems, at least those with sufficiently
high eccentricities.

\subsection{Magnetically Confined Wind Shocks}

Where strong, large-scale magnetic fields exist on early-type stars, 
wind streams from opposite hemispheres are 
channeled toward the magnetic equator, where they collide head-on, 
leading to strong shocks and associated X-rays 
\citep{babel97}.  MCWS X-ray emission can be 
differentiated from EWS X-rays by its temperature distributions 
\citep{wojdowski05}, which can be dominated by plasma with 
temperatures of 30-50 million K.  For oblique dipole 
configurations, the X-rays can be rotationally modulated due to 
differential occultation of the shock-heated plasma by the star
\citep{gagne97}.

The prototype MCWS source is $\theta^1$~Orionis~C, O7~Vp, the illuminating 
star of the Orion Nebula, which has a 1.5~kG oblique dipole field, 
measured with Zeeman spectropolarimetry \citep{donati02, wade06}.
{\it Chandra} grating spectroscopy shows a high X-ray luminosity 
(corresponding to $L_{\rm X} > 10^{-6} L_{\rm bol}$) and peak X-ray emission 
measure $\sim 33$~MK.  The X-rays vary (by roughly 20\%) 
in a manner consistent with the rotationally modulated magnetic field 
configuration, and these X-ray properties are well explained by MHD 
simulations of the confined wind of this star \citep{gagne05a}.

Early B stars of the chemically peculiar B2p class show very strong 
dipole magnetic fields, though they have much weaker winds than the
O star $\theta^1$~Ori~C. Some, like the prototype $\sigma$~Ori~E, have modest, 
relatively hard X-ray emission \citep{sanz-forcada04,skinner08}, while 
other B2p stars have no X-ray emission or emission at levels below the 
detection threshold.  Even the stronger X-ray sources among these Bp 
stars have X-ray luminosities of only
$L_{\rm X} \approx 10^{30}$~ergs~s$^{-1}$~cm$^{-2}$
\citep{drake94}.

The ability to channel a wind depends both on the strength of the 
magnetic field and (inversely) on the density of the wind.  These 
scalings are verified in detail by MHD simulations \citep{uddoula02,gagne05a}. 
Some O stars with strong winds and weak fields, such as 
$\zeta$~Ori, have X-ray emission consistent with the EWS scenario 
\citep{cohen06}. 

On the other hand, the other well-studied magnetic O stars such as HD~108, O7~Ifpe,
and HD~191612, 06.5~IIIf, have $L_{\rm X}/L_{\rm bol}\approx-6.2$ and $kT\approx0.3$~keV.
They are more X-ray active than typical EWS sources, but substantially cooler than
$\theta^1$~Ori~C \citep{donati06a,naze07,naze10}.
Also curious is the early B star, $\tau$~Scorpii, B0.2~V, 
which has a highly structured (non-dipole) field \citep{donati06b},
strong and hard X-ray emission, but no evidence for wind confinement of 
the X-ray emitting plasma \citep{cohen03,ignace10}.

At this point, there are too few O and early B stars with 
well-characterized magnetic fields (aside from several dozen Bp stars), 
and their X-ray properties are too heterogeneous, to make definitive 
statements about what X-ray signatures will be seen from magnetic OB 
stars in Carina.  However, we can say that it is possible that 
strong X-ray emission may be produced in Carina O stars via magnetically confined
shocks if conditions are right. Overall, though, the general X-ray 
properties from the MCWS mechanism -- elevated X-ray luminosities, hard 
X-ray spectra, and time variability -- overlap with those 
from the CWS mechanism, though the variability in CWS systems is orbital,
whereas the variability in MCWS is often rotationally modulated.

\subsection{Unseen Pre--Main-Sequence Companions}

The three X-ray emission mechanisms described above convert the kinetic energy of
the supersonic wind into shocks and thermal X-ray emission. As the
mass-loss rate and terminal wind speed diminish towards later spectral
type, embedded wind shocks, colliding wind shocks, and 
magnetically confined wind shocks should be much weaker.

In star-forming regions, at spectral subtype $\sim$B2, $L_{\rm X}$ falls below
a few $10^{30}$~ergs~s$^{-1}$ \citep{caillault94,daniel02}, 
and the upper end of the low-mass PMS X-ray
luminosity function begins to outshine the massive stars.
It is now generally understood that hard, variable X-ray emission from
main-sequence mid-B to late-A type stars is produced by unseen,
lower-mass companions \citep{gagne95,briggs03}.
In most cases, the X-rays provide the only clue to the presence of cool companions.
In a reverse application, \citet{evans11} use X-ray detections in the CCCP and BV photometry 
to identify unknown B stars in Tr~16.
 
The situation for the PMS intermediate-mass stars, i.e., the Herbig Ae/Be stars,
is quite different. Many are copious, often hard X-ray emitters,
despite the fact that they have weak winds and radiative outer envelopes,
so they are not expected to have wind shocks, or solar-type magnetic activity.
\citet{skinner04} consider low-mass PMS companions and an intrinsic, rotational shear dynamo
to explain the X-rays from Herbig Ae/Be stars, but so far the mechanism is not known.

\subsection{Other X-ray Emission Mechanisms}

In this paper, we assume that the soft X-ray emission from most O stars is produced in shock-heated,
collisionally ionized, thermal plasmas. Specifically, plasmas whose densities are high enough to be modeled by 
equilibrium ionization codes like APEC.

However, other emission mechanisms have been proposed to explain the X-ray properties of massive stars.
Most fundamentally, \citet{pollock07} suggests that the X-rays from O supergiants like $\zeta$~Orionis,
O9.7~Ib, are produced in the far wind where densities are low, and where the Coulumb collisional mean-free path
is many stellar radii, so that the shocked wind is far from collisional or thermal equilibrium.
In the \citet{pollock07} scenario, {\em ions}, not electrons, ionize ions, and the shocks are collisionless;
shock heating occurs via plasma processes and magnetic fields. The major difficulty with the 
\citet{pollock07} paradigm is that the X-ray spectra of O stars appear to be produced by
collisionally ionized, thermal plasmas with electron-ion bremsstrahlung continua \citep[c.f.,][]{raasen08}.

The low-resolution ACIS CCD spectra from the CCCP cannot directly address the fundamental nature of the
shock heating; we use the APEC model to estimate plasma temperatures and emission measures because the APEC model
fits the CCD spectra well with only a few free parameters.
When time variability or high-temperature components are observed,
we examine secondary processes like colliding wind shocks, magnetically confined wind shocks, or coronal emission
from unseen, lower-mass, pre--main-sequence companions, though other mechanisms have been proposed.

For example, the rigidly rotating magnetosphere model \citep{townsend05,townsend08} and its successor,
the rigid-field hydrodynamic model \citet{townsend07}, have been used to model 
the X-ray, UV and H$\alpha$ emission of the He-strong stars like $\sigma$~Ori E.
This model may have applications to other B stars with strong magnetic fields like $\tau$~Sco, B0.2~V.

For the O9.5~V spectroscopic binary $\theta^2$~Ori~A, \citet{schulz06} proposed an interaction
between the magnetospheres of the spectroscopic primary and secondary at periastron to explain
its x-ray spectrum and time variability. Magnetic fields have yet to be seen on either component, though.

\citet{cassinelli08} proposed a model for OB stars in which the thin fast wind forms bow shocks on dense,
slowly moving clumps. The adiabatic shocks produce peak emission measure near $\log T = 6.5$, with some
emission out to $\log T = 7.5$. It is unlikely that this model can operate in stars with strong magnetic
fields like $\tau$~Sco or $\theta^1$~Ori~C, but it might be viable in non-magnetic OB stars,
though the effects of radiative cooling should be added to test the model further.

\citet{waldron09} proposed a highly accelerated diamagnetic plasmoid model to explain the production
of high-temperature plasma very close to the photospheres of some O stars. This is essentially the
opposite mechanism proposed by \citet{cassinelli08}. In the diamagnetic plasmoid model, the blobs are
ejected from the surface and plow into the cooler, slowly moving plasma at the base of the wind.

The origin of magnetic fields on massive stars is still not fully known. They could be generated by
subsurface convection or via differential rotation, or they could be fossil fields.
In the \citet{spruit02} dynamo model, and its adaptation by \citet{mullan05,mullan06},
differential rotation and Tayler instabilities in the interior magnetic field lead to surface magnetic fields.
Finally, we note that some of the known OB stars in Carina could harbor neutron star companions,
producing strong, accretion-driven X-ray emission
\citep[e.g.,][]{davidson73}.

The discovery of magnetic fields on some massive stars, in combination with colliding winds
from massive binaries, leads to a number of complicated but intriguing X-ray emission mechanisms.
In this paper, we will focus on the X-ray diagnostics -- variability, high $L_{\rm X}/L_{\rm col}$ or
high $kT$ --
that may signal the presence of magnetic fields or colliding wind shocks in the Carina massive stars.

\section{Carina OB Star Optical Catalog}

The goal of this study is to characterize the X-ray properties of the O and early-B stars
with reliably determined spectral types in the $\sim 1$ square degree CCCP field, and
to correlate those X-ray properties with effective temperature, bolometric luminosity,
and binary properties. The optically selected sample is biased towards stars that are bright from 3900--4800 \AA,
and those that are close to the well-studied clusters Trumpler 14, 15 and 16, Bochum 10 and 11, and Collinder 228.
This sample complements the X-ray selected massive-star candidates of \citet{povich11a}, 
though these
are also biased towards luminous stars because of the correlation between $L_{\rm X}$ and $L_{\rm bol}$.
The optically selected sample is biased to stars with $A_V < 3$~mag while the candidate 
massive stars of \citet{povich11a} are mostly obscured stars with $A_V > 3$~mag.

We began with the
\dataset[Vizier/B/mk]{``Catalogue of Stellar Spectral Classifications"}
\citep[][and references therein]{skiff09} selecting stars with spectral type O2-B3.
To these we added stars in \citet{tapia03, smith06, degioia01, walborn95, masseyandjohnson93},
including two B5 stars, and a B7 star.
For some individual stars, mostly spectroscopic binaries and Ofc stars, we used the updated spectral types of 
\citet{morgan55, morrisonandconti80, morrell88, rauw00, rauw01, freyhammer01, albacete02, niemela06, rauw09, gosss10, walborn10, sota11}.

The spatial distribution of the 200 OB stars is shown in Figure~1,
where the clusters Trumpler 15, 14, and 16 are visible from N to S at
$-59\arcdeg 24\arcmin$, $-59\arcdeg 36\arcmin$, and $-59\arcdeg 42\arcmin$, respectively.
Many of the confirmed massive stars seen in Fig.1 are coincident with the low-mass populations
in the rich clusters (Tr~14, Tr~15, Tr~16, and Coll 228), as expected. 
However, a significant fraction lie well outside the clusters, supporting the idea that star
formation in Carina has been ongoing for a considerable length of time, 
allowing stars to drift or be ejected from their natal groupings.
This is discussed further by \citet{feigelson11,povich11a,povich11b}. 

Johnson UBV photometry was gleaned primarily from the photoelectric catalog of 
\dataset[Vizier/II/168]{``Homogeneous Means in the UBV System"} and its update
\dataset[Vizier/II/169]{``UBV Photoelectric Cat: Data 1986-1992"} \citep{mermilliod94}, 
and the CCD survey of Trumpler 14 and 16 by \citet{masseyandjohnson93}. For a few close visual binaries,
we obtained UBV from \citet{massey01, vazquez96, forte81, forte76}.

We note that a number of other CCD and photoelectric photometry studies have
been published on the Carina hot stars, particularly for those in Trumpler 14 and 16.
For example, the compilation of \citet{reed05},
\dataset[Vizier/V/125]{``Photometry and Spectroscopy for Luminous Stars"}
lists $V$, $U-B$, and $B-V$ from a number of reliable photometric surveys.
Unfortunately, only 101 of the 200 OB stars in the CCCP are listed in the \citet{reed05} catalog.
Some surveys like \citet{degioia01} and the {\it Tycho} catalog measured $B$ and $V$. 
After analyzing color-magnitude diagrams from a number of these studies, we chose the compilations
of \citet{mermilliod94} and \citet{masseyandjohnson93} as our primary references because they provided 
consistent UBV measurements for all but one star: HD~92937, B2.5~II:.

For the 70 O stars, effective temperature was interpolated on the basis of spectral type, using the
observational $T_{\rm eff}$ scale of \citet{martins05}. For consistency, we used the O-star synthetic colors 
of \citet{martins06} to estimate $(B-V)_0$ and bolometric correction. We note that the Martins et al.
work uses solar abundance, wind- and line-blanketed, non-LTE model atmospheres, that, for OB stars, lead to
lower $T_{\rm eff}$ and $L_{\rm bol}$ for a given observed spectral type than did previous work.

For the 130 B stars, consistent sets of NLTE synthetic colors, effective temperatures, and bolometric
corrections were not available, though NLTE effects are expected to be smaller in the lower-luminosity Carina
B dwarfs. We used the intrinsic colors of \citet{wegner94} to estimate $(B-V)_0$. For the B stars, we used
the bolometric corrections and effective temperature scale of \citet{bessell98}.

\subsection{Reddening and Bolometric Luminosity}

This study and the study of \citet{naze11} rely on two methods for determining $L_{\rm bol}$:
(1) the spectral energy distribution (SED) fitting method of \citet{povich11a}, and 
(2) the more traditional method employing color excess and bolometric correction. 
We recall that in the color excess method,
$M_{\rm bol} = V - A_V - {\rm DM} + {\rm BC}_V$, $R = A_V/E(B-V)$ and $E(B-V) = (B-V) - (B-V)_0$,
where ${\rm BC}_V$ is the $V$-band bolometric correction and DM is the distance modulus.

Of the 200 OB stars in our spectroscopic sample, 182 have high-quality UBV, 2MASS JHK and {\it Spitzer} IRAC 
photometry in the VelaÐCarina Point-Source Archive \citep[PI: S. Majewski,][]{povich11b}.
\citet{povich11a} used the SED fitting method of \citet{robitaille07}, 
which applies an ISM absorption model to the revised ATLAS9 stellar atmosphere models of \citet{castelli04}.
Though we recognize that the ATLAS9 LTE model atmospheres do not account for non-LTE wind and line blanketing effects,
\citet{povich11a} show that LTE and non-LTE atmospheres produce similar broad-band optical/infrared SEDs, 
and hence very similar estimates of $A_V$ and $L_{\rm bol}$.

Recognizing that a number of stellar and ISM parameters can adequately fit a 10-point SED, \citet{povich11a} 
generate a family of solutions for each SED. The 182 OB stars in the validation sample have the advantage of 
having well-determined spectral types, and hence $T_{\rm eff}$, thereby significantly reducing the number of 
statistically acceptable solutions, and significantly constraining $A_V$ and $L_{\rm bol}$.

The SED fitting depends on the adopted reddening law,
which effectively specifies the scaling between $A_V$ and the absorption in each photometric band. In the
traditional color excess method, this comes down to choosing $R_V=A_V/E(B-V)$. As described in \citet{povich11a},
the estimates for $R_V$ in Carina have varied from an ISM value of 3.1 or 3.2 \citep{turner80a,turner80b}
to above 5.0 \citep{herbst76}.

Table 1 of \citet{walborn95} illustrates the interplay between $R$ and DM in the Carina clusters.
Throughout the CCCP, we adopt  ${\rm DM} = 11.81$, corresponding to $d=2.3$~kpc, the distance to
$\eta$ Carinae derived from the Doppler velocity and proper motion of the expanding Homunculus nebula
\citep{smith06}.
\citet{walborn95} find that for this distance modulus, $R=4$ provides the best fit to the O stars in Tr 16,
though star-to-star, and cluster-to-cluster variations probably do exist.
Similarly, \citet{povich11a} find better agreement between the $L_{\rm bol}$ estimates using
$B-V$ and those derived from the optical-infrared SED method using $R_V=4$ than using $R_V=3.1$ or $R_V=5$.
Hence, we and \citet{povich11a} adopt an average $R_V=4$ and $d=2.3$~kpc to all the stars in Carina.
We do not attempt to account for star-to-star variations in the reddening law.

We note that, based on Walborn's analysis of six O stars in Tr 14, $R_V=4$ implies ${\rm DM} = 12.33$ if the Tr~14 and Tr~16 are coeval (and thus the O stars in both clusters have the same average luminosity).
A more likely solution, however, is that Tr~14 is younger than Tr~16 and the main-sequence O stars
are sub-luminous relative to the average class V calibration.

For 181 of the 200 stars in the Carina OB database, we compute $L_{\rm bol}$ using both the SED
(tag name {\sc loglbol\_sed}) and color excess ({\sc loglbol\_bv}) methods. 
We note that one star, MJ~501=Tr~16~74, B1 V, has unreliable $B-V$.
For these 181, the offset in $A_V$ (SED - BV) is $\Delta A_V = 0.1$ with a standard deviation $\sigma=0.25$ mag.
This is a lower bound on the systematic uncertainty in $A_V$.
The offset in $\log L_{\rm bol}$ (SED - BV) is $\Delta \log L_{\rm bol} = 0.04$ with a standard deviation $\sigma=0.14$ mag.

We note though that CMFGEN model atmospheres \citep{hillier01} produce more EUV, NIR, and MIR emission for
a given $T_{\rm eff}$ and $\log g$ than non-LTE model atmospheres, especially for O supergiants. 
Because we wish to exploit the full set of optical, near- and mid-infrared photometry,
and to maintain consistency with the rest of the CCCP, we use the SED-derived $L_{\rm bol}$ values
whenever possible, despite this deficiency. Future updates to the OB catalog will contain
new $L_{\rm bol}$ estimates, based on updated SED-fitting codes.

The FITS table headers and the CDS header files list the name, unit, and fortran format code for all columns
in the OB catalog. The first 38 columns are devoted to the optical/infrared database and include
the following columns (or tags): IAU J2000 name, default designation (label), HD name, CPD name
(from the 1919 list of Cannon, Pickering \& Draper), LS or ALS name \citep{reed05},
Feinstein Trumpler 14/15/16, Collinder 228, or Bochum 10/11 designation, and
\citet{masseyandjohnson93} designations. J2000 R.A. and Decl. are listed in decimal degrees and sexagesimal
hours and degrees. Also listed are spectral type, spectral type reference, binary flag, notes, $V$, $U-B$, $B-V$,
$Q$ \citep{masseyandjohnson93}, and UBV reference. We list the derived quantities $(B-V)_0$, $E(B-V)$, BC,
and $A_V$, $\log L_{\rm bol}$ (solar units), and $\log T_{\rm eff}$ as determined via the color excess method,
the SED fitting method, and the adopted value.
We note that the label tag contains the commonly used designation used in the text, tables and figures,
but that the other designations in the electronic files are precisely those used by {\it Simbad},
and by association other databases, for rapid searching.

Table 1 shows a cross section of the optical database for the most luminous stars,
sorted by $\log L_{\rm bol}$. Figure~2 shows the resulting H-R diagram.
We overlay the {\em average} main sequence of \citet{martins05b} from their sample of 1-5 Myr-old stars. The B-star
main sequence is from \citet{dejager87} \citep[see also][]{povich11a}.

The zero-age main sequence (ZAMS) is clearly delineated in Fig.\ 2,
as is the extent of the terminal-age main sequence (TAMS). The supergiants and bright giants are shown with white dots.
The uncertainty in $L_{\rm bol}$ is dominated by random photometric errors (0.1 mag) and by
systematic uncertainties, especially in the distance modulus and $R=A_V/E(B-V)$, on the order of 0.3 mag.
As noted earlier, we use a global reddening law; we do not account for possible star-to-star variations in $R$.
Despite these uncertainties, much of the spread above the ZAMS appears to indicate a real spread in age.
We note that the most luminous OB star in Carina, QZ Car, is a quadruple system \citep{parkin11}.
$T_{\rm eff}$ for the candidate Herbig Be star SS73~24 was estimated from its optical-infrared SED (see \S6.4.1).

All but two of the 70 O stars were detected with at least 3 source counts in the CCCP:
Tr14-27, O9 V, Tr15-18, O9 I/II:(e:). Tr15-18 also has the highest $A_V$ of all 200 OB stars;
it may have avoided detection because of its high column density.
While $A_V=5.66$~mag for Tr15-18 is high among the known OB stars,
it is typical of the X-ray selected candidate OB stars identified by \citet{povich11a}.
Finally, we note that Tr15-19, O9 V, and Tr15-21, B0~III, are well below the ZAMS; they may be a background stars,
not members of Trumpler~15 (Tr~15).

To estimate the completeness of the optical OB catalog, we consider the populations of OB stars with measured spectral types,
and the 94 X-ray-detected stars without measured spectral types discovered by \citet{povich11a}.
Using an initial mass function (IMF) exponent $\Gamma = -1.3$ \citep{kroupa02},
normalized to the 47 LBV, WR, and O2-O8.5~V stars with mass $M > 20 M_{\odot}$, we expect to find
24 stars in the mass range $15-20 M_{\odot}$, compared to 27 O9 and O9.5 stars, and 
130 stars in the mass range $7-15 M_{\odot}$, compared to 123 main-sequence B0-B2.5 stars.
Note, we are considering only the primary stars. This suggests that, if the number of early-mid O stars is complete,
then the number of late-O and early-B stars is also nearly complete. 

However, the 94 stars identified by \citet{povich11a} with $L_{\rm bol} > 10^4 L_{\odot}$ could have main-sequence
spectral types as early as O4. I.e., the \citet{povich11a} result implies that a substantial fraction of early-O to early-B
stars do not have measured spectral types. To estimate this fraction more precisely, we note that 140 stars in our
optical catalog have $L_{\rm bol} > 10^4 L_{\odot}$, and that 76\% are detected in the CCCP.
Applying this detection fraction to the 94 stars in Table 3 of \citet{povich11a}
suggests approximately 124 stars with $L_{\rm bol} > 10^4 L_{\odot}$. I.e., 
an additional $\sim 30$ luminous stars were not detected with {\it Chandra}. This suggests 
a total population of 140 + 94 + 30 = 264 luminous stars; our optical catalog is thus $\sim 53\%$ complete.

Thus if the \citet{povich11a} result is confirmed, then a significant fraction (up to half) 
of the OB stars in the CCCP region have yet to be spectroscopically identified. These stars
have higher than average $A_V$ and generally reside outside the well-studied clusters in Carina.
Similarly, \citet{wright10} find that a large fraction of the B-star population in the
Cyg OB2 star-forming region is not identified.

\section{Carina OB Star X-ray catalog}

The 200 OB star optical positions were matched to the list of 14\,368 CCCP sources
\citep[see][for a detailed description of the catalog matching]{broos11a}.
The astrometric systems in the two catalogs were well aligned
($\Delta\alpha = 0.005\arcsec$ and $\Delta\delta = 0.012\arcsec$),
and the optical and X-ray positions reported for individual stars were in good agreement
(the median offset was $r\approx0.19\arcsec$).
This initial pass through the CCCP X-ray data produced 118 matches (out of 200),
including 67 of 70 O stars, and 51 of 130 B stars.

In order to estimate count upper limits for the 82 OB stars not detected in the automated {\em ACIS Extract}
procedure described by \citet{broos11a}, we determined optimal source and background regions for the undetected
OB stars at their optical positions, accounting for the {\it Chandra} point spread function and the presence
of nearby X-ray point sources.

The full {\em ACIS Extract} procedure was run again on the combined list of 14\,368 sources, 82 undetected OB stars,
and a number of photometrically selected candidate mid-to-late B stars \citep{evans11}.
The {\em ACIS Extract} procedure used in the CCCP produces source, background and net counts in three bands:
hard (2-8 keV), soft (0.5-2 keV) and total (0.5-8 keV). The procedure also estimates the probability $P_{\rm B}$
in these three bands that the source counts were produced by fluctuations in the observed local background.

Although they were not identified in the automated source detection procedure,
an additional 11 stars met the detection criteria used in the CCCP catalog:
at least 3 extracted counts and $P_{\rm B} \leq 0.01$ in a least one of the three bands \citep{broos11a}.
They are:
Coll228-66, O9.5 V, 
LS 1745, B2 III, 
HD 305515, B1.5 Vsn:, 
Tr16-13, B1 V, 
HD 305534, B0.5 V: + B1 V:, 
Tr16-29, B2 V, 
LS 1866, B2 V, 
Tr16-33, B2 V, 
HD 93342, B1 Iab-Ib, 
Coll228-81, B0.5 V, 
HD 93723, B3 III.
Thus, the only undetected O stars are
Tr14-27, O9~V, 
Tr15-18, O9~I/II:(e:).
Thus the OB catalog contains 129 X-ray matches (out of 200), including 68 of 70 O stars,
and 61 of 127 B0-B3 stars. We note that the three stars with spectral types later than B3
were not detected in X-rays.

For the undetected stars, the 90\% count rate upper limit, {\sc netcounts\_hi\_t} 
is used to derive a photon flux upper limit corrected for the mean effective area in the 0.5-8 keV band,
the effective exposure time, and the PSF fraction of the selected source extraction region
\citep[using eqn.\ 1 in][]{broos11a}.

The 200 OB stars have been divided into three groups: 
(1) 63 undetected stars with $P_{\rm B} > 0.01$ for which we calculate a photon flux upper limit,
(2) an additional 51 X-ray detected OB stars with 3-50 counts, for which we also determine
median energy, $E_{\rm med} =~${\sc medianenergy\_t} and the 0.5-8 keV absorbed (uncorrected) energy flux,
$f_{\rm X} =~${\sc energyflux\_t} \citep[see eqns.\ 1 and 2,][]{broos11a}, and
(3) 78 stars with at least 50 net counts for which \citet{naze11} also derive
$kT$ and $N_{\rm H}$ by fitting the ACIS spectra in XSPEC with a
one- or two-temperature VAPEC emission model, and a two-component {\sc tbabs} absorption model.

The results of the XSPEC fitting procedure are described fully by \citet{naze11}, but the absorption
modeling deserves mention here. For all 200 OB stars, we have estimated $A_V$ and converted
that absorption into an ISM column density using
$N_{\rm H}/A_V = 1.6\times10^{21}~{\rm cm}^2~{\rm mag}^{-1}$
\citep{vuong03,getman05}\footnote{We note that the $N_{\rm H}/A_V$ ratio is robust
to very high $A_V$, but that the scatter in the $N_{\rm H}$ versus $A_V$ diagram
is quite large, especially at high $A_V$.}.

In the XSPEC fits, \citet{naze11} fix the first ISM column density parameter, and 
allow a second column density parameter to vary. Many OB-star CCD spectra 
appear to require this extra absorption, presumably caused by absorption
of shocked emission in the overlying wind.

The absorbed X-ray fluxes, used in this paper
were computed in XSPEC with the first absorption parameter set to $N^{\rm ISM}_{\rm H}$, and the 
second absorption parameter at its best-fit value. The {\em unabsorbed}
fluxes were computed using the best-fit emission model, the ISM absorption
parameter set to zero, and the second absorption parameter at its best-fit value.
This way the unabsorbed flux represents the flux emerging from
the far-wind of the star, corrected for ISM absorption.
$L_{\rm X}$ is the corresponding 0.5-8 keV unabsorbed X-ray luminosity assuming $d=2.3$~kpc,
{\sc fitluminosity\_tc}.

We note that \citet{naze11} derive the absorption-corrected 0.5-10~keV X-ray luminosity, 
related to tag name {\sc fcto}, assuming the same distance.
We note that 0.5-8.0~keV is the total band of the CCCP,
and 0.5-10~keV is the default energy range of previous {\it XMM-Newton} surveys
\citep[e.g.,][]{sana06b}.
The two X-ray luminosities are very similar, with an average offset of 0.008 in the log,
and RMS deviation of 0.033 in the log.

We note that the four X-ray brightest stars were piled up in the CCCP ACIS-I data: 
HD~93129A, O2~If${\ast}$, HD 93205, O3~V~+~O8~V, HD~93250, O4~III(fc),
and QZ~Car, O9.7~I~+~O8~III. A more extensive set of QZ~Car data have been fully 
analyzed by \citet{parkin11}. In the X-ray catalog, we use the average
QZ~Car parameters in Table~4 of \citet{parkin11}. For HD 93129A and HD 93205
\citet{naze11} performed their two-temperature XSPEC analysis on the pileup-corrected
spectrum \citep[for more details see][]{broos11a}. For HD 93250, we used a series of {\it Chandra} ACIS-S observations;
these data and the XSPEC spectral analysis are described in Appendix A.

Throughout this paper, we use $kT_{\rm avg}$, the emission-measure weighted
mean $kT$ for the 78 OB stars with XSPEC fit parameters. For the 53 stars with
one-temperature XSPEC fits, $kT_{\rm avg} = kT_1$. For the 24 stars with
two-temperature XSPEC fits, and for QZ Car's three-temperature fit,
$kT_{\rm avg} = \Sigma kT_i n_i / \Sigma n_i$, where $n_i$ is the 
normalization parameter of the $i$th temperature component.
The volume emission measure is $4\pi d^2\Sigma n_i$.

Thus, in addition to the many parameters produced by the extended {\it ACIS Extract} procedures
described by \citet{broos11a}, our OB X-ray database contains an additional 30 parameters to describe the results
of the two-temperature XSPEC fits of \citet{naze11}.
Merging the X-ray parameters with the optical parameters for each star
produces a set of tables with 200 rows (one row for each OB star).
We used the following column names (i.e., IDL structure tags) extensively to create the figures and tables in this paper:
{\sc label}, {\sc sptype}, {\sc ktmean}, {\sc loglbol}, {\sc logt}, {\sc probks\_single}, {\sc probks\_merge},
{\sc probnosrc\_t}, {\sc netcounts\_t}, {\sc netcounts\_hi\_t}, {\sc fitluminosity\_tc},
{\sc medianenergy\_t}, {\sc energyflux\_t}, where {\sc \_t} refers to the total 0.5-8 keV band, 
and {\sc \_tc} refers to total band, absorption-corrected.

The full set of optical/X-ray tables are available electronically as a machine-readable tables from the CDS,
a FITS binary table file, and an IDL save file\footnote{The FITS and IDL save files are available from the authors upon request.}. 
The catalog can also be queried from the Vizier database at the CDS.

\section{X-ray Spectral Results}

Tables 2 and 3 summarize the X-ray properties for the single O stars
and the known O+O binaries, respectively, sorted by $\log L_{\rm X}/L_{\rm bol}$.
Table 4 summarizes the X-ray spectral and timing analysis results for the 28 X-ray brightest B stars,
sorted by $\log f_{\rm X}$. 
From Tables~2, 3, and 4 we find:

\begin{enumerate}
\item The great majority of single and binary O stars have
$-6.8 < \log L_{\rm X}/L_{\rm bol} < -8.0$ and
have soft X-ray spectra characterized by $kT_{\rm avg} < 0.8$~keV.
\item A small number of single O stars have hard X-ray spectra characterized by $kT_{\rm avg} \gtrsim 1$~keV:
HD~93250, O4~III(fc), MJ 496, O8.5~V, and MJ 449 = LS~1865, O8.5~V((f)).
\item The spectroscopic O+O binaries with $P_{\rm orb} > 10$~days
like QZ Car, HD 93403 and HD 93343
show hard X-ray spectra characterized by $kT_{\rm avg} > 1$~keV.
\item The O+O binaries with $P_{\rm orb} < 10$~days
like HD 93205, HD 93161A, FO~15 and five other known short-period O+O binaries in Tr16 have softer X-ray spectra
characterized by $kT_{\rm avg} < 1$~keV.
\item The shortest-period binary in Carina, FO15, O5.5 Vz + O9.5 V, has low
$\log L_{\rm X}/L_{\rm bol} \approx -7.65$.
\item All but one of the B stars with $\log L_{\rm X} \gtrsim 31$ have $kT_{\rm avg} \gtrsim 1$~keV
\item All but one of the B stars with $\log L_{\rm X} < 31$ or $\log f_{\rm X} < -14$
have relatively soft X-ray spectra, with $kT_{\rm avg} < 0.6$~keV.
\end{enumerate}

The $L_{\rm X}/L_{\rm bol}$ patterns have been discussed in detail by \citet{naze11}.
We recall that the ISM-corrected $L_{\rm X}$ is computed for the 78 stars with more than 50 counts in XSPEC, whereas
$f_{\rm X}$, which is not corrected for ISM absorption, was computed for all the X-ray detected stars in
{\it ACIS Extract} (except HD 93250, HD 93129A, HD 93403 and QZ Car).
Noting that $f_{\rm bol} = {L_{\rm bol}} / {4\pi d^2}$,
in Figure~3 we show $\log f_{\rm X}$ versus $\log f_{\rm bol}$ using the
same symbol definitions as Figs.\ 1 and 2 to highlight the location of early and late, single and binary
O stars, B stars, and hard and soft X-ray spectra.

In Figs.\ 3 and 4, the 51 X-ray detected stars with 3-50 counts are shown as filled
diamonds; 49 are B stars (pink).
We categorize the OB stars as {\em hard} (with $kT_{\rm avg} > 1.8$~keV, shown as triangles in scatter plots),
{\em medium-energy} (with $kT_{\rm avg}$ in the range 1-1.8 keV, squares), and
{\em soft} (with $kT_{\rm avg} < 1$~keV, circles). 
In Figs.\ 3 and 4a, the upper dotted lines represent
$\log f_{\rm X}/f_{\rm bol} = {-7}$ and the lower dashed lines represent
$\log L_{\rm X}/L_{\rm bol} = {-7.23}$, the mean of $\log L_{\rm X}/L_{\rm bol}$
with $kT_{\rm avg} < 1$~keV.

The upper and lower panels of Figure~5 show the number distributions of
$\log L_{\rm X}$ and $\log f_{\rm X}$, respectively, for O and B stars:
early-O binaries are black, early-O single stars are blue, late-O binaries are dark green,
late-O single stars are light green, and B stars are pink.

Fig.\ 3 shows two populations of O stars (green and blue/black symbols):
(i) a population of relatively cool X-ray sources clustered around $\log f_{\rm X}/f_{\rm bol} \approx Ð7.55$,
though there are some hotter sources in this group, and
(ii) a group of sometimes hotter X-ray sources with higher $\log f_{\rm X}/f_{\rm bol}$.
This trend is emphasized in Figure 4, which plots 
$\log L_{\rm X}/L_{\rm bol}$ versus $kT_{\rm avg}$ and  $\log f_{\rm X}/f_{\rm bol}$ 
versus median energy $E_{\rm med}$ for the same samples as Fig.\ 3. 
The dotted and dashed vertical lines coincide with the dividing lines between 
soft, medium and hard symbol definitions used in Figs.\ 1-4. 
The soft, low $L_{\rm X}/L_{\rm bol}$ stars are clumped in the lower left of both panels of Fig.\ 4. 
The horizontal dotted and dashed lines correspond to  $\log L_{\rm X}/L_{\rm bol} = -7$ and $-7.23$. 
This latter value is the mean of $\log L_{\rm X}/L_{\rm bol}$ for the stars with $kT_{\rm avg} < 1$~keV
\citep{naze11}. 

The upper and lower panels of Figure 5 show the distributions of $\log L_{\rm X}$ and $\log f_{\rm X}$, respectively,
for the samples shown in Figs.\ 3 and 4.
The upper and lower panels of Figure~6 show the distributions of $\log L_{\rm X}/L_{\rm bol}$ and
$\log f_{\rm X}/f_{\rm bol}$, respectively, for the samples shown in Figs.\ 3-5. 
The histograms suggest two distributions, as do Figs.\ 3 and 4: a group of stars centered around 
$\log L_{\rm X}/L_{\rm bol} < -6.8$ and $kT_{\rm avg} < 1$~keV,
and a group with somewhat higher $\log L_{\rm X}/L_{\rm bol}$ and
$kT_{\rm avg} \geq 1$~keV.
We note a similar pattern for the B stars.

The upper and lower panels of Figure 7 show the distributions of $kT_{\rm avg}$ and median energy $E_{\rm med}$,
respectively, for the samples in Figs.\ 3-6. The upper panel of Fig.\ 7 reflects the trend seen in the top panel of Fig.\ 4: 
a large distribution of OB stars with $kT\approx0.5$~keV, and 
a small, broad distribution of stars with $kT > 1$~keV. 
The lower panel of Fig.\ 7 shows no clear segregation; 
$E_{\rm med}$ is a less useful measure of intrinsic X-ray temperature, 
because higher absorption also produces higher median energy.

Figs.\ 3, 4, 6 and 7a suggest two populations of OB stars.
The first is a low activity group with $\log L_{\rm X}/L_{\rm bol} < -6.8$ and $kT < 1$~keV.
We suggest that for most of these stars, conventional embedded wind shocks produce most of the X-ray emission.
We suggest that the stars with $kT_{\rm avg} > 1$~keV require an additional X-ray emission mechanism
to produce the enhanced activity and hotter X-ray shocks. We will explore these results in more detail for 
single O stars, astrometric binaries, O+O binaries, and B stars in \S 6 and \S7 below.

Table~4 lists the names, spectral types, and selected X-ray properties of the 28 X-ray brightest early-B stars, sorted by $f_{\rm X}$.
We note that 14 of the 15 X-ray brightest early-B stars in Table 4 have $\log L_{\rm X}/L_{\rm bol} > -7$, that 14 of those 15
have $kT_{\rm avg} \gtrsim 1$~keV, and that 7 of the top 12 are probably variable on long time scales, though
none show the characteristic rapid rise and slower decay of strong PMS flares (see Figs.\ 9 and 10).
For the B stars (in pink), Fig.\ 4b shows a wide range of median energy and $f_{\rm X}/f_{\rm bol}$, with only
a slight positive correlation.

We address the origin of the X-ray emission among the 130 early B stars by comparing the X-ray flux distributions
of the early B stars and the X-ray detected low-mass stars.
Figure~8 shows histograms of log photon flux for
61 X-ray detected early B stars (in pink), 69 B-star upper limits (in brown), and the 14\,250 CCCP sources not associated
with massive stars (in gray hatching). The vast majority of the general CCCP source population is associated with
low-mass PMS stars \citep{feigelson11}. The PMS histogram has been normalized to the 130 B stars for comparison.
We note that $f_{\rm X}$ cannot be computed for the undetected sources, so we plot the 0.5-8 keV photon flux
(or the 90\% confidence upper limit), corrected for the time-averaged PSF fraction and effective area at the
location of each star.

At first glance, Fig.\ 8 suggests that, overall, the X-ray detected B stars have higher flux than
the X-ray detected low-mass PMS stars. For example, the 61 X-ray detected B stars have mean X-ray flux 
$f_{\rm X} = 4.6\times10^{-15}$~ergs~cm$^{-2}$~s$^{-1}$, with a standard deviation
$\sigma=2.7\times10^{-15}$~ergs~cm$^{-2}$~s$^{-1}$. 
The upper limits span one order of magnitude in photon flux, illustrating the non-spatial uniformity of the CCCP survey
\citep[c.f.][]{townsley11a}. Fig.\ 8 suggests
that the CCCP is only complete to a photon flux of $\sim 10^{-3}$ counts~ks$^{-1}$~cm$^{-2}$.

Adjusting for this bias, two results emerge from Fig.\ 8:
(i) the B stars with log photon flux below $-2.3$, corresponding to $\log f_{\rm X} < -14$~ergs~cm$^{-2}$~s$^{-1}$,
appear to have the same flux distribution as the low-mass PMS stars, with a companion fraction $\sim 50\%$, and
(ii) the distribution of B stars with $\log f_{\rm X} > -14$~ergs~cm$^{-2}$~s$^{-1}$ cannot be explained by a
distribution of typical, low-mass PMS companions.

Result (i) is consistent with the results of \citet{evans11}: most X-ray detected B stars probably harbor coronal
PMS companions.
Result (ii) suggests either 
(1) the brightest dozen or so B stars formed with hyperactive companions, or
(2) that their X-rays are produced by some intrinsic shock mechanism. 

These stars are: SS73~24, Be pec, Tr16-64, B1.5~Vb, Tr16-10, B0~V, Tr16-5, B1~V, Tr14-28, B2~V,
HD~93501, B1.5~III:, Coll228-68, B1~Vn, Tr14-124, B1~V, HD 93190, B0~IV:ep, Tr14-18, B1.5~V,
LS~1813, B2~V, Tr16-11, B1.5~V, Tr14-19, B1~V, and Tr14-29, B1.5~V.

\section{X-ray Timing Analysis}

We have looked for time variability in the CCCP event data of the 129 X-ray detected OB stars in three ways:
(1) we visually examined the sequenced and stacked corrected photon flux light curves,
(2) we used the merged Kolmogorov-Smirnov test probability, $P_{\rm KS} =~$ {\sc probks\_merge}
in the CCCP data products \citep{broos10}, and
(3) we used a maximum likelihood procedure to divide the event data into blocks of approximately constant count rate.
The latter, an IDL procedure {\sc mlb\_acis} developed by E. Flaccomio, was used by
\citet{wolk05} to identify flares in the {\it Chandra} Orion Ultradeep Project (COUP).
An X-ray source is variable if an event list produces more than one maximum likelihood block.
In {\sc mlb\_acis}, the significance thresholds are calibrated as a function of counts
using a large set of simulated constant count rate sources.
For each star, the number of maximum likelihood blocks,
assuming at least 5 counts per block, and 95\% confidence,
are listed in the last column of Tables~2, 3, and 4.

We note that the maximum likelihood procedure can find two maximum likelihood blocks for a constant source,
if that source is observed in two OBSIDs at significantly different off-axis angles.
This was not an issue in the study of
\citet{wolk05} because the COUP observations were all centered on the Trapezium cluster -- stars appeared at
the same off-axis angle. We note that the photon flux curves produced by {\it ACIS Extract} 
are corrected for effective area and do not suffer this bias.

The sequenced photon flux light curves of the visually identified sources are shown in Figures~9 and 10.
Some very bright sources like HD~93403 in the lower left panel of Fig.\ 9, show significant long- and short-term
variability, and others like FO~15 in the upper left panel of Fig.\ 9 show clear long-term variability. None of the 
OB stars showed the characteristic rapid rise and slower decay of strong coronal flares \citep{wolk05}.

\section{Notes on Individual Massive Star Systems}

In this section we examine the X-ray, optical, infrared, and radio properties of the most X-ray luminous OB stars,
dividing them into four categories based on spectral type and binary separation: 
astrometric O binaries, spectroscopic O+O binaries, 
apparently single O stars, including weak-wind stars,
and B stars.

\subsection{Single O-type Stars}

Table~2 and Fig.\ 4 suggest that most single O stars have low $L_{\rm X}/L_{\rm bol}$ and low $kT$. Below we
consider the three single O stars with $\log L_{\rm X}/L_{\rm bol} > -7$ and $kT > 1$~keV: 
HD~93250, MJ~496 = Tr16-22, and MJ~449 = LS~1865.

\subsubsection{HD~93250, O4~III(fc)}

The most X-ray luminous O star in Carina, and possibly the most enigmatic, is the Ofc star HD~93250,
located east of Tr 14, and north of Tr 16. Though long classified as O3.5~V((f+)) \citep{walborn02,skiff09},
\citet{walborn10} have reclassified it as O4~III(fc). They write, 
``The Ofc category consists of normal spectra with \ion{C}{3} $\lambda\lambda$4647-4650-4652 emission
lines of comparable intensity to those of the Of defining lines \ion{N}{3} $\lambda\lambda$4634-4640-4642.''

HD~93250 was observed far off-axis in a series of 
HETG/ACIS-S observations of $\eta$~Car (see Appendix A).
The merged ACIS-S spectrum in Figure~11 shows strong \ion{Si}{13} and \ion{Si}{14} emission lines
and strong Fe K$\alpha$ emission at 6.7~keV, mostly from \ion{Fe}{25},
confirming the presence of extremely hot plasma ranging in temperature from 6-40~MK.
Its mean 0.5-8 keV X-ray luminosity is $L_{\rm X} \approx 1.5\times10^{33}$~ergs~s$^{-1}$,
similar to the value found by \citet{sanchawala07a}.

Figure\ 12 shows the $L_{\rm X}$ light curve of HD~93250 based on the XSPEC fits to the
individual ACIS-I and ACIS-S datasets. In all but one OBSID, its X-ray luminosity is marginally
consistent with a single value, $L_{\rm X} \approx 1.45\times10^{33}$~ergs~s$^{-1}$.
During the ACIS-I OBSID 4495 when the source was marginally piled up,
$L_{\rm X} = 1.7\pm0.1\times10^{33}$~ergs~s$^{-1}$, suggesting $\sim20\%$ time variability on
time scales of months (between observations), not unlike the variability seen in a series of 21
{\it XMM-Newton} spectra of HD~93250 \citep{rauw09}.

Based on the spectral fits presented in Appendix~A, we use the wind mass column to derive a
mass-loss rate $\dot{M}=1.4\pm0.5\times10^{-6} M_{\odot}$~yr$^{-1}$. This is a factor of 3-4 smaller than the
value estimated by \citet{puls96}, who did not allow for wind clumping,
yet somewhat larger than the recent estimate of \citet{martins05b},
$\dot{M}=6\pm2\times10^{-7} M_{\odot}$~yr$^{-1}$.

\citet{oskinova07} have shown that the conventional treatment of wind 
clumping commonly used in stellar atmosphere codes \citep[e.g.,][]{martins05b}
leads to an underestimate of empirical mass-loss rates by a factor of few.
The mass-loss rate for HD~93250 derived from the CCCP spectral fits appears to be generally
consistent with the findings of \citet{oskinova07}.

Along with HD~93129A, HD 93250 is one of only two non-thermal radio O stars in Carina, and
\citet{debecker07} list it as a ``suspected binary'' on the basis of its X-ray and non-thermal radio
emission. \citet{rauw09} note however that the only radio detection of HD~93250 is at 8.6~GHz
\citep{leitherer95}. The radio flux is assumed to be non-thermal because, if it were thermal, 
the 8.6~GHz flux would suggest a mass-loss rate $\dot{M} > 10^{-5} M_{\odot}$~yr$^{-1}$.

Though HD~93250 shows all the signatures of a colliding wind binary, an optical companion
has yet to be detected. \citet[][and Nelan \& Walborn 2010, private communication]{nelan04}
found no evidence of a luminous astrometric companion in their {\it HST} FGS data, and
\citet{rauw09} detected no statistically significant radial velocity variations
($\sigma = 1.3$~km~s$^{-1}$ using seven absorption lines over six epochs).

As \citet{rauw09} point out, we are left with two plausible scenarios:
(i) a wide, massive binary that has evaded detection, or
(ii) magnetically confined wind shocks around a single magnetic O4 star.

Magnetically confined wind shocks are difficult to reconcile with strong non-thermal 
radio emission. Though non-thermal radio emission would be produced in magnetized
wind shocks, most of that emission would occur within a few stellar radii of the visible 
photosphere, well inside the radio photosphere.
At 8.6 GHz, the wind of the O4 giant is optically thick out to hundreds of stellar radii
\citep[e.g.,][]{debecker04}.

Given the similarity between the primary of HD~93403, O5~III(fc), and 
HD~93250, O4~III(fc), in both its optical and X-ray spectral properties,
and given the observed correlation between shock temperature and binary period (see Fig.\ 13 and \S 6.3),
we speculate that HD~93250 is a wide binary with an orbital period of 30-60 days.
If the primary also possesses a strong magnetic field, high-S/N X-ray grating spectra will
be needed to untangle the origin of the strong X-ray shocks.

Detections at two or more radio frequencies are needed to measure
HD~93250's spectral index, and to confirm its non-thermal radio emission.
In addition, spectro-polarimetric monitoring is needed to look for surface 
magnetic fields and measure the primary's rotation period.
Ground-based optical interferometry \citep[e.g.,][]{patience08} is needed
to look for a luminous companion inside the $\sim20$~AU limit imposed by the {\it HST} FGS.

\subsubsection{Tr16-22 = MJ~496, O8.5 V}

The single O8.5~V dwarf MJ~496=Tr~16~22 was detected with {\it Chandra} by
\citet{albacete-colombo08} and \citet{sanchawala07a} and with {\it XMM-Newton} by 
\citet{naze09} and \citet{antokhin08}.
\citet{morrell01} use Tr16-22 as a spectroscopic template for the O8~V secondary in
the colliding wind binary HD~93205.
Compared to the results in Table~3, \citet{albacete-colombo08} derive similar, though slightly lower,
$kT_1$, $kT_2$, $L_{\rm X}$, and $L_{\rm bol}$, and list Tr16-22 as a probable binary.
Similarly, \citet{evans03} single out Tr16-22 as a possible colliding wind binary. 
We note however that Tr16-22's $L_{\rm X}$ and $\log L_{\rm X}/L_{\rm bol}$
are higher than those of any late-O + late-O binary in Carina.
As such, Tr16-22 is a good candidate for magnetically confined wind shocks.
High-S/N spectro-polarimetric monitoring are needed to detect and measure
its magnetic geometry. 

\subsubsection{MJ~449 = CPD-59~2610 = LS~1865, O8.5 V((f))}

The last of the apparently single O stars with very hot X-ray shocks is MJ~449 = LS~1865, O8.5 V((f)).
We find $\log L_{\rm X} = 31.42\pm0.20$,
$\log L_{\rm X}/L_{\rm bol} = -6.90$,
$kT_1= 0.59_{-0.3}^{+0.1}$, and $kT_2 > 3.1$~keV,
with comparable emission measure in the two thermal components \citep{naze11}.
Though MJ~449 is not shown in Fig.~10, it is most likely time variable, with $P_{\rm KS} = 16\%$,
and three maximum likelihood blocks.
Like Tr16-22, MJ~449 is a candidate for magnetically confined wind shocks,
though its lower $\log L_{\rm X}/L_{\rm bol}$ suggests that other emission mechanisms
may be at work, including coronal emission from a PMS companion.

\subsection{Astrometric Binaries}

We first examine the orbital properties and radio emission of the four astrometric binaries detected 
with {\it HST}. \citet{nelan04} used a series of {\it HST} Fine Guidance Sensor observations (in high angular resolution mode 1r)
to look for luminous companions around O and early B stars in Tr~14 and 16. The observations were 
typically sensitive to binary separations as low as $\sim10$~mas, or $\sim23$~AU at $d=2.3$~kpc.
We consider the four O stars in Table~1 of \citet{nelan10}.
We note that Tr16-31=MJ 484, B0.5~V is a wide binary with 
a separation of $352\pm2$~mas ($\sim810$~AU); it is not detected in the CCCP.

\subsubsection{HD 93129A, O2~If$\ast$}

HD~93129A is the earliest O star in Carina, and one of the earliest known in the Galaxy.
The O3.5~V((f)) star HD 93129B is located $2.74\arcsec$ to the southeast
($6300$~AU at the distance of Carina) -- HD~93129A and B are clearly resolved in the {\it Chandra} data.
\citet{nelan04,nelan10} discovered a luminous companion,
HD~93129Ab, at a separation of  $0.053\arcsec\pm0.003\arcsec$
($\approx122$~AU at 2.3 kpc), and $\Delta V = 0.9\pm0.05$~mag.
\citet{nelan04} estimate the primary's mass to be
$M_{\rm Aa} \approx 110 M_{\odot}$, and the secondary's mass to be $M_{\rm Ab}\approx 70 M_{\odot}$,
implying an approximate spectral type o3.5~v for Ab, very similar to HD~93129B.
An unpublished analysis by \citet{maiz-apellaniz10} indicates that the Aa/Ab system may be approaching
periastron in $\sim2020$.

HD~93129A is a strong non-thermal radio source \citep{benaglia06}. 
\citet{benaglia10} resolve the emission at 2.37~GHz into what appears to be a bow shock structure,
consistent with a wind collision zone at the interface of Aa and Ab.
Wide, massive binaries like HD~93129A are sometimes strong non-thermal radio sources because
the wind collision zone is outside the radio photosphere of both stars \citep[c.f.,][]{debecker07}.

Although hard X-ray emission and non-thermal radio emission are usually associated with colliding wind shocks
in WR+O binaries, the {\it Chandra} ACIS-I data do not allow us to establish the location of the X-ray shocks within
the HD 93129A system. We note that the system's $\log L_{\rm X}/L_{\rm bol} = -6.85$ is higher than
typical single, early-O stars in Table 2, and the spectrum requires a 1.9~keV emission
component with about 12\% of the total emission measure.

A {\it Chandra} high-energy transmission grating spectrum of HD~93129A
shows broad, asymmetric line profiles, and moderately weak He-like forbidden lines of \ion{Si}{13} and \ion{Mg}{11}
that suggest that most of the X-rays are produced in wind shocks a few $R_{\star}$ from the photosphere
\citep{cohen11}. The most likely explanation for HD~93129A's X-ray and radio properties is 
that both mechanisms are operating: embedded wind shocks close to Aa and Ab, 
and colliding wind shocks between Aa and Ab.

\subsubsection{HD~303308, O4.5 V((fc))}

HD~303308 is the fifth most X-ray luminous O binary in Carina with a soft spectrum ($kT\approx 0.25$~keV) and
$\log L_{\rm X}/L_{\rm bol} \approx -6.89$. \citet{nelan10} find a binary separation of $15\pm2$~mas, corresponding to
$\sim 35$~AU at $d=2.3$~kpc, the closest of their five astrometric binaries. The apparent V magnitude difference was 
$\Delta V=1.0\pm0.3$. \citet{sota11} find a spectral type of O4.5 V((fc)).

From the standpoint of its X-ray emission, HD~303308 is quite
typical of other early-mid type O stars, though we cannot rule out some colliding wind emission
in the zone between the two stars. We note that HD~303308 showed some evidence
of time variability, with four maximum likelihood blocks, and $P_{\rm KS}=20\%$.

\subsubsection{Tr16-9 = MJ 481, O9.5 V}

\citet{nelan10} find a binary separation of $16\pm2$~mas for Tr16-9 = MJ 481, corresponding to
$\sim 35$~AU at $d=2.3$~kpc, the second closest of their five astrometric binaries.
MJ 481 shows low-level time variability, $\log L_{\rm X}/L_{\rm bol} = -7.17$,
and a higher-than-average $kT_1 = 0.70_{-0.1}^{+0.2}$~keV \citep{naze11}. Like HD~303308,
we cannot rule out some colliding wind emission.

\subsection{O+O Spectroscopic Binaries}

The O+O binaries listed in Table 3 are among the best studied in Carina and among them
HD 93403, HD 93205, HD 93343, Tr16-112, and QZ Car have been proposed as colliding wind binaries
\citep{rauw09,morrell01,parkin11}.

We note however the wide variety of post-shock temperatures in these systems in Table~3.
We find that the longer-period O+O spectroscopic binaries like
HD~93403 (15.09 d), QZ Car (20.72 d), and HD~93343 (44.15 d) have high $kT$ shocks,
and the shorter-period systems like HD~93205 (6.08 d), Tr16-112 (4.02 d), and FO 15 (1.41~d),
show only soft X-ray shocks. Figure~13 plots $kT_{\rm avg}$
versus orbital period for the 11 spectroscopic binaries in Table~3.
The orbital periods are from the compilations of \citet{rauw09,nelan10}.

For multiple systems like QZ Car, we use the orbital period of the primary
in Fig.\ 13, because the primary pair accounts for most of the X-ray luminosity.
For example, in QZ Car the A1/A2 pair with the O9.7~I primary accounts for $\sim 74\%$ of the system's
X-ray flux \citep[see Table 5,][]{parkin11}. For Tr16-110, O7~V + O8~V+ O9~V, a $2+1$ spectroscopic
binary, the primary and secondary orbital periods are 3.63 and 5.03 days, respectively \citep[Table~7,][]{rauw09}.
We use $P = 3.63$ in Fig.\ 13, though both periods are short; either period yields the same correlation.
Similarly, for Tr16-104, an eclipsing O7~V + O9.5~V binary bound to a B0.2~IV star, we use the 2.15-d primary
orbital period. The orbital period of the B0.2~IV star is either 285~d or 1341~d \citep{rauw01}. We ignore the
possible contribution of wind shocks with the distant B0.2~IV star.

Fig.\ 13 shows a correlation between orbital period and $kT$,
with a Pearson's linear correlation coefficient of 0.95.
The correlation hinges on the three long-period systems with $kT_{\rm avg} > 1$~keV:
HD~93403, QZ~Car, and HD~93343. 
A larger sample of O+O binaries is needed to confirm a correlation between
orbital period or binary separation and average shock temperature.
We note that the wide binaries $\eta$~Car with $P_{\rm orb}\approx 5.54$~yr
\citep{gull09}, and WR~25 (WN6h + O) with $P_{\rm orb}\approx 207.7$~d \citep{gamen08},
are consistent with this correlation.

If such a correlation is confirmed, then the presence of a close, luminous companion
(separation $\lesssim 5R_{\star}$) may significantly reduce the observed post-shock
temperatures, and thus the wind speed at the shock, whether those shocks are
produced throughout the wind (EWS) or in the wind collision zone (CWS).
Put another way, there may not be enough room to allow the winds to ramp up to terminal speed.
In systems like FO 15 with low $L_{\rm X}/L_{\rm bol}$ the emission measure and, 
hence, the density in the X-ray shocks may also be reduced.

Including the astrometric binaries in our analysis, it appears that, as the binary separation increases
beyond a few tens of AU, the colliding wind shocks become weaker, as expected from adiabatic expansion
\citep[e.g.,][]{antokhin04}, and the X-rays are dominated by embedded wind shock emission produced within
a few $R_{\star}$. We emphasize that for most O+O binaries, both mechanisms 
-- embedded wind shocks and colliding wind shocks -- may be at work.

\subsubsection{HD~93403, O5 III(fc) + O7~V}

The 15.093-d O5~III(fc) + O7~V eccentric ($e=0.234$) binary HD~93403 is the most X-ray active O+O binary in Carina,
with $\log L_{\rm X} \approx33.1$~ergs~s$^{-1}$ and $\log L_{\rm X}/L_{\rm bol} \approx -6.4$.
In the CCCP data, HD~93403 was observed in OBSIDs 9484, 9486 and 9891
\citep[for the observation log see Table 1 of][]{townsley11a}.
\citet{naze11} fit the combined ACIS-I spectrum with $\log N_{\rm H}^{\rm ISM} = 21.49$~cm$^{-2}$, and find
$N_{\rm H}^{\star} = 21.5_{-0.07}^{+0.06}$, $kT_1 = 0.62_{-0.01}^{+0.02}$ and $kT_2 = 1.6_{-0.05}^{+0.06}$,
with approximately equal emission measure in each temperature component.

We note that \citet{walborn10} recently reclassified the primary of HD~93403 as O5~III(fc)
on the basis of its \ion{C}{3} emission lines.
Using the common ephemeris of \citet{rauw00}, $P=15.093$~d and $T_0 = 51355.14$~MJD,
OBSIDs 9486 and 9491 occurred from phase $0.15$ to $0.28$, and 
OBSID 9484 occurred from phase $0.44$ to $0.48$, near apastron,
with similar exposure times in each phase range.
We note that throughout this study we use the {\it Chandra} convention
${\rm MJD}={\rm HJD} - 2400000.5$, rather than the more common convention
${\rm HJD} - 2450000.0$ to express $T_0$.
The lower left panel of Fig.\ 9 shows the sequenced light curve of HD~93403.

In the most comprehensive examination of HD~93403's X-ray light curve to date,
\citet{rauw02} use a series of {\it ROSAT} HRI and {\it XMM-Newton} MOS1 photometry and PN spectra
to show  a clear orbital modulation of its X-ray luminosity with orbital phase, with X-ray maximum 
occurring at $\phi = 0.0$ (periastron).
The lower mean count rate at the beginning of the HD~93403 light curve,
corresponding to $0.44 < \phi < 0.48$, suggests an X-ray minimum near apastron.

The short-term variability seen throughout the three {\it Chandra} observations,
though small, is statistically significant. The rapid variability amplitude varies from $5-15\%$,
and is comparable to the 20\% long-term variability amplitude reported by \citet{rauw02}.

\citet{rauw02} used a greater distance to Carina, $d = 3.2$~kpc,
and a higher ISM column density, $\log N_{\rm H}^{\rm ISM} = 21.56$~cm$^{-2}$.
They found a 0.5-2.5~keV $\log L_{\rm X} \approx 33.57$~ergs~s$^{-1}$.
Applying a distance $d = 2.3$~kpc yields $\log L_{\rm X} \approx 33.28$~ergs~s$^{-1}$ in the 0.5-2.5~keV band
for the mean of the four {\it XMM} observations.

Using \citet{rauw02}'s higher ISM column density, we find a CCCP 0.5-2.5~keV
$\log L_{\rm X} \approx 33.19$~ergs~s$^{-1}$ during the CCCP observations in 2008 August,
only 19\% lower than the {\it XMM-Newton} observations in 2000 December and 2001 February.
Overall the CCCP spectral and timing analysis of HD~93403 confirms the results of \citet{rauw02}.

\subsubsection{HD 93205, O3.5~V((f)) + O8~V}

\citet{morrell01} present the definitive study on HD 93205 from an extensive set of ground-based spectroscopic,
{\it IUE} far ultraviolet, and {\it ROSAT} X-ray observations. They find a $6.0803\pm0.0004$-d orbital period,
primary and secondary masses of $52-60 M_{\odot}$ and $22-25 M_{\odot}$,
an orbital inclination of about $55\arcdeg$, 
an orbital eccentricity $e=0.37$, an apsidal period of $185\pm16$~yr,
and phase-locked X-ray variability, with X-ray maximum near periastron.

HD~93205's CCCP light curve in Fig.\ 9 shows a slow steady decline over the 1 day observation,
which began and ended at MJD 53977.794 and 53978.817.
From the combined ephemeris in Table~4 of \citet{morrell01} in which $T_0 = 50498.589$~MJD,
OBSID 4495 began at orbital phase $\phi=0.21\pm0.04$ and ended at phase $\phi=0.38\pm0.04$,
during which the average X-ray parameters were $\log L_{\rm X} \approx 32.55$~ergs~s$^{-1}$,
$\log L_{\rm X}/L_{\rm bol} = -6.82$, $kT_1 = 0.23\pm0.01$, and $kT_2 =0.74\pm0.03$,
with most of the emission measure in the cooler component \citep{naze11}.

This is the end of the declining phase in the {\it ROSAT} HRI light curve in Fig.\ 8 of \citet{morrell01}.
Our results thus support the X-ray periodicity of \citet{morrell01}.
Its high $L_{\rm X}/L_{\rm bol}$ is similar to other binaries with early-O primaries.
As with the other short-period binaries,
the colliding wind shock temperatures do not exceed 1~keV.

\subsubsection{Tr16-112 = LS 1874 = CPD-59 2641 = MJ~535, O5.5-O6 V(n)((fc)) + B2~V}

\citet{rauw09} presented a detailed multiplicity study of three X-ray luminous Carina O stars
as part of the X-Mega project\footnote{http://lheawww.gsfc.nasa.gov/users/corcoran/xmega/xmega.html}: 
Tr16-112, HD~93343 and HD~93250.
Based on a long series of medium- and high-resolution spectra of Tr16-112,
they obtained spectral types of O5.5-6~V((f)) and B2~V, for the primary and secondary.
\citet{sota11} revised the primary spectral type to O5.5-O6 V(n)((fc)).

\citet{rauw09} find an orbital period $P = 4.0157$~d, eccentricity $e=0.15 \pm 0.01$,
inclination $i = 54\arcdeg^{+4}_{-3}$,
primary and secondary masses $M_1 = 41.0\pm2.5 M_{\odot}$ and $M_2 = 11.7\pm0.6 M_{\odot}$,
and periastron passage time $T_0 = 54559.810\pm0.060$ MJD.
They estimate that the primary fills $\frac{3}{4}$ of its Roche lobe.

In the CCCP data, Tr16-112 was observed in OBSID 6402 in 2006 August during which time
it showed no signs of variability, with $P_{\rm KS} = 53\%$ and only one maximum likelihood block
(95\% confidence). OBSID 6402 began and ended at MJD 53977.78 and 53978.83,
corresponding to orbital phases 0.06 and 0.32 in Tr16-112.
That is close to periastron passage, though we note that the orbit is not highly eccentric.
This short-period binary has $kT_{\rm avg}\approx0.35$~keV.

\subsubsection{HD~93343, O8~V + O7-8.5~V}

\citet{rauw09} present the first comprehensive analysis of the double-lined spectroscopic binary HD~93343.
They find a set of sharp lines associated with the less massive O8~V secondary moving in opposite
phase to the higher-mass O8~V primary. They note that the spectral type of the secondary could be in the
range O7-O8.5~V. They find a period of $\approx 44.15$ days, though they were not able to establish
an absolute ephemeris.

For HD~93343 we note
$\log L_{\rm X} = 31.66\pm0.24$,
$\log L_{\rm X}/L_{\rm bol} \approx -6.98$,
$kT_1 = 0.70_{-0.09}^{+0.13}$, and $kT_2 > 6.5$~keV,
with comparable emission measure in each temperature component.
HD~93343 showed no evidence of time variability in OBSID 6402.

\citet{antokhin08} found $kT_1 = 0.29^{+0.04}_{-0.05}$, $kT_2 = 2.8^{+1.5}_{-1.4}$, and
$\log L_{\rm X}\approx32.2$, a factor of 3 more luminous in the 0.4-10~keV band with {\it XMM} 
than was found in the same band with {\it Chandra} \citep{naze11}, with most of the
added luminosity in the {\it XMM} fit in the softer component.

Further optical spectroscopic monitoring is needed to establish a firm ephemeris for HD~93343.
Further {\it Chandra} observations would provide a better estimate of the shock temperatures,
and a third estimate of $L_{\rm X}$ to confirm the variability. An X-ray maximum near
periastron would signal a colliding-wind binary. Spectropolarimetric monitoring could
reveal the presence of large-scale magnetic fields on one or the other star.

For now, we consider HD~93343 a CWS X-ray source, so we include it in Fig.\ 13.
We note though that the hard shocks seen on HD~93343 with {\it Chandra} are not predicted
by \citet{pittard09} in their colliding wind binary model {\sc cwb3},
an O6~V + O8~V binary with a 10.7-d period, the closest model to HD~93343.
The predicted temperature distributions are somewhat consistent with the
\citet{antokhin08} result, however.

\subsubsection{The Short-Period Eclipsing Binary FO~15, O5.5 Vz + O9.5 V}

The double-lined eclipsing binary FO~15 was first typed by \citet{forte81} and
studied in detail by \citet{niemela06}. Its circular period is $1.413560 \pm 0.000003$~d
with $T_0 = 53159.0\pm0.2$ MJD, $i\approx80\arcdeg$, and primary and secondary masses
$M_1 = 30.4\pm1.0 M_{\odot}$ and $M_2 = 15.8\pm1.0 M_{\odot}$.
The stellar radii are $R_1\approx7.5 R_{\odot}$ and $R_2\approx 5.3 R_{\odot}$ and
the Roche lobe radii are just $8.3 R_{\odot}$ and $6.15 R_{\odot}$.
The semi-major axis is $a = 19.1 R_{\odot}$, implying the photospheres are separated by $\sim3 R_1$.

The ``z" spectral type indicates strong \ion{He}{2} $\lambda 4686$ absorption suggesting the primary is on the ZAMS
\citep[c.f.,][]{walborn97}. Located in the south pillars \citep[see][]{povich11b}, FO~15 is a ZAMS-age
O+O binary, with a separation of only $3 R_{\star}$, in a region of active star formation.
Moreover, \citet{niemela06} note the absence of wind features in both stars at most phases.
They suggest that both O dwarfs may be weak-wind stars \citep{martins05b}.

The top left panel of Fig.\ 9 shows the sequenced light curve of FO~15, showing a steady decline from
the beginning of the first observation through the end of the second. 
OBSID 6402 began and ended at MJD 53977.78 and 53978.83,
corresponding to orbital phases $\phi=0.231\pm0.001$ and $0.978\pm0.001$.
OBSID 9483 began and ended at MJD 54706.97 and 54707.68, 
corresponding to orbital phases $\phi=0.083\pm0.002$ and $0.591\pm0.002$.
Note that OBSIDs 6402 and 9483 were separated by nearly 729 days, i.e., 515 orbital cycles.

FO~15 is noteworthy because it has the lowest $L_{\rm X}/L_{\rm bol}$ of any O binary in the CCCP:
$\log L_{\rm X}/L_{\rm bol} = -7.65$. It is relatively soft with $kT=0.5\pm0.3$~keV, yet it is clearly time variable.
It is very young, it has the shortest orbital period in our sample, and it is one of only four weak-wind systems in Carina \citep[see also][]{naze11}.

The soft X-ray spectrum of FO~15 (indicating low-velocity shocks) and
low $L_{\rm X}/L_{\rm bol}$ (indicting low shock densities) suggest that in very close,
short-period binaries the winds may not have reached $v_{\infty}$ when they collide,
or alternatively, the luminous companion may inhibit wind acceleration and mass-loss
via radiative braking \citep{gayley97}
as a result of non-radial radiative forces
(from the distorted shape of the star).

\subsection{B Stars}

B stars provide a transition in the X-ray properties of massive stars.
Early B stars are often detected in X-rays, though their detection fraction
is typically less than 50\%. Like most late-O stars,
B stars are often soft, low $L_{\rm X}/L_{\rm bol}$ sources \citep{cohen97,evans03}, but
in some cases they are particularly hard and strong X-ray sources
(e.g. $\tau$~Sco, B0.2~V), which has a complex magnetic geometry \citep{donati06b}.
The Carina Nebula Complex objects provide an excellent sample to explore this interesting spectral region further.

\citet{evans11} examine a group of photometrically selected, candidate late-B stars in Tr~16.
For this group, \citet{evans11} find no stars with $\log L_{\rm X} > 31$ and suggest that the X-ray
detected late-B stars in Tr~16 probably harbor unseen, lower-mass,
late-type pre--main sequence companions.

With regard to the B stars, the main results of this study are that
(i) many B stars with moderate X-ray fluxes ($\log f_{\rm X} < -14$~ergs~cm$^{-2}$~s$^{-1}$,
corresponding approximately to $\log L_{\rm X} < 31$~ergs~s$^{-1}$) probably
harbor unseen, coronal PMS companions, and 
(ii) that most of the B stars with very high X-ray fluxes ($\log f_{\rm X} > -14$~ergs~cm$^{-2}$~s$^{-1}$,
corresponding approximately to $\log L_{\rm X} > 31$~ergs~s$^{-1}$, {\em cannot} be
accounted for by unseen, coronal companions.

Because their winds are too weak to produce much
embedded wind shock emission, these active B stars may be similar
to the magnetic B stars like $\tau$~Scorpii, B0.2~V \citep{cohen03, donati06b, petit11}
or the He-strong B2~Vp stars like $\sigma$~Ori~E \citep{shore90,sanz-forcada04,townsend05}.

Thus, the X-ray active B stars at the top of Table~4 merit detailed follow-up ground-based
spectroscopic, photometric, and spectro-polarimetric monitoring. Below we discuss the two most remarkable
B stars in the CCCP survey in more detail.

\subsubsection{The Candidate Herbig Be Star SS73 24}

The most X-ray active B star in Carina is the emission-line star Hen~3-485 = Tr 16 MJ 640 = SS73~24 = WRA~642,
with $\log L_{\rm X} = 31.70$ and $\log L_{\rm X}/L_{\rm bol} = -5.65$.
SS73~24 was classified by Sanduleak and Stephenson (1973)
as Be! pec,  remarking in the footnote to their Table 1, ``Our plates show numerous weak emission lines
some of which appear to be \ion{Fe}{2} or [\ion{Fe}{2}]."
SS73 24 is located close to the Treasure Chest Cluster 
\citep{smith05}
in the southern portion of the Carina Nebula, 
and is a strong infrared emission source (see Fig.\ 14).

Based on its emission lines, spectral type, and strong infrared excess, Miroshnichenko (2007) classified
Hen~3-485 = SS73~24 as a dust-forming B[e] star, and listed it as a candidate FS~CMa star. He proposed, 
``that these objects are binary systems that are currently undergoing or have recently undergone a phase
of rapid mass exchange, associated with a strong mass loss and dust formation.
A new name, FS CMa stars, and classification criteria are proposed for the unclassified B[e] stars."
Based on its X-ray and infrared emission, and its association with the star-forming Treasure Chest cluster,
we show that SS73~24 is most likely a young, accreting Herbig Be star, not an evolved B[e] binary.

In Fig.\ 14a we show the UBV/JHK/IRAC/MIPS $0.3-24~\mu$m SED of SS73~24,
though we note that the Spitzer-Vela Survey IRAC $3.5-8~\mu$m
photometry is mildly saturated. Moreover, its $U-B=-0.71$ and $B-V=0.70$
colors indicate relatively strong emission-line veiling, which we do not model.

The UBV/JHK/MIPS photometry were fit using the
\citet{robitaille06} YSO models. The SED fitting method described by \citet{povich11b}
generates a set of best-fit models from a large grid of star/disk/envelope models.
The set of most likely models yields the following parameters:
$L_{\rm bol} = 6\,000 \pm 2\,000 L_{\odot}$, $T_{\rm eff} = 25\,000\pm 1\,400~{\rm K}$, and
$A_V = 2.0\pm0.2~{\rm mag}$, consistent with a 9-11 $M_{\odot}$ B2 dwarf, a warm accretion disk,
and a cooler circumstellar envelope (blue, green, and red curves, respectively, in Fig.\ 14b).

SS73~24 was observed for 69~ks during OBSID 6578 and 9483, separated by $\sim 400$ days.
It showed no strong short-term variability during either observation, and no long-term variability
between observations. Its grouped ACIS spectrum was fit with a single-temperature,
solar-abundance APEC emission model, and two {\sc tbabs} absorption components:
one fixed at $N_{\rm H}=3.2\times10^{21}~{\rm cm}^{-2}$ representing ISM absorption with $A_V = 2.0~{\rm mag}$,
the other column density (a free parameter) to represent circumstellar absorption from the disk/envelope.
We note that SS73~24 is the most X-ray luminous B star in Carina with $\log L_{\rm X}\approx 31.7$~ergs~s$^{-1}$,
and one of the hardest with $kT\sim 3~{\rm keV}$, comparable to only a few other known Herbig Be stars
\citep{hamaguchi00, zinnecker94, stelzer06}.

It should be noted that the origin of hard X-rays on Herbig~Be stars
is not known. \citet{vink05} show that the flaring X-ray Herbig Be star MCW~297 has an astrometric binary
companion (a lower-mass PMS star) that may the source of the hard X-ray flare.
That said, SS73~24 is not flaring in the CCCP data and its $L_{\rm X}$ is higher than any other non-flaring
PMS star in the CCCP.
 
\subsubsection{HD~93501, B1.5~III:}

The B1.5~III: star HD~93501=Coll 228 96 was detected by \citet{levato90} as SB2, though no orbital
parameters were published. The star is noteworthy because of its extremely hard X-ray spectrum,
that we model as very hot thermal plasma ($kT > 6$~keV, 90\% confidence lower bound),
with little excess column density.
It's spectrum can also be fit a power-law with index $\Gamma = 1.3\pm 0.2$.
Assuming HD~93501 is a member of the Collinder 228 cluster at $d = 2.3$~kpc, its
$\log L_{\rm X}\approx 31.1$~ergs~s$^{-1}$. It was observed in three {\it Chandra} OBSIDs
with no indication of long- or short-term variability. HD~93501's very hard spectrum,
high $L_{\rm X}$, and lack of variability is not typical of coronal PMS stars.
HD~93501's X-ray spectrum is unique among the 78 OB stars with more than 50 counts.

\section{Summary and Discussion}

We have assembled an optical/X-ray catalog of the 200 known O and early-B stars in the 
{\it Chandra} Carina Complex Project to study the X-ray emission mechanisms on young, massive stars.
Like \citet{naze11}, we find that most O stars in Carina approximately follow the well-established
relation between $L_{\rm X}$ and $L_{\rm bol}$, and generally have soft X-ray spectra,
with characteristic temperatures $kT < 0.8$~keV. 

There are numerous examples of O and B stars, though, that do not follow this trend.
We examine their X-ray properties with an eye to classifying their primary underlying 
X-ray emission mechanism as
(1) embedded wind shocks,
(2) colliding wind shocks,
(3) magnetically confined wind shocks,
(4) low-mass, coronal PMS companions, or
(5) some other mechanism, possibly related to magnetic fields.

Fig.\ 4 provides a good overview of the X-ray/optical data: 
$\approx\frac{2}{3}$ of the O and early-B stars lie
on the cool side of the $L_{\rm X}/L_{\rm bol}$ versus $kT$ diagram.
Most of these stars have X-ray properties consistent with embedded wind shock sources,
with a characteristic average $kT\approx0.5$~keV and $\log L_{\rm X}/L_{\rm bol} \approx -7.2$.

Some of the well-known spectroscopic binaries in Table~3 also lie on the cool side of Fig.\ 4a,
suggesting perhaps a universal X-ray emission mechanism for all O stars.
However, Fig.\ 13 suggests a different trend. It shows a correlation between
orbital period and $kT$, with a Pearson's linear correlation coefficient of 0.95.
The correlation, which hinges on the longest-period binaries, requires confirmation
using a larger sample of binaries.

If the trend is confirmed, we suggest that in short-period binaries like HD~93205 and FO~15,
the winds have not reached $v_{\infty}$ when they collide or, alternatively, the luminous companion
may inhibit wind acceleration via radiative braking or some other mechanism \citep{gayley97}.
In some very wide systems like HD~93129A, O2~If*, it appears that embedded wind shocks
(close to the O2 photosphere of Aa) and colliding wind shocks in the far wind collision
zone between Aa and Ab are both at work.

Then there is the curious case of HD~93250, the O4~III(fc) star.
We argue in \S6.2.1 \citep[see also][]{rauw09} that HD~93250's strong X-ray and radio emission are
evidence for colliding wind shocks. Given the $kT$ versus $P_{\rm orb}$ trend seen in Fig.\ 13, we suggest that
it could be an O+O binary with a period $>30$ days, and/or a magnetic O star, and that further observations
are needed: multi-frequency radio interferometry to measure its spectral index, ground-based optical interferometry
to look for a luminous companion, and optical spectro-polarimetry to look for surface magnetic fields.

We are left with two additional candidate magnetic O stars with high $L_{\rm X}/L_{\rm bol}$ and $kT$:
Tr16-22 = MJ~496, O8.5~V, and CPD-59~2610 = MJ~449 = LS~1865, O8.5~V((f)).
So far, magnetic fields have not been reported on any Carina OB stars.

Among the B stars, Fig.\ 8, shows a large population of B stars, some X-ray detected, some not,
whose photon fluxes and upper limits are consistent with a distribution of unseen, low-mass, coronal PMS stars.
Superposed on this distribution is a group of B stars whose high-$L_{\rm X}$ cannot be explained by a
distribution of ordinary coronal, PMS companions. The situation here is reminiscent of the X-ray active
Herbig Be stars: either their companions are hyperactive compared to other low-mass PMS stars, or there is
some intrinsic magnetic mechanism in these B stars. The dozen or so B stars near the top of Table~4 could,
for example, be related to the magnetic B stars like $\tau$~Sco, or the He-strong Bp stars.

The B star with the highest $L_{\rm X}$ and the OB star with the highest $L_{\rm X}/L_{\rm bol}$, 
SS73~24, is located adjacent to the pillar in the Treasure Chest Cluster (see Fig.\ 15), and it shows a prominent 
mid-infrared excess (Fig.\ 14). We model its $0.3-24~\mu$m SED as an accretion disk + envelope + B2 photosphere.
SS73~24 appears to be the first Herbig Be star in Carina.

\acknowledgments
This research has made extensive use of the SIMBAD and {\it Vizier} databases, operated at CDS, Strasbourg, France.
This work was partially supported by SAO/{\it Chandra} X-ray Observatory grants GO8-9014X and G09-0019A (PI: Gagn\'e).
The CCCP is supported by {\it Chandra} X-ray Observatory grant GO8-9131X (PI: Townsley) and by the
ACIS Instrument Team contract SV4-74018 (PI: Garmire), issued by the {\it Chandra} X-ray Center, which
is operated by the Smithsonian Astrophysical Observatory for and on behalf of NASA under contract NAS8-03060.
AFJM is grateful to NSERC (Canada) and FQRNT (Qu\'ebec) for financial assistance.
NRW acknowledges the support of the Space Telescope Science Institute, operated by the Association of Universities for Research in Astronomy, Inc., under NASA contract NAS5-26555.
MSP is supported by an NSF Astronomy and Astrophysics Postdoctoral Fellowship under award AST-0901646.
The authors wish to thank the anonymous referee for many helpful suggestions.

{\it Facilities:} \facility{CXO (ACIS)}.

\appendix
\section{Appendix: {\it Chandra} Spectral Analysis of HD 93250}

In addition to the CCCP ACIS-I observation of Tr14 (OBSID 4495), HD 93250 was observed far off-axis in seven additional
grating observations of $\eta$~Car. Even though HD~93250 is located $\sim 7.5\arcmin$ NNW of $\eta$~Car,
the satellite roll angle placed HD~93250 on one of the ACIS-S chips during many of the $\eta$~Car monitoring
observations. Even though the PSF is quite large at $7.5\arcmin$ off-axis, HD~93250 is the only bright source there.
Pileup was mitigated by the effects of comatic aberration and vignetting at those off-axis angles, and the reduced
effective area in zeroth order. The {\it Chandra} observation log is shown in Table 5.

The data were reduced in the usual way \citep[see][]{broos11a} and spectra were extracted using the {\it CIAO}
tool {\tt psextract} using a $12\arcsec$ source circle, and a nearby source-free background region, away from 
the dispersed HEG and MEG spectra of $\eta$~Car. The spectra were grouped using the {\it ACIS Extract} tool 
{\tt ae\_group\_spectra}. The ACIS-S spectra were merged with the {\it CIAO} tool {\tt acisspec} 
for a combined exposure time of 238~ks (see Table~5).
We note that \citet{evans04} performed a similar wind-absorption analysis of the OBSID 6402 ACIS-I 
spectrum of HD~93250, and found two temperature components at 0.55 and 2.33~keV.

To exploit the high S/N of the merged ACIS-S spectrum, we fit a rather complex XSPEC spectral model.
As before, we modeled the cold, neutral ISM with {\sc tbabs}, freezing $\log N_{\rm H}^{\rm ISM} = 21.48$.
Assuming that some of the emission arises from embedded wind shocks close to HD~93250, and that the harder emission
forms in colliding wind shocks where the wind is less dense, we used a {\sc windtabs} absorption component 
\citep{leutenegger10} for the cool emission component. The main advantage of {\sc windtabs} over {\sc tbabs} is that 
(i) it is not a simple slab model -- the emission and absorption is distributed radially throughout the wind
via a specified onset shock radius and wind velocity law, and
(ii) the absorption model accounts for the ionization of the wind.
The only free parameter in {\sc windtabs} is the characteristic mass column in g~cm$^{-2}$,
\[ \Sigma_{\star} = \frac{\dot{M}}{4 \pi R_{\star} v_{\infty}}. \]

To estimate the distribution of absorbing material in the wind of HD 93250 we adopt wind parameters
$R_0 = 1.5 R_{\star}$ (the onset radius for X-ray shocks) and $\beta=0.7$ (the wind velocity law)
that are consistent with recent X-ray line-profile measurements of O supergiants
\citep{cohen10, cohen06}. We note that a direct finite-disk corrected CAK \citep{castor75}
model with fixed $\bar{Q}$ and $\alpha$ also produces $\beta\approx0.75$.

For the APEC emission model and the {\sc windtabs} and {\sc tbabs} absorption models we adopt the solar abundance 
values of Asplund et al. (2009). Though we expect differing levels of CNO processed material in the atmospheres and 
winds of the Carina O stars, we note that the CNO lines and absorption edges occur longward of 20 \AA\ where the
{\it Chandra} ACIS-I spectra are extremely weak. The ACIS spectra are truncated below 0.5 keV (above 25 \AA).

The best-fit {\sc tbabs(windtabs(apec)+apec+apec)} model is shown in Fig.\ 11 in red. 
The ACIS-I and ACIS-S spectra were then fit individually to measure $L_{\rm X}$ during each OBSID.
In Fig.\ 12, we merged the last three spectra, OBSID 7341, 7342 and 7189 from 2006 June.

The best-fit mass column $\Sigma_{\star}=0.03\pm0.01$~g~cm$^{-2}$ yields a time-averaged
ISM-corrected X-ray luminosity $\log L_{\rm X} = 33.18$. We note that this is very close to 
the value found by \citet{naze11} fitting only the ACIS-I data without {\sc windtabs}, $\log L_{\rm X} = 33.19$.
The benefit of using {\sc windtabs} is that it provides a physically meaningful estimate of the wind mass column.

Using a terminal wind speed $v_{\infty}=3\,000$~km~s$^{-1}$ \citep{martins05b} and a radius $R=17.7 R_{\odot}$ 
implied by $\log L_{\rm bol}/L_{\odot} =  5.95$ and $\log T_{\rm eff} = 4.62$, the best-fit mass column yields
a mass-loss rate $\dot{M}=1.4\pm0.5\times10^{-6} M_{\odot}$~yr$^{-1}$.

\clearpage

\begin{figure}
\includegraphics[scale=0.7]{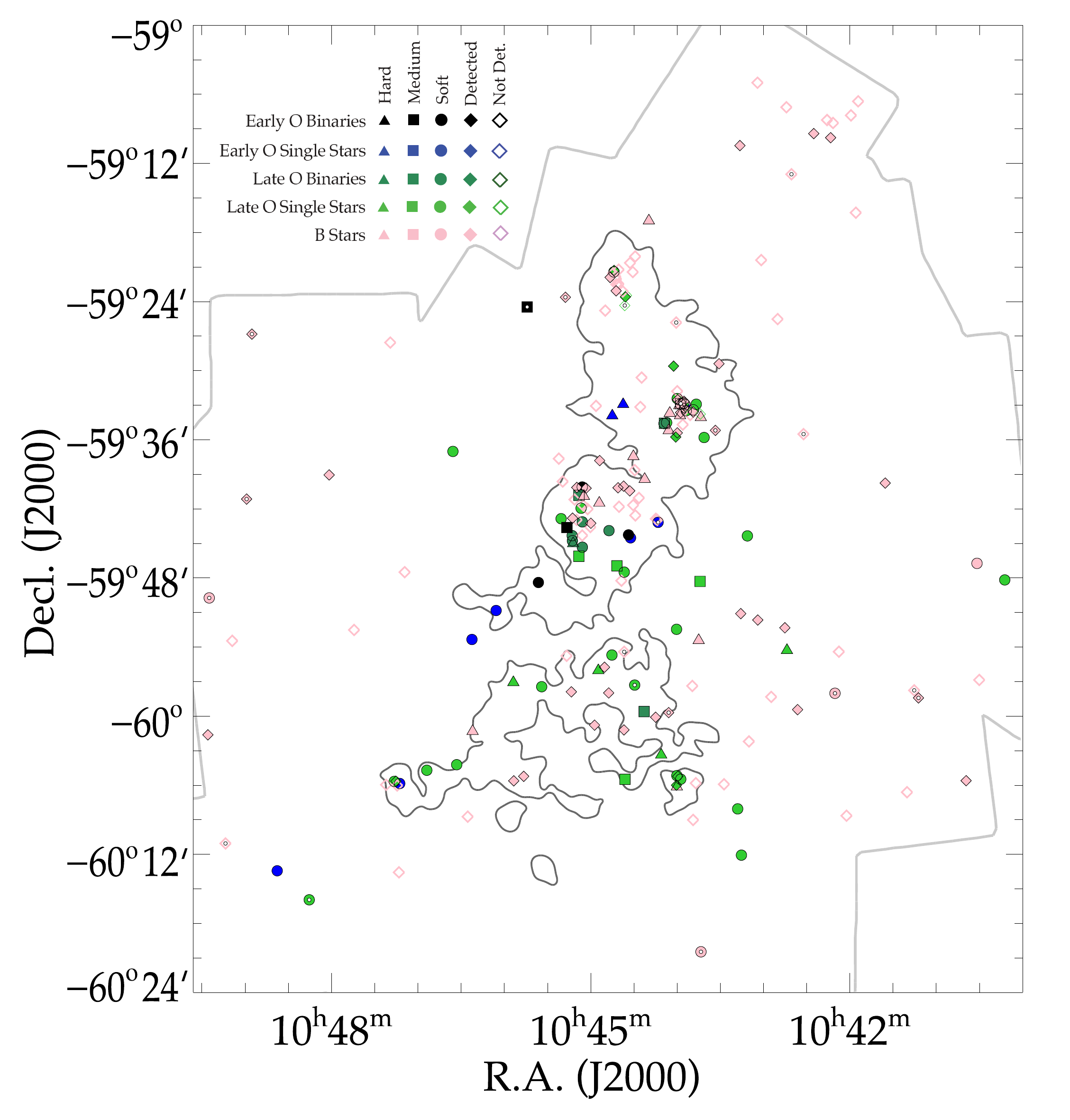}
\caption{
Spatial distribution of the 200 OB stars with determined spectral types in Carina.
Stars detected in the CCCP are shown as filled symbols.
Early-O binaries are black, early-O single stars are blue, late-O binaries are dark green,
late-O single stars are light green, and B stars are pink.
For the 78 OB stars with more than 50 counts
hard, medium, and soft X-ray stars are shown as triangles, squares, and circles, respectively.
The 51 X-ray detected stars with fewer than 50 counts 
are shown as filled diamonds. The open diamonds represent undetected stars (mostly B stars, in pink).
The grey outline shows the boundaries of the CCCP and the contours show boundaries of the large-scale
clustering of low-mass stars, reproduced from Figure~1 of \citet{feigelson11}.
See \S4 for symbol definitions.
}
\label{f1}
\end{figure}

\begin{figure}
\includegraphics[scale=0.75]{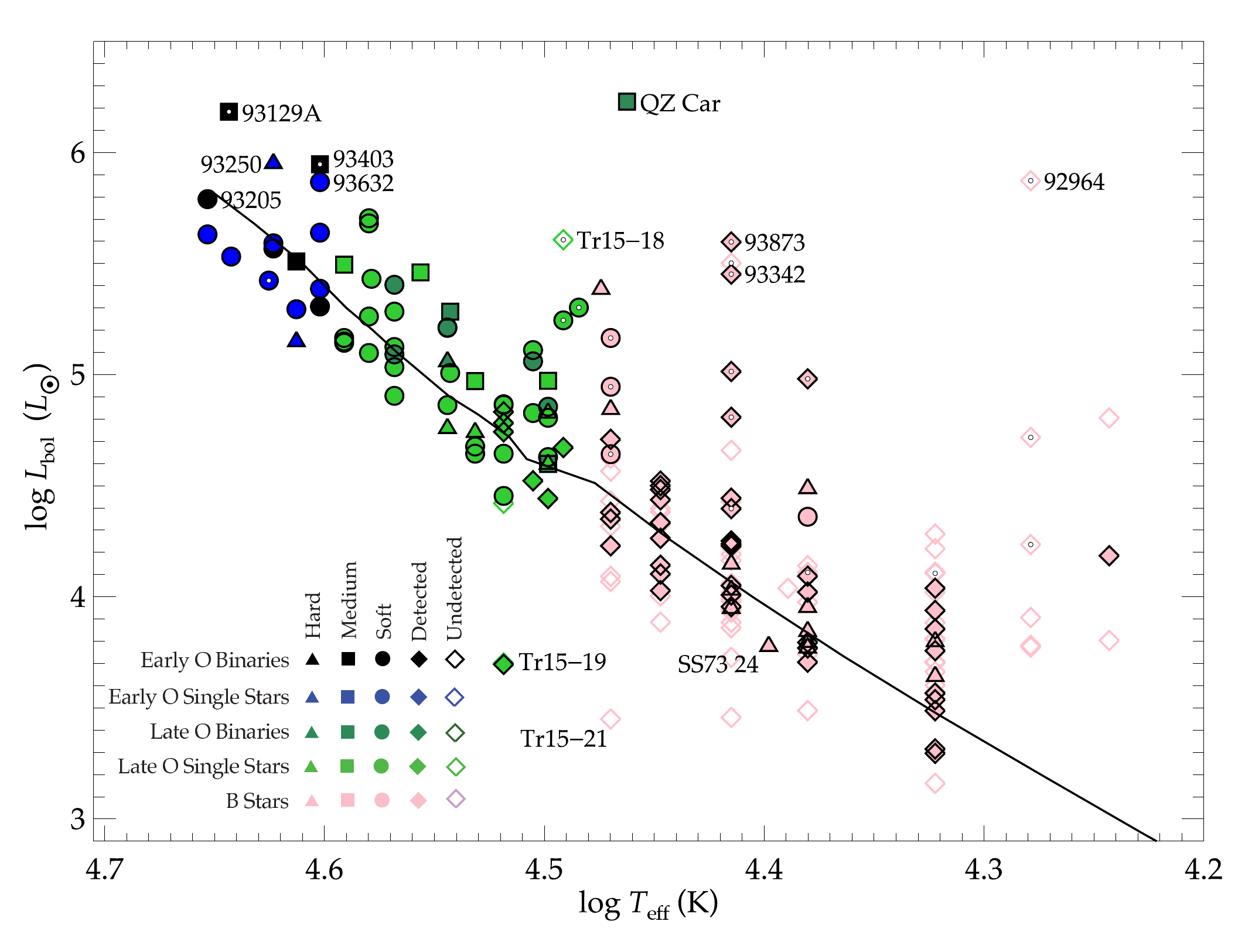}
\caption{ 
HR diagram of the 200 OB stars in the CCCP field of view. See Fig.~1 and \S4 for symbol definitions.
The most luminous stars are labeled.
}
\label{f2}
\end{figure}

\begin{figure}
\includegraphics[scale=0.6]{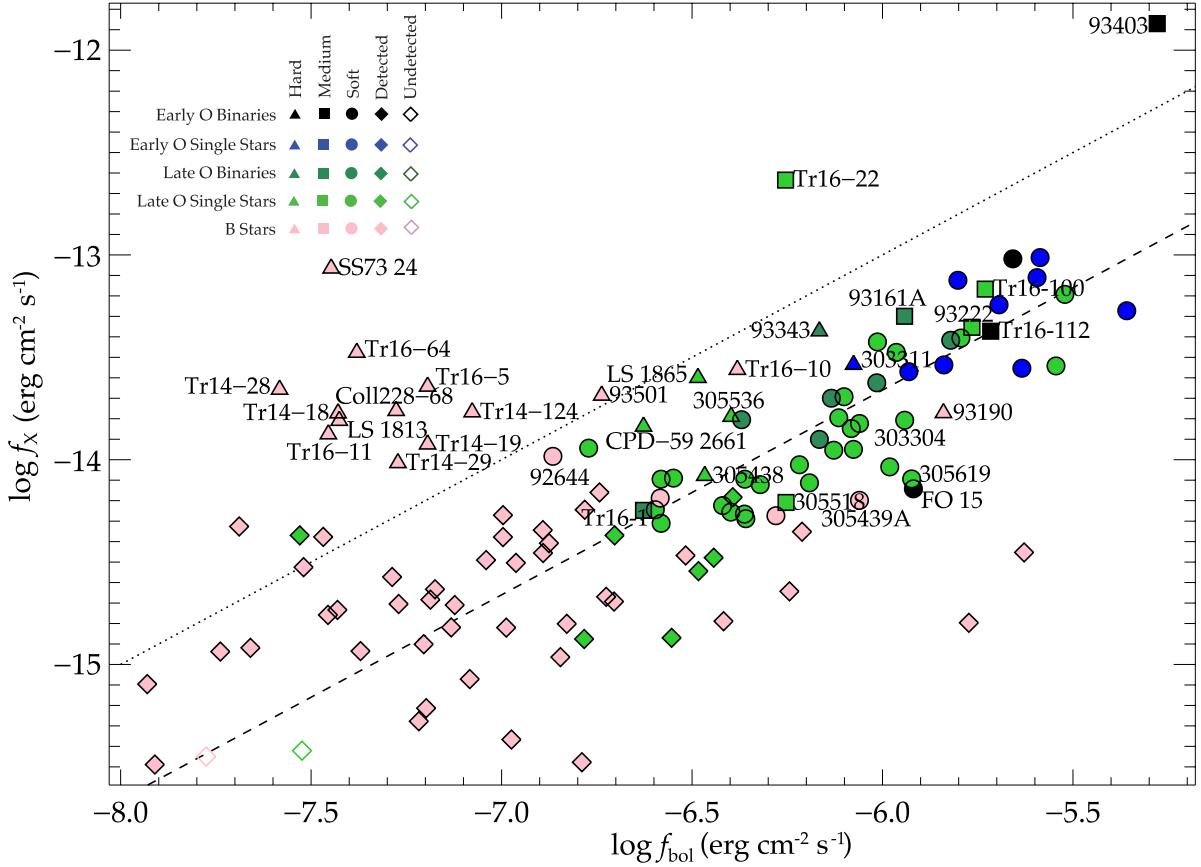}
\caption{ 
0.5-8.0 keV X-ray flux at Earth versus bolometric flux for the full sample of X-ray detected OB stars.
See Fig.~1 for symbol definitions. Note: the ACIS-I CCCP survey data of HD~93250, HD~93129A, QZ~Car, and HD~93205
were piled up; $f_{\rm X}$ was not computed in {\em ACIS Extract} for these four bright stars.
The upper dotted and lower dashed lines represent
$\log f_{\rm X}/f_{\rm bol} = -7$ and $-7.23$, respectively.}
\label{f3}
\end{figure}

\begin{figure}
\includegraphics[scale=0.55]{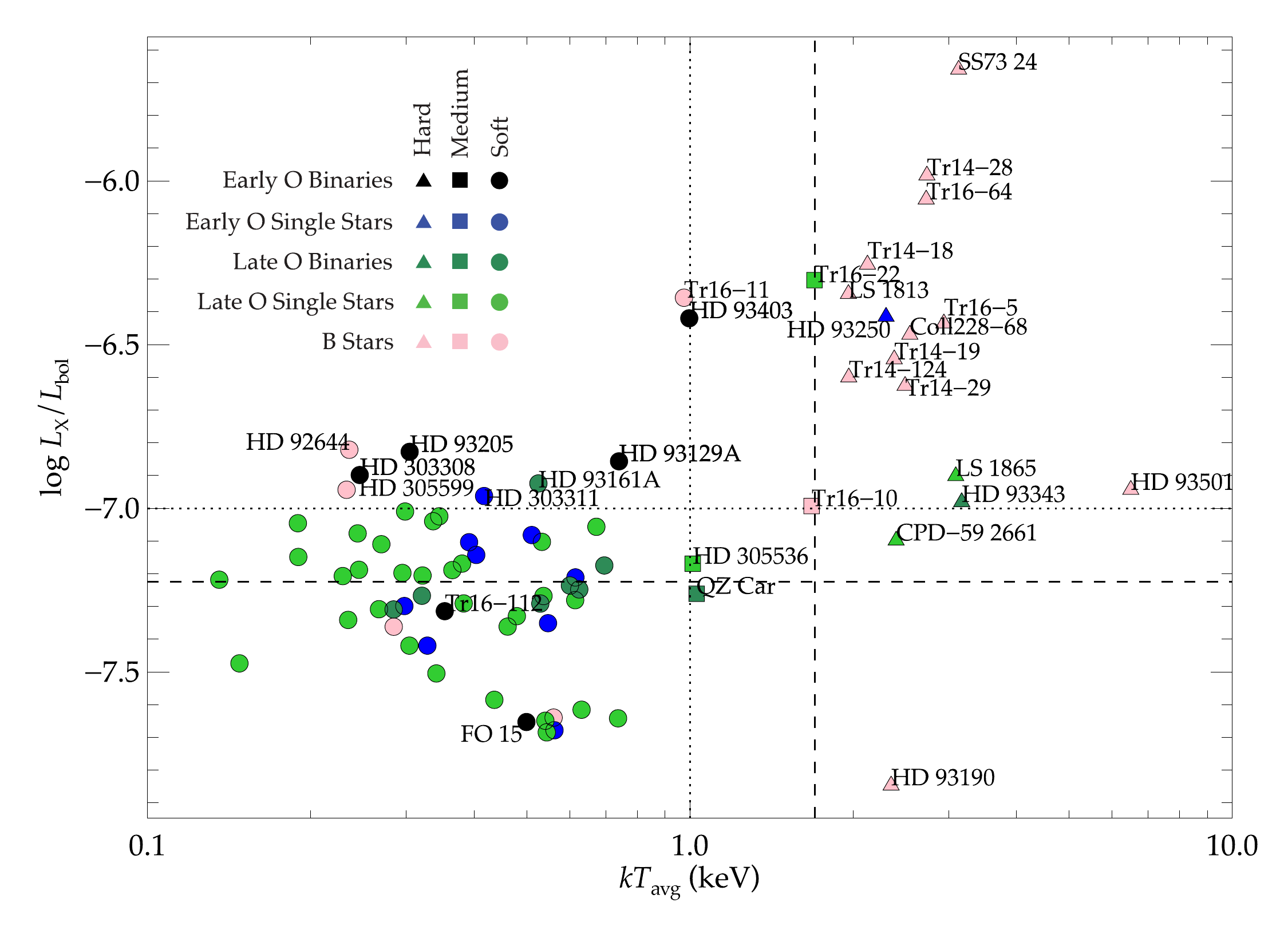}
\includegraphics[scale=0.55]{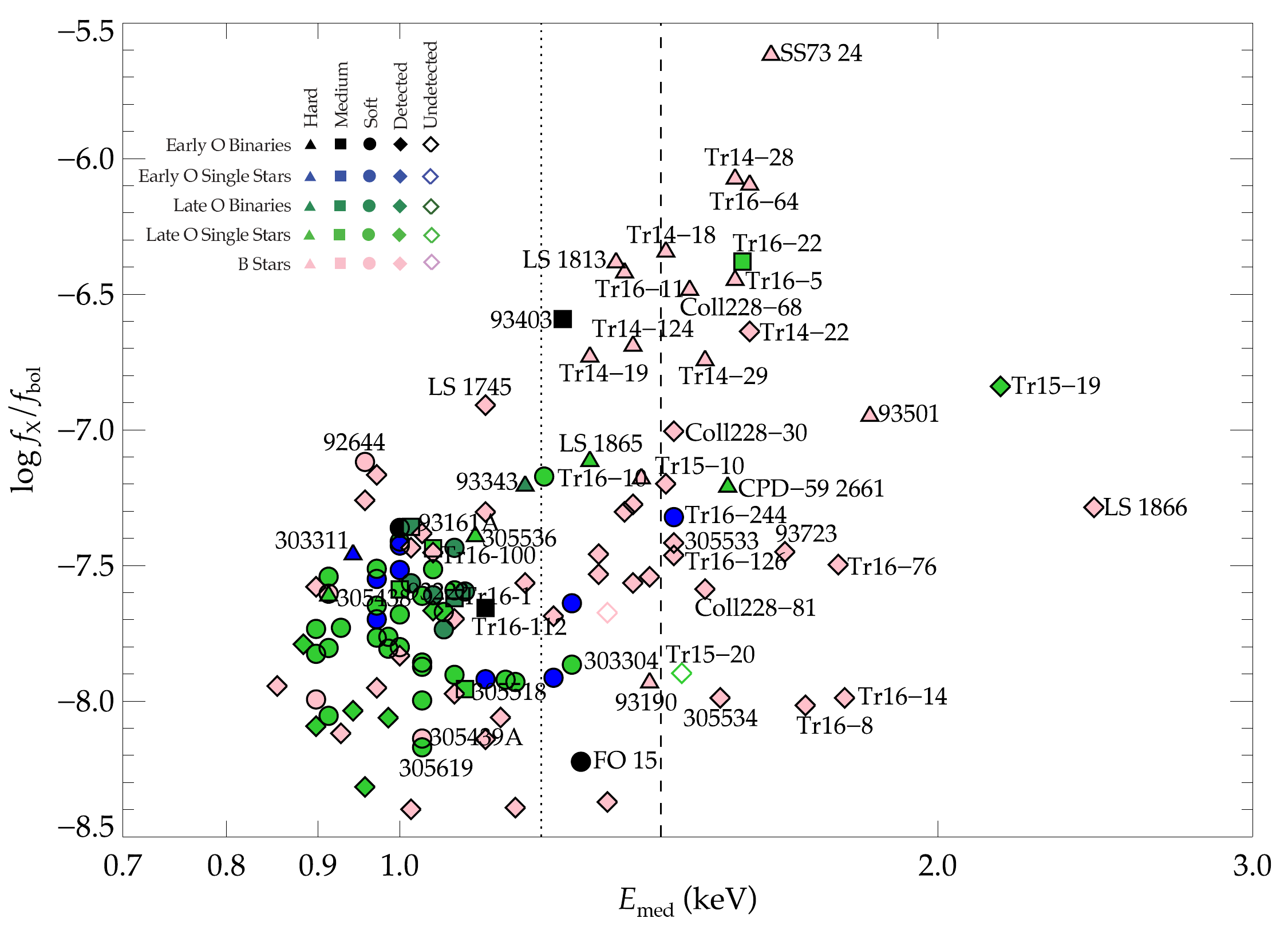}
\caption{
{\it Top panel:} $\log L_{\rm X}/L_{\rm bol}$ versus $kT_{\rm avg}$ for the 78 OB stars with XSPEC parameters and more than 50 net ACIS counts. See Fig. 1 for symbol definitions. 
The upper dotted and lower dashed lines represent
$\log L_{\rm X}/L_{\rm bol} = -7$ and $-7.23$, respectively. For stars with $kT_{\rm avg} < 1$~keV,
the mean of $\log L_{\rm X}/L_{\rm bol}$ is $-7.23$.
{\it Bottom planel:} $\log f_{\rm X}/f_{\rm bol}$ versus $E_{\rm med}$ for the full sample of X-ray detected OB stars. See Fig.~1 for symbol definitions.
}
\label{f4}
\end{figure}

\begin{figure}
\includegraphics[scale=0.44]{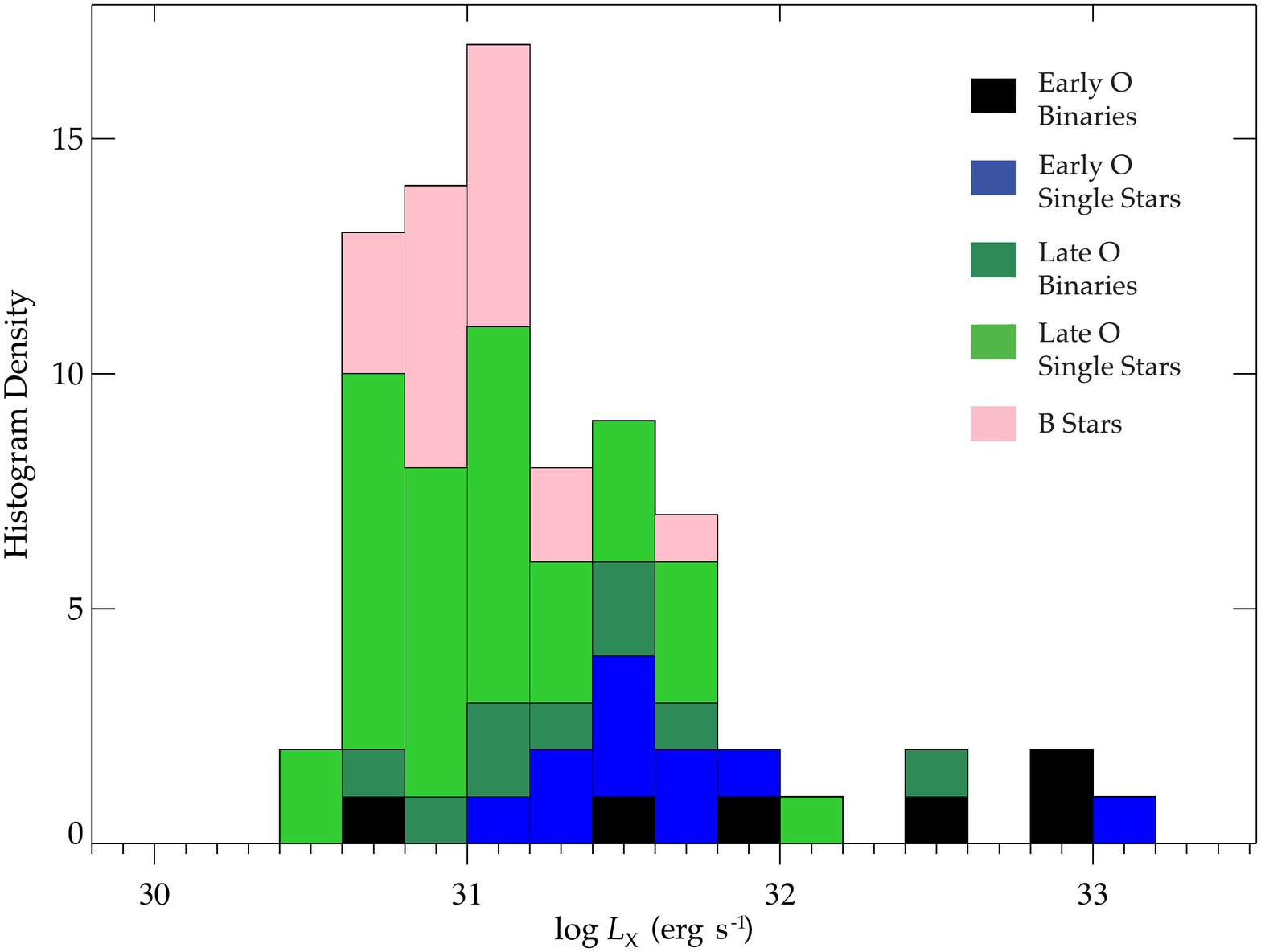}
\includegraphics[scale=0.44]{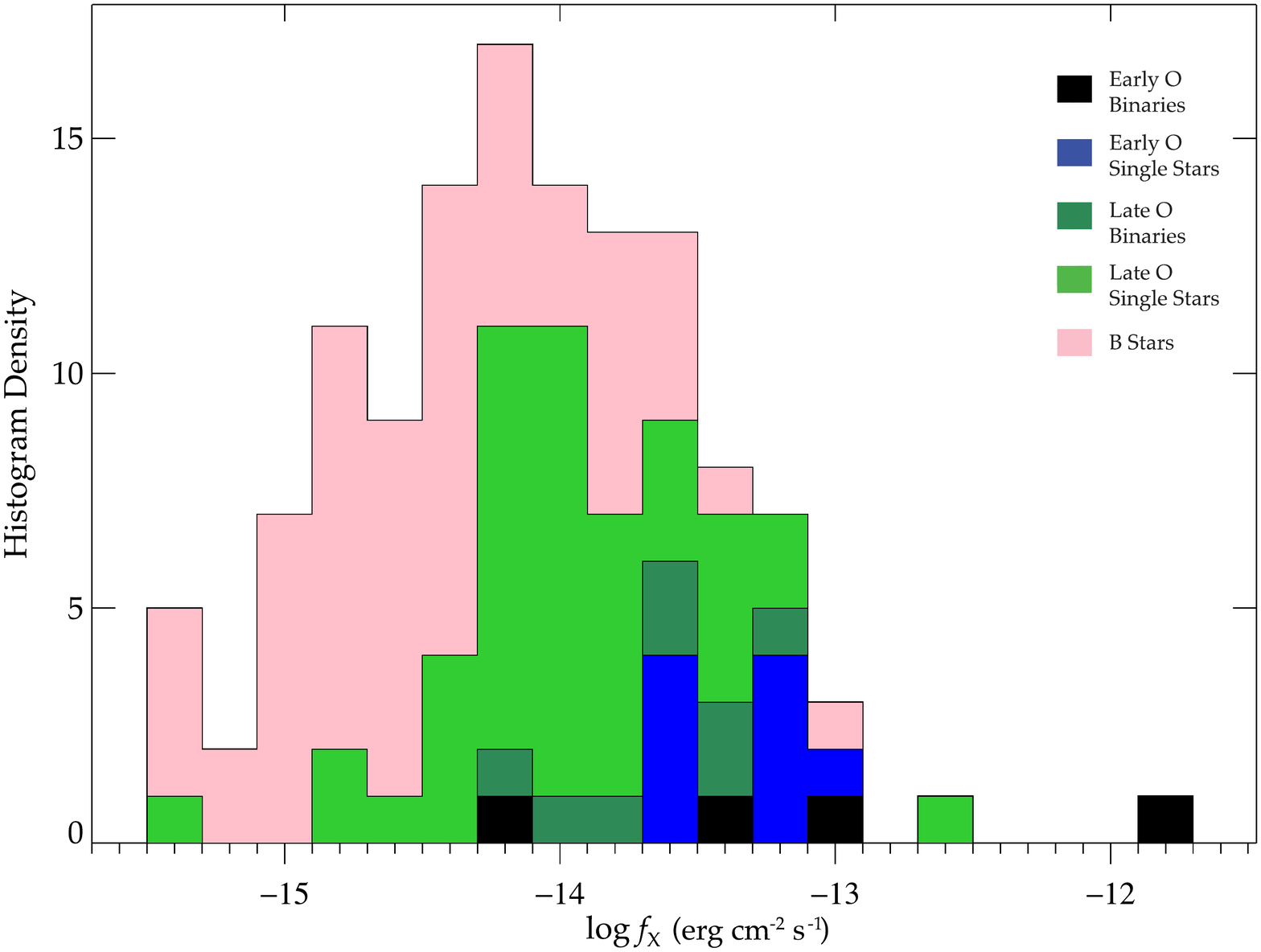}
\caption{
{\it Top panel:} $\log L_{\rm X}$ histogram for the 78 OB stars with XSPEC parameters and more than 50 net ACIS counts.
{\it Bottom panel:} $\log f_{\rm X}$ histogram for the full sample of X-ray detected OB stars.
As in Figs.~1-4, early-O binaries are shown in black, early-O single stars in blue, late-O binaries in dark green,
late-O single stars in light green, and B stars in pink.
}
\label{f5}
\end{figure}

\begin{figure}
\includegraphics[scale=0.55]{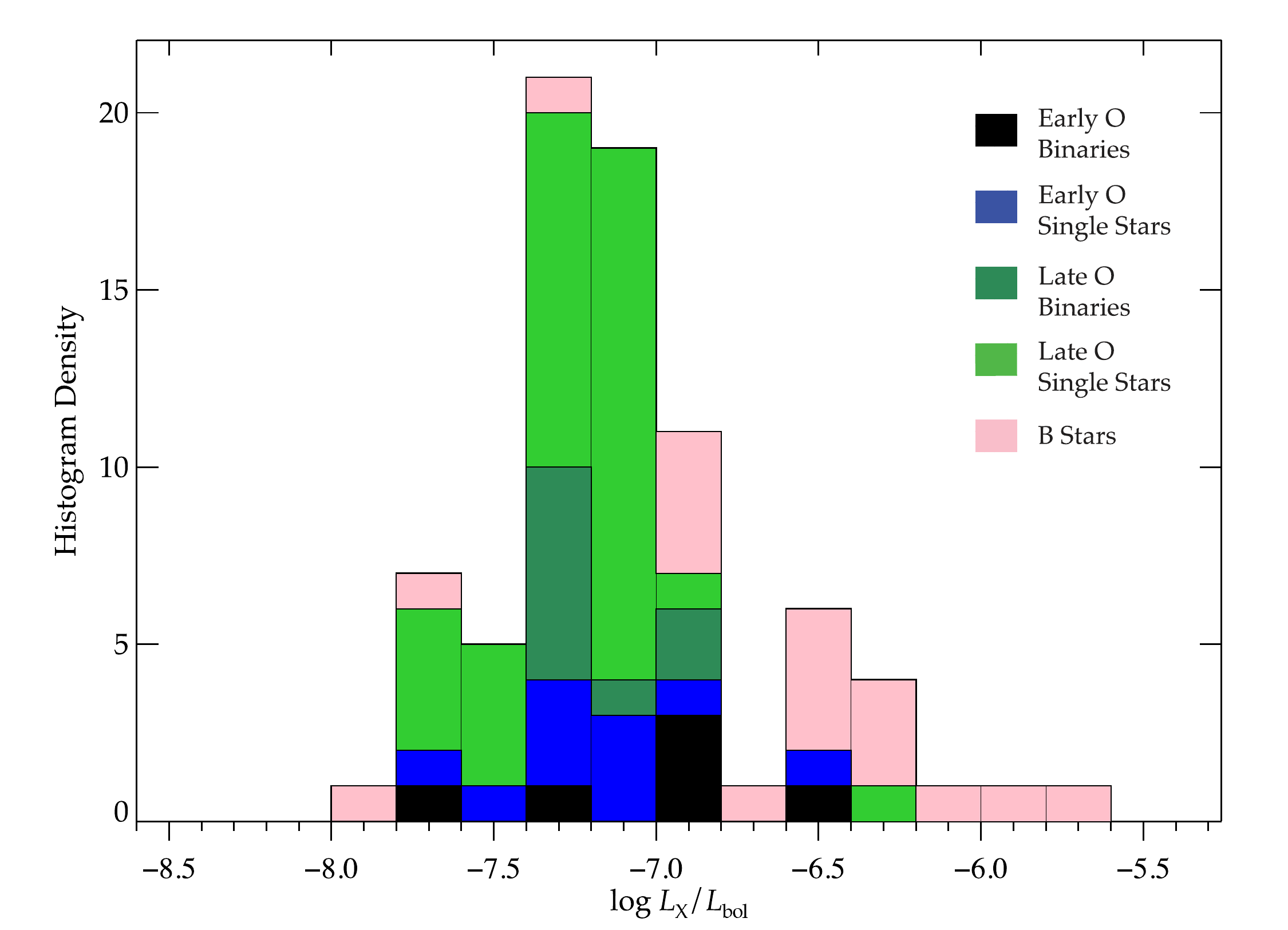}
\includegraphics[scale=0.44]{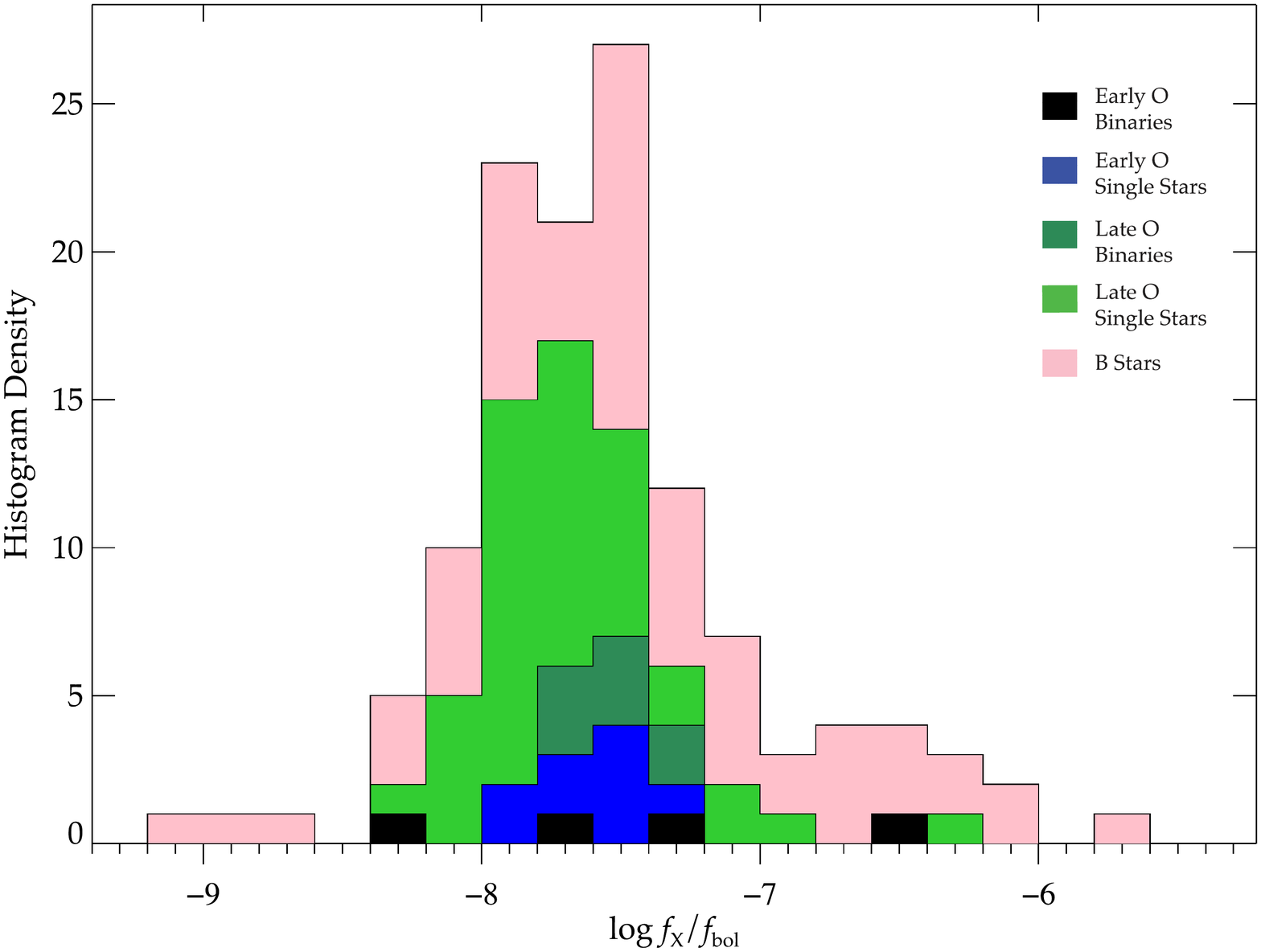}
\caption{
{\it Top panel:} $\log L_{\rm X}/L_{\rm bol}$ histogram for the 78 OB stars with XSPEC parameters and more than 50 net ACIS counts.
{\it Bottom panel:} $\log f_{\rm X}/f_{\rm bol}$ histogram for the full sample of X-ray detected OB stars.
}
\label{f6}
\end{figure}

\begin{figure}
\includegraphics[scale=0.41]{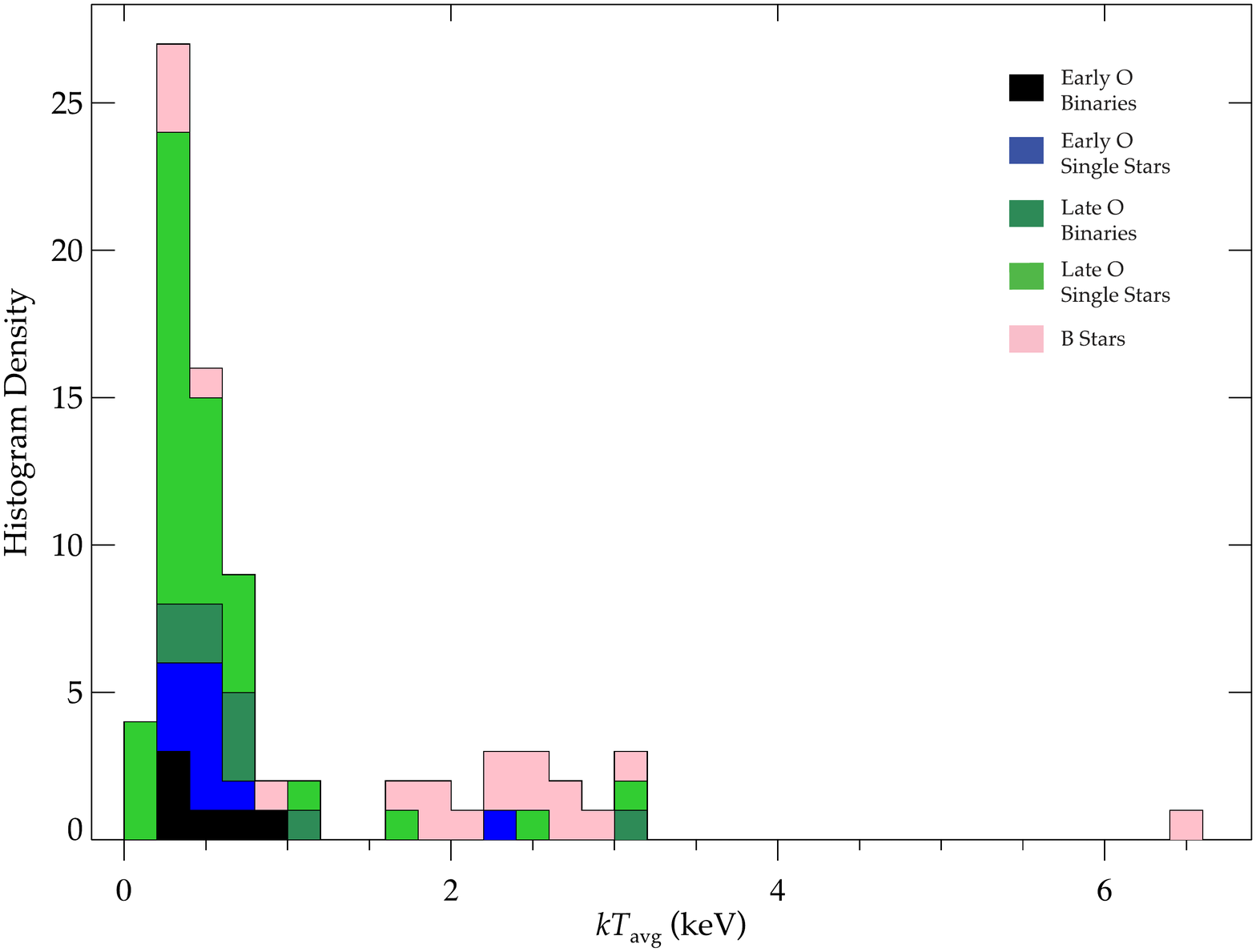}
\includegraphics[scale=0.41]{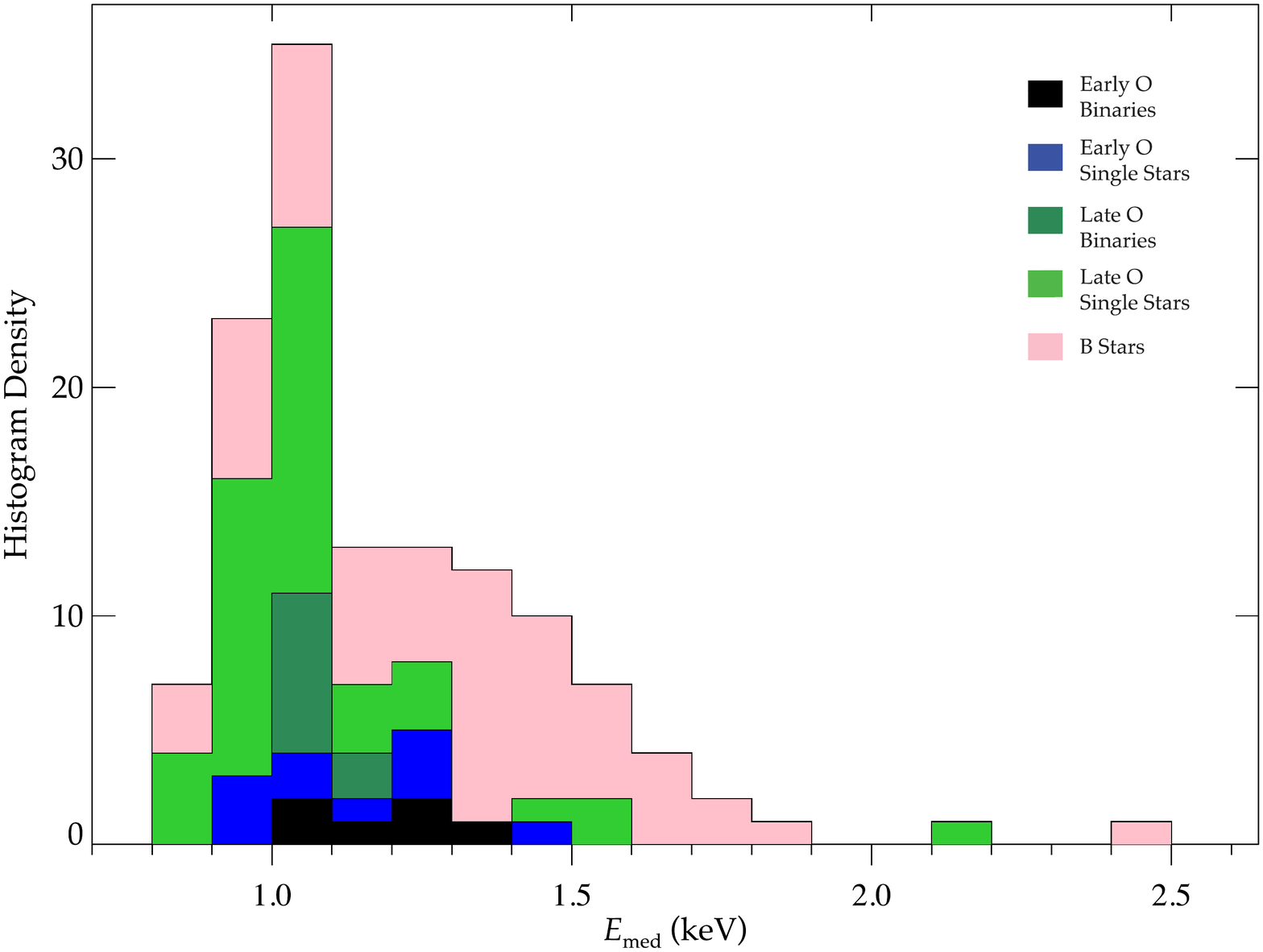}
\caption{
{\it Top panel:} $kT_{\rm avg}$ histogram for the 78 OB stars with XSPEC parameters and more than 50 net ACIS counts.
{\it Bottom planel:} $E_{\rm med}$ histogram for the full sample of X-ray detected OB stars.
For the 51 OB stars with fewer than 50 counts and no XSPEC fits,
{\em hard} corresponds approximately to $E_{\rm med} > 1.5$~keV.
}
\label{f7}
\end{figure}

\begin{figure}
\includegraphics[scale=0.6]{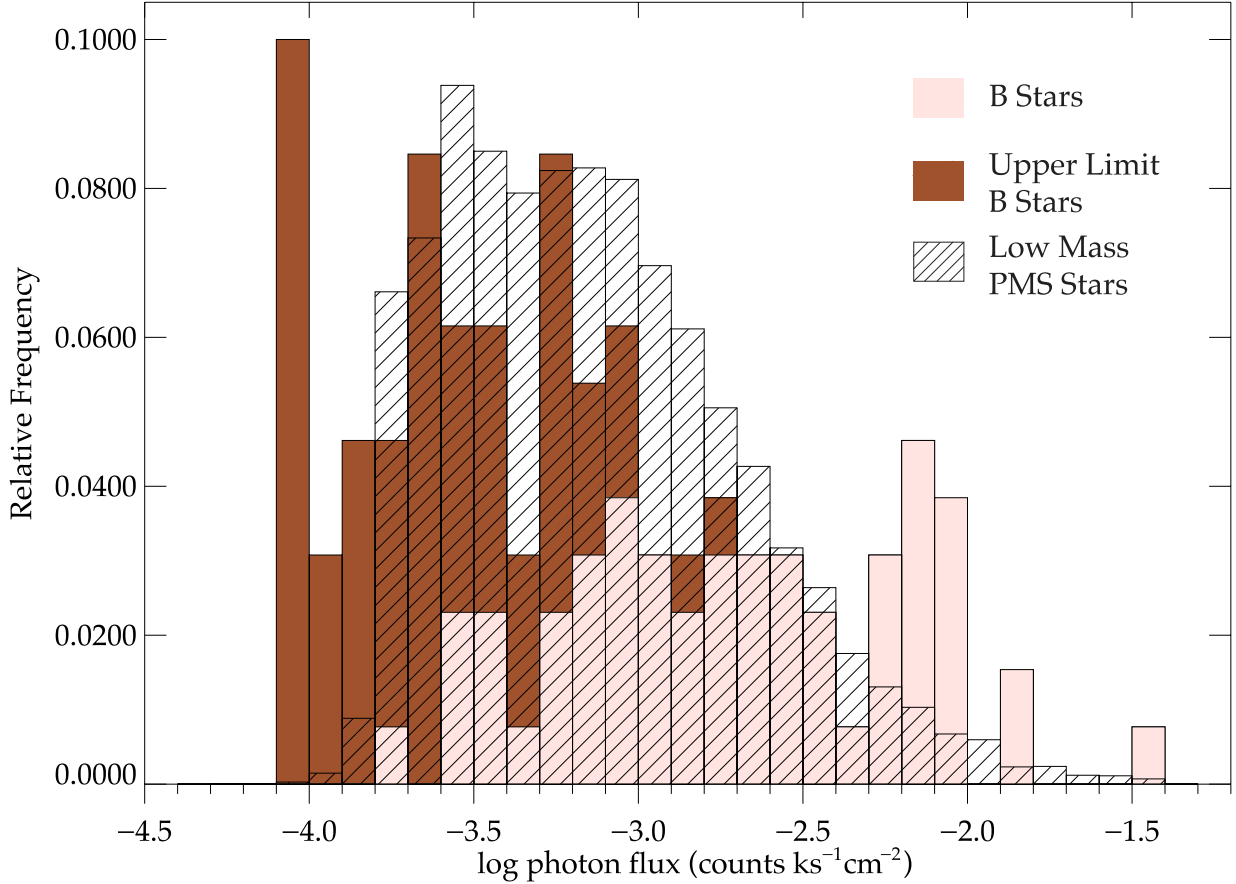}
\caption{
Photon flux histogram of the 61 X-ray sources with B-star primaries (in pink),
the 69 B-star upper limits (in brown), and the relative frequency of the 14\,250 CCCP sources (in gray)
with measured photon flux, and not associated with a known massive star (WR or OB star).
Most of the 14\,250 are associated with lower-mass, coronal,
pre--main-sequence stars. This figure and Table 4 suggest two B star populations:
a group with $\log$ photon flux below $-2.3$ (corresponding to $\log f_{\rm X} < -14$)
whose X-ray properties are similar to those of the general coronal PMS population,
and an active group with $\log$ photon flux above $-2.3$ (corresponding to $\log f_{\rm X} > -14$),
the majority of whose X-rays are probably {\em not} produced by coronal PMS stars.
}
\label{f8}
\end{figure}

\begin{figure}
\includegraphics[scale=0.7]{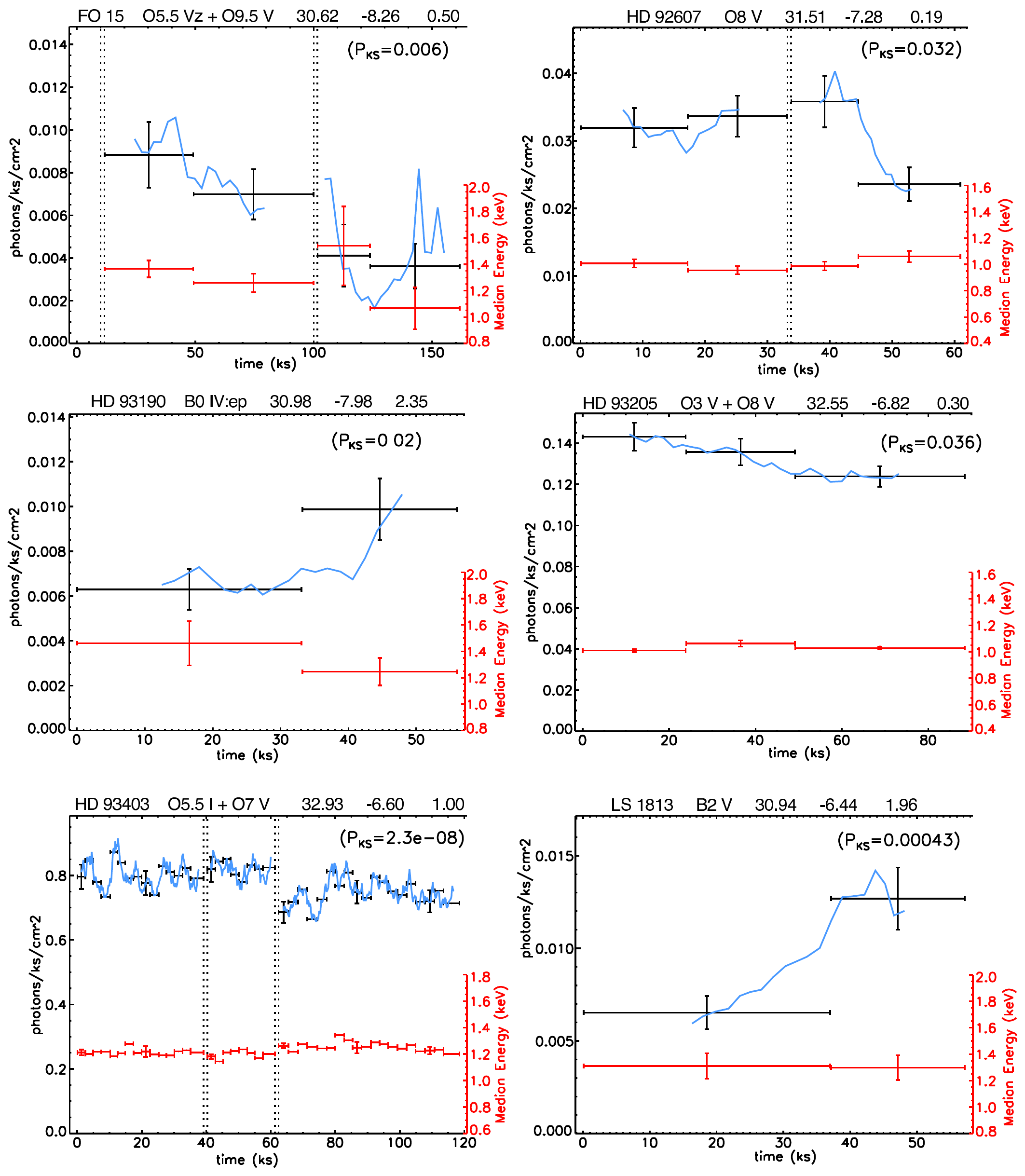}
\caption{
Sequenced photon flux light curves of FO 15 (O5.5 Vz + O9.5 V), HD 92607 (O8 V), HD 93190 (B0 IV:ep), HD 93205 (O3 V + O8 V),
HD 93403 (O5.5 I + O7 V), and LS~1813 (B2 V). 
Listed are $\log L_{\rm X}$, $\log L_{\rm X}/L_{\rm bol}$, $kT_{\rm avg}$, and probability of constancy, $P_{\rm KS}$.
The blue curves without error bars are the running count rates, corrected for effective area (in photons~ks$^{-1}$~cm$^{-2}$).
The corrected count rates (in black) and median energy (in red) are binned to show variability.
The vertical dotted lines indicated large time gaps between adjacent observations.
}
\label{f9}
\end{figure}

\begin{figure}
\includegraphics[scale=0.8]{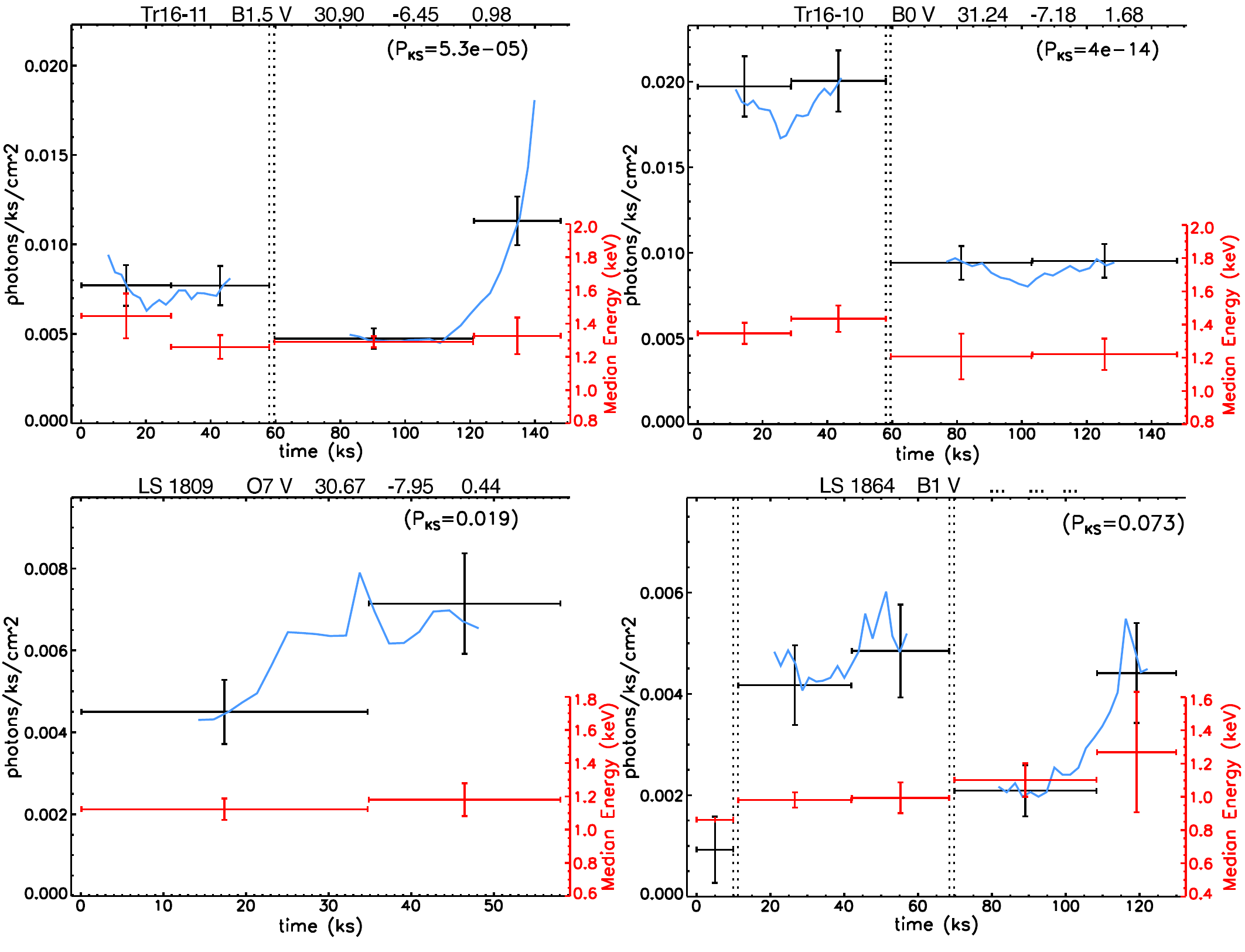}
\caption{
Sequenced light curves of Tr16-11 (B1.5 V), Tr16-10 (B0 V), LS~1809 (O7 V), LS~1864 (B1 V).
Listed are $\log L_{\rm X}$, $\log L_{\rm X}/L_{\rm bol}$, $kT_{\rm avg}$, and probability of constancy, $P_{\rm KS}$.
The blue curves without error bars are the running count rates, corrected for effective area (in photons~ks$^{-1}$~cm$^{-2}$).
The corrected count rates (in black) and median energy (in red) are binned to show variability.
The vertical dotted lines indicated large time gaps between adjacent observations.
}
\label{f10}
\end{figure}

\begin{figure}
\includegraphics[scale=0.6]{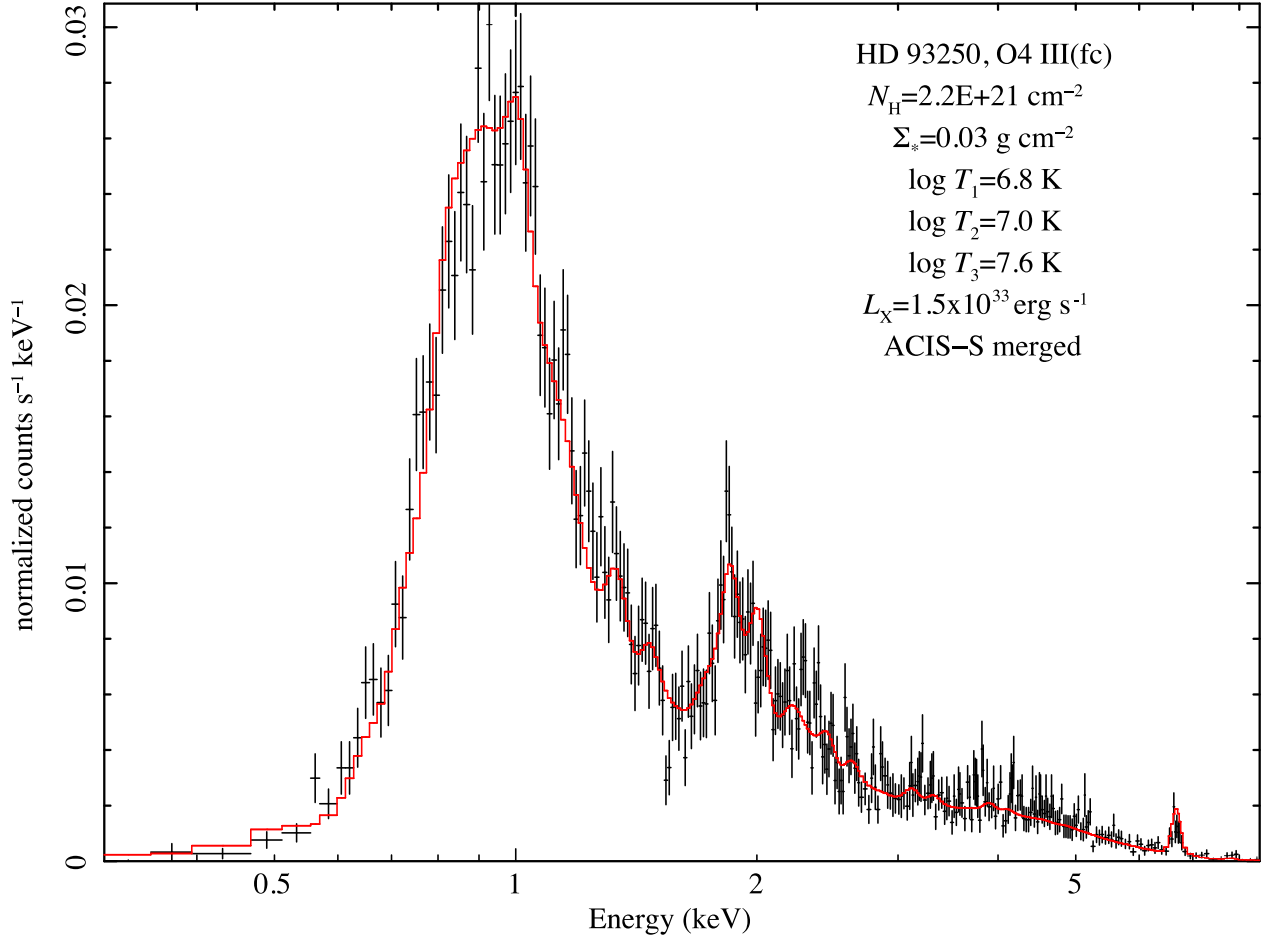}
\caption{
Merged off-axis zeroth-order ACIS-S spectrum of HD 93250, O4 III(fc) and best-fit {\sc tbabs*(windtabs(apec)+apec+apec)}
parameters. In this model all three emission components are absorbed by cold, neutral ISM gas,
and the 0.5~keV embedded wind shocks are absorbed by warm, partially ionized, solar-abundance plasma
distributed throughout the wind. The best-fit mass column is $\Sigma_{\star}=0.03\pm0.01$~g~cm$^{-2}$.
}
\label{f11}
\end{figure}

\begin{figure}
\includegraphics[scale=0.6]{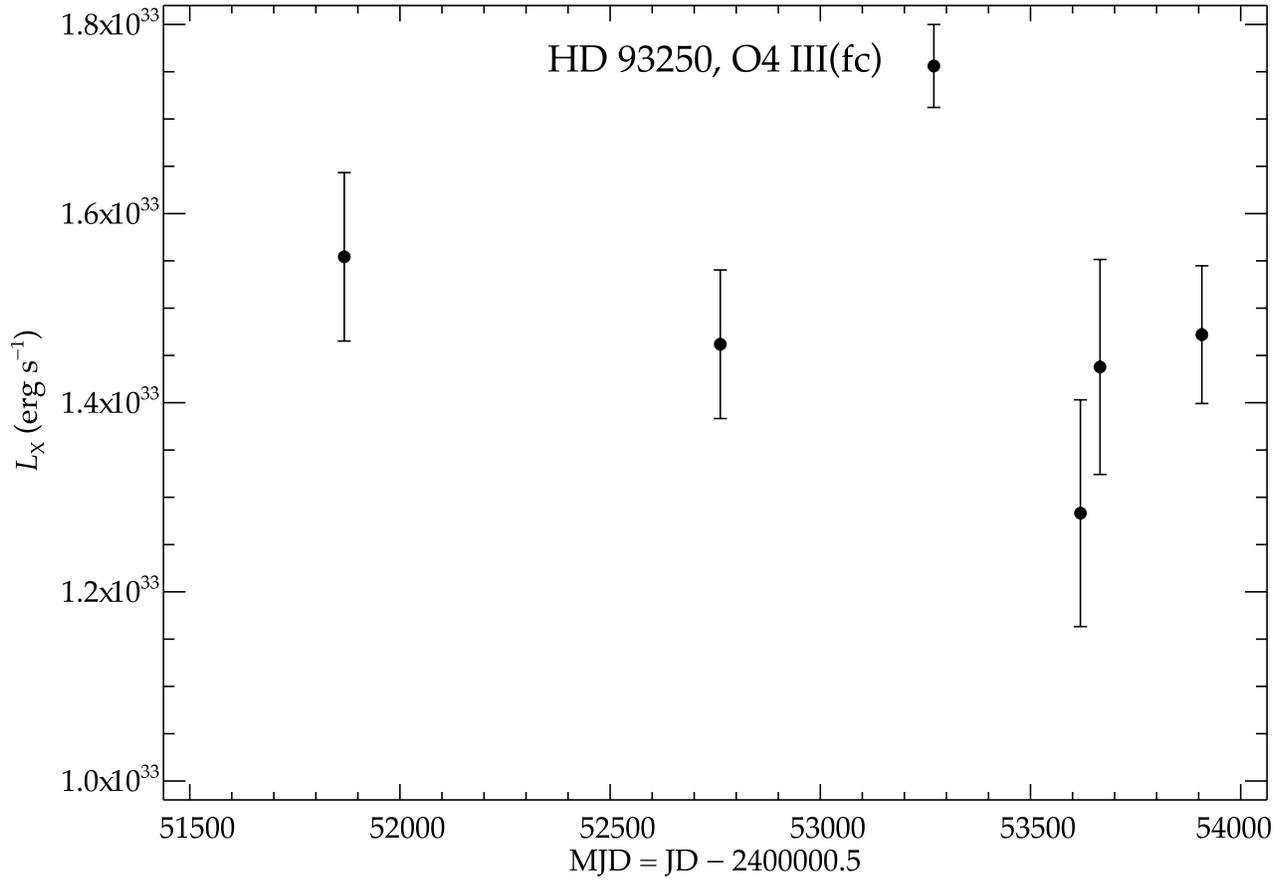}
\caption{
$L_{\rm X}$ light curve of HD 93250, O4 III(fc), assuming the same model as in Fig.\ 11,
indicating significant long-term variability.
}
\label{f12}
\end{figure}

\begin{figure}
\includegraphics[scale=0.75]{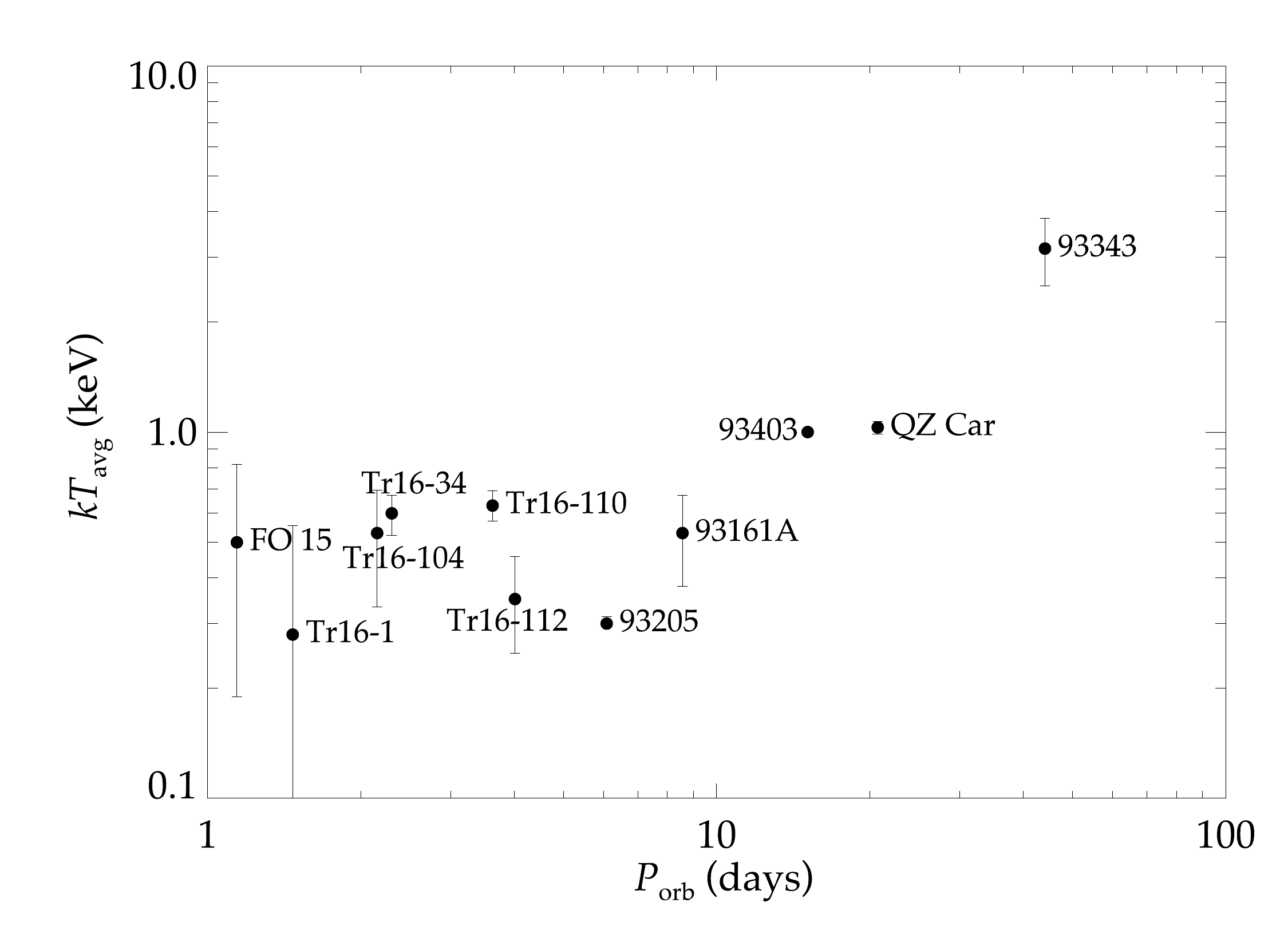}
\caption{
$kT_{\rm avg}$ with 90\% confidence limits versus primary orbital period 
for the 11 spectroscopic binaries in Table~3.
}
\label{f13}
\end{figure}

\begin{figure}
\includegraphics[scale=0.85]{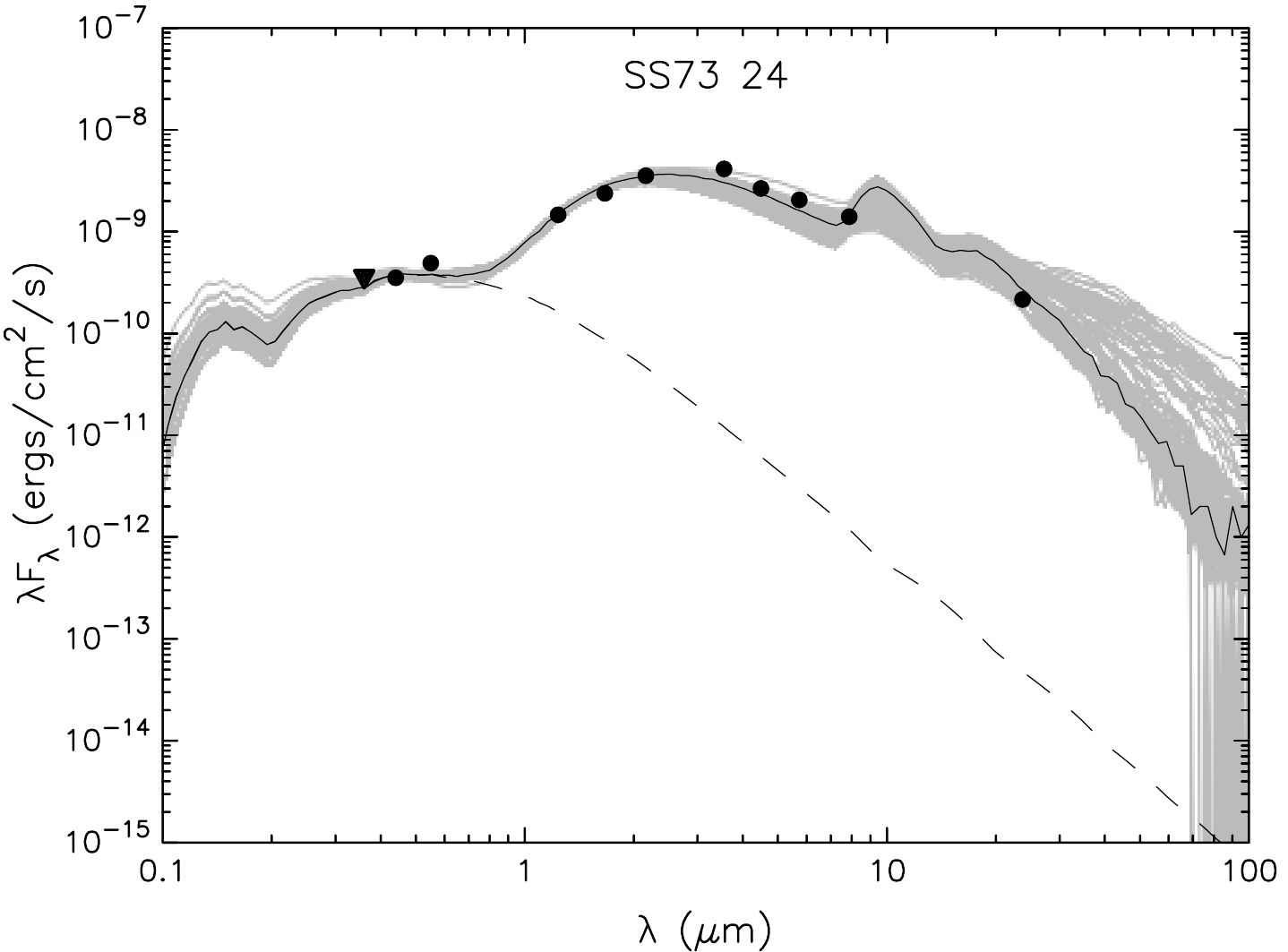}
\includegraphics[scale=0.45]{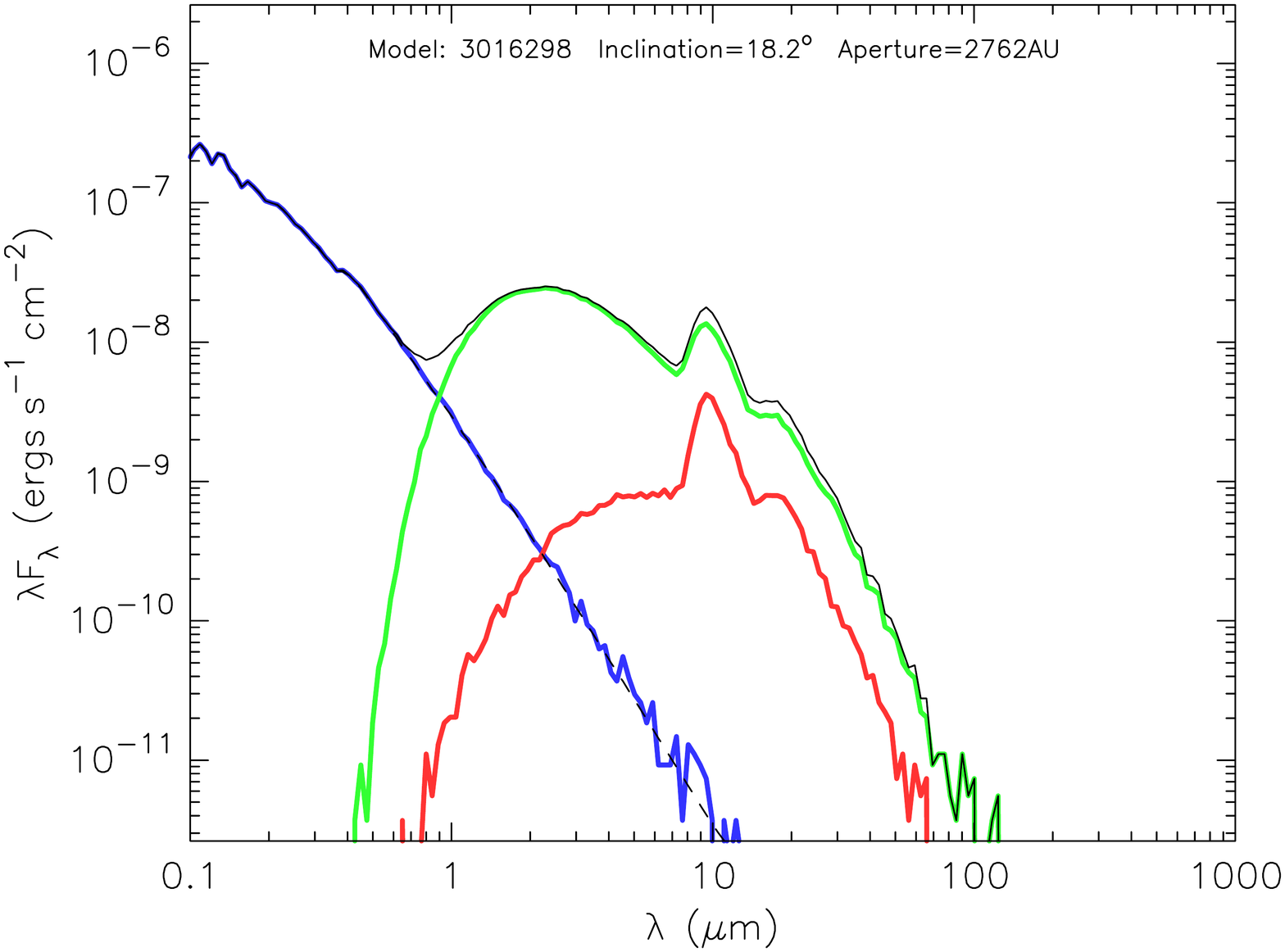}
\caption{
{\em Upper panel:} Optical-infrared SED of SS73 24 and the best-fit (black) Robitaille et al. (2006)
YSO model (star + disk + envelope). {\em Lower panel:} SS73 24's SED suggests a
$6\pm2\times10^3 L_{\odot}$ photosphere with $T_{\rm eff}=25\pm1.4$~kK (dashed, blue line),
a large disk (solid, green line), and cooler envelope (solid, red line).
}
\label{f14}
\end{figure}

\begin{figure}
\includegraphics[scale=0.865]{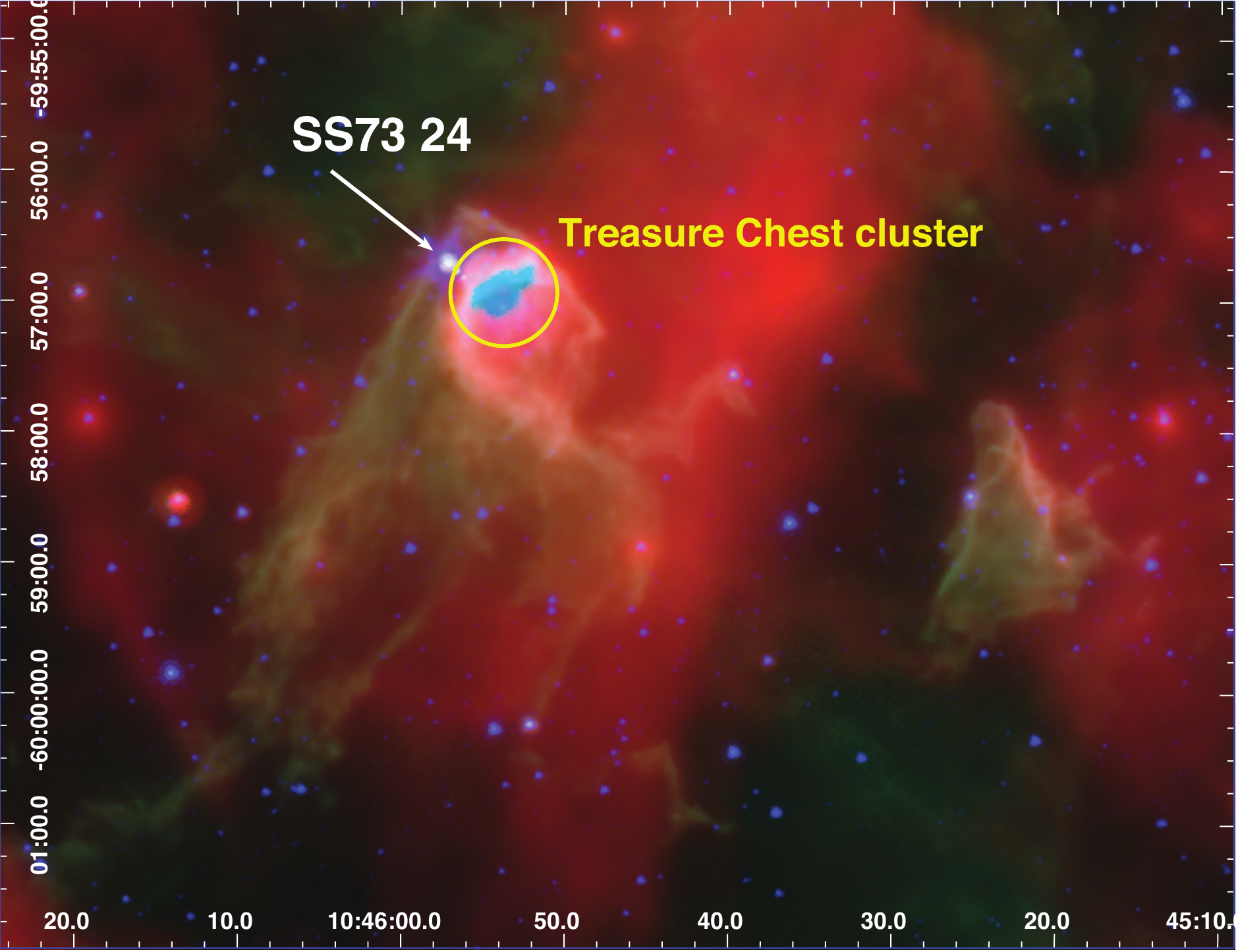}
\caption{
{\it Spitzer} IRAC + MIPS 24~$\mu$m image of SS73 24 and the nearby Treasure Chest cluster.
SS73 24's proximity to the most active star-forming portion of the Treasure Chest suggests it is a very
young Herbig Be star.
}
\label{f15}
\end{figure}

\clearpage

\begin{landscape}
\begin{deluxetable}{llllcclccccc}
\tabletypesize{\scriptsize}
\tablecaption{OB Star Catalog: Most Luminous Stars}
\tablewidth{0pt}
\tablehead{
\colhead{Star Name} & \colhead{Feinstein} & \colhead{MJ} & \colhead{Other Name} & \multicolumn{2}{c}{RA. (J2000) Decl.} &
\colhead{Spectral Type} & \colhead{$\mathrm{V}$} & \colhead{$\mathrm{U-B}$} & \colhead{$\mathrm{B-V}$} & \colhead{$A_\mathrm{V}$} & \colhead{$\log \frac {L_{\rm bol}}{L_{\odot}}$}
}
\startdata
QZ Car       & Coll 228 33  & \nodata      & HD 93206     & 10 44 22.92  & -59 59 35.9  & O9.7~I~+~O8~III              &    6.30  &   -0.81 &    0.14 &    2.21 &    6.23 \\
HD 93129A    & Tr 14 1      & MJ 177       & CPD-58 2618A & 10 43 57.46  & -59 32 51.4  & O2 If*                       &    7.26 &   -0.81  &    0.25  &    2.32 &    6.18 \\
HD 93250     & Tr 16 180    & \nodata      & CPD-58 2661  & 10 44 45.02  & -59 33 54.7  & O4 III(fc)                   &    7.41 &   -0.85 &    0.17  &    1.87 &    5.95 \\
HD 93403     & \nodata      & \nodata      & CPD-58 2680  & 10 45 44.11  & -59 24 28.2  & O5 III(fc) + O7 V            &    7.27 &   -0.76 &    0.22  &    1.95 &    5.95 \\
HD 92964     & Bo 10 44     & \nodata      & CPD-58 2581  & 10 42 40.58  & -59 12 56.6  & B2.5 Ia                      &    5.38 &   -0.66 &    0.26  &    1.61 &    5.87 \\
HD 93632     & Bo 11 1      & \nodata      & CPD-59 2696  & 10 47 12.64  & -60 05 50.9  & O5 I-IIIf                    &    8.36 &   -0.73 &    0.30  &    2.72 &    5.87 \\
HD 93205     & Tr 16 179    & MJ 342       & CPD-59 2587  & 10 44 33.75  & -59 44 15.4  & O3.5 V((f)) + O8 V           &    7.76 &   -0.94 &    0.08  &    1.60 &    5.79 \\
HD 93160     & Coll 232     & MJ 229       & CPD-58 2631C & 10 44 07.26  & -59 34 30.5  & O6 III                       &    7.88 &   -0.68 &    0.07  &    2.01 &    5.70 \\
HD 93130     & Coll 228 1   & MJ 208       & CPD-59 2556  & 10 44 00.38  & -59 52 27.5  & O6 III                       &    8.11 &   -0.75 &    0.19  &    2.20 &    5.68 \\
HD 93843     & \nodata      & \nodata      & CPD-59 2732  & 10 48 37.77  & -60 13 25.6  & O5 III(fc)                   &    7.32 &   -0.94 &   -0.04\phd  &    1.17 &    5.64 \\
HD 93128     & Tr 14 2      & MJ 157       & CPD-58 2617  & 10 43 54.41  & -59 32 57.4  & O3.5 V((fc))                 &    8.81 &   -0.79 &    0.19  &    2.21 &    5.63 \\
Tr15-18      & Tr 15 18     & \nodata      & \nodata      & 10 44 36.36  & -59 24 20.3  & O9 I/II:(e:)                 &   11.28\phn  &   -0.09 &    1.11 &    5.66 &    5.61 \\
HD 93873     & \nodata      & \nodata      & CPD-58 2747  & 10 48 55.21  & -59 26 48.3  & B1 Ia                        &    7.74 &   -0.50 &    0.45  &    2.60 &    5.60 \\
HD 305525    & Coll 228 98  & MJ 661       & CPD-59 2665  & 10 46 05.70  & -59 50 49.4  & O4 V                         &   10.03\phn &   -0.22 &    0.57  &    3.64 &    5.59 \\
HD 303308    & Tr 16 7      & MJ 480       & CPD-59 2623  & 10 45 05.92  & -59 40 06.1  & O4.5 V((fc))                 &    8.19 &   -0.82 &    0.14  &    1.72 &    5.57 \\
HD 93129B    & \nodata      & \nodata      & CPD-58 2618B & 10 43 57.65  & -59 32 53.8  & O3.5 V((fc))                 &    8.84 &   -0.79 &    0.23  &    2.16 &    5.53 \\
Tr16-112     & Tr 16 112    & MJ 535       & CPD-59 2641  & 10 45 16.51  & -59 43 37.0  & O5.5-O6 V(n)((fc)) + B2 V-III          &    9.28 &   -0.69 &    0.29  &    2.68 &    5.51 \\
Coll228-12   & Coll 228 12  & MJ 366       & CPD-59 2592  & 10 44 36.76  & -59 54 24.9  & B1 Ib                        &    9.52 &   -0.21 &    0.68  &    4.03 &    5.50 \\
Tr16-100     & Tr 16 100    & MJ 380       & CPD-59 2600  & 10 44 41.77  & -59 46 56.3  & O6 V                         &    8.65 &   -0.77 &    0.19  &    2.18 &    5.50 \\
HD 93222     & Coll 228 6   & \nodata      & CPD-59 2590  & 10 44 36.25  & -60 05 28.9  & O7 V((f))                       &    8.10 &   -0.89 &    0.05  &    1.74 &    5.46 \\	
HD 93342     & \nodata      & \nodata      & CPD-58 2674  & 10 45 17.57  & -59 23 37.5  & B1 Iab-Ib                    &    9.09 &   -0.40 &    0.60  &    3.52 &    5.45 \\
HD 93161B    & \nodata      & \nodata      & \nodata      & 10 44 09.09  & -59 34 35.4  & O6.5 V(f)                    &    8.60 &   -0.77 &    0.23  &    2.10 &    5.43 \\
Tr16-244     & Tr 16 244    & MJ 257       & \nodata      & 10 44 13.20  & -59 43 10.3  & O3/4 If                      &   10.78\phn &   -0.32 &    0.68  &    4.00 &    5.42 \\
Tr16-110     & Tr 16 110    & MJ 517       & CPD-59 2636  & 10 45 12.88  & -59 44 19.2  & O7 V + O8 V + O9 V           &    9.31 &   -0.64 &    0.29  &    2.74 &    5.40 \\
HD 93204     & Tr 16 178    & MJ 340       & CPD-59 2584  & 10 44 32.34  & -59 44 31.0  & O5.5 V((fc))                 &    8.48 &   -0.88 &    0.09  &    1.70 &    5.39 \\
HD 93190     & \nodata      & \nodata      & CPD-58 2637  & 10 44 19.61  & -59 16 59.0  & B0 IV:ep                     &    8.57 &   -0.80 &    0.32  &    2.38 &    5.39 \\
FO 15        & \nodata      & MJ 596       & \nodata      & 10 45 36.32  & -59 48 23.4  & O5.5 Vz + O9.5 V             &   12.05\phn &   -0.20 &    0.86  &    4.96 &    5.31 \\
HD 305619    & \nodata      & \nodata      & CPD-59 2727  & 10 48 15.54  & -60 15 56.9  & O9.7 Ib                      &    9.43 &   -0.55 &    0.44  &    3.05 &    5.30 \\
Coll228-97   & Coll 228 97  & MJ 691       & CPD-59 2673  & 10 46 22.46  & -59 53 20.5  & O5 V(n)((f))                        &   10.42\phn &   -0.48 &    0.41 &    3.28 &    5.29 \\
HD 93161A    & \nodata      & \nodata      & \nodata      & 10 44 08.82  & -59 34 34.5  & O8 V + O9 V                  &    8.56 &   -0.75 &    0.20 &    1.94 &    5.28 \\
HD 303304    & \nodata      & MJ 709       & CPD-58 2697  & 10 46 35.69  & -59 37 00.6  & O7 V                         &    9.71 &   -0.62 &    0.33 &    2.85 &    5.28 \\
HD 93146     & Coll 228 65  & \nodata      & CPD-59 2555  & 10 44 00.15  & -60 05 09.9  & O7 V((f))                       &    8.44 &   -0.91 &    0.02 &    1.46 &    5.26 \\
HD 305523    & Coll 228 32  & MJ 336       & CPD-59 2580  & 10 44 29.48  & -59 57 18.2  & O9 II                        &    8.50 &   -0.83 &    0.18 &    1.97 &    5.24 \\
HD 92607     & \nodata      & \nodata      & CPD-59 2404  & 10 40 12.42  & -59 48 10.0  & O8 V                         &    8.23&   -0.87  &    0.00 &    1.34 &    5.21 \\
Tr16-34      & Tr 16 34     & MJ 516       & CPD-59 2635  & 10 45 12.71  & -59 44 46.0  & O8 V + O9.5 V                &    9.27 &   -0.73 &    0.23 &    2.37 &    5.21 \\
Tr14-20      & Tr 14 20     & MJ 115       & CPD-58 2611  & 10 43 46.70  & -59 32 54.8  & O6 V                         &    9.65 &   -0.63 &    0.20 &    2.35 &    5.17 \\
HD 305439A   & \nodata      & \nodata      & CPD-59 2479A & 10 42 10.34  & -59 58 00.8  & B0 Ia                        &    9.56 &   -0.52 &    0.51 &    2.98 &    5.17 \\
HD 303311    & Tr 16 98     & MJ 351       & CPD-58 2652  & 10 44 37.45  & -59 32 55.2  & O5 V                         &    9.03 &   -0.86 &    0.13 &    1.59 &    5.15 \\
HD 305532    & Coll 228 38  & MJ 593       & CPD-59 2650  & 10 45 34.04  & -59 57 26.8  & O6 V                         &   10.19 &   -0.70 &    0.32 &    2.80 &    5.15 \\
HD 303316    & \nodata      & MJ 22        & CPD-59 2518  & 10 43 11.19  & -59 44 21.1  & O6 V                         &    9.64 &   -0.76 &    0.26 &    2.30 &    5.14 \\
HD 305524    & Coll 228 7   & MJ 404       & CPD-59 2602  & 10 44 45.23  & -59 54 41.6  & O7 V((f))                    &    9.32 &   -0.75 &    0.24 &    2.12 &    5.12 \\
HD 93249     & Tr 15 1      & \nodata      & CPD-58 2659  & 10 44 43.88  & -59 21 25.0  & O9 III                       &    8.36 &   -0.75 &    0.14 &    1.53 &    5.11 \\
Tr14-8       & Tr 14 8      & MJ 192       & CPD-58 2620  & 10 43 59.93  & -59 32 25.4  & O6.5 V                       &    9.45 &   -0.78 &    0.05 &    2.00 &    5.10 \\
Tr16-104     & Tr 16 104    & MJ 408       & CPD-59 2603  & 10 44 47.30  & -59 43 53.2  & O7 V((f)) + O9.5 + B0.2 IV   &    8.82 &   -0.79 &    0.14 &    1.55 &    5.09 \\
Tr16-23      & Tr 16 23     & MJ 484       & CPD-59 2626  & 10 45 05.79  & -59 45 19.6  & O9 III                       &   10.00 &   -0.61 &    0.37 &    2.91 &    5.06 \\
HD 93343     & Tr 16 182    & MJ 512       & CPD-59 2633  & 10 45 12.21  & -59 45 00.4  & O8 V + O7-8.5 V              &    9.60 &   -0.75 &    0.24 &    2.33 &    5.06 \\
LS 1809      & \nodata      & \nodata      & CPD-58 2608  & 10 43 41.24  & -59 35 48.2  & O7 V                         &   10.44 &   -0.40 &    0.50 &    3.08 &    5.03 \\
HD 305520    & Coll 228 4   & \nodata      & CPD-59 2560  & 10 44 05.86  & -59 59 41.5  & B1 Ib                        &    8.70 &   -0.71 &    0.18 &    2.00 &    5.01 \\
Tr14-9       & Tr 14 9      & MJ 165       & \nodata      & 10 43 55.41  & -59 32 49.3  & O8 V                         &    9.73 &   -0.68 &    0.32 &    2.42 &    5.01 \\
\enddata
\end{deluxetable}
\end{landscape}

\begin{deluxetable}{lclccccrc}
\tabletypesize{\scriptsize}
\tablecaption{Single O Stars: XSPEC and time variability parameters}
\tablewidth{0pt}
\tablehead{
\colhead{Star} & \colhead{ACIS} & \colhead{Spectral} & \colhead{$\log L_{\rm X}$\tablenotemark{\sharp}} & \colhead{$\log \frac{L_{\rm X}}{L_{\rm bol}}$} & 
\colhead{$kT_{\rm avg}$} & \colhead{$\log \frac{f_{\rm X}}{f_{\rm bol}}$\tablenotemark{\spadesuit}} & \colhead{$P_{\rm KS}$} & 
\colhead{MLB\tablenotemark{\dagger}} \\
\colhead{Name} & \colhead{Name} & \colhead{Type} & \colhead{(erg~s$^{-1}$)} &
\colhead{} & \colhead{(keV)} & \colhead{} & \colhead{(\%)\tablenotemark{\star}} & \colhead{}
}
\startdata
Tr16-22            & 104508.23-594607.0 & O8.5 V             &   32.25 &   -6.30 &    1.70 &   -6.39 &   14 &    1 \\
HD 93250\tablenotemark{\ddagger} & 104445.04-593354.6 & O4 III(fc) & 33.12 & -6.41 &  2.30 & \nodata &   20 &    1 \\
LS~1865            & 104454.70-595601.8 & O8.5 V((f))        &   31.42 &   -6.90 &    3.09 &   -7.12 &    6 &    3 \\
HD 303311          & 104437.47-593255.3 & O5 V               &   31.77 &   -6.96 &    0.42 &   -7.47 &   33 &    1 \\
LS 1821            & 104357.46-600528.3 & O8.5 V             &   31.25 &   -7.01 &    0.30 &   -7.55 &   74 &    1 \\
Tr16-100           & 104441.80-594656.4 & O6 V               &   32.06 &   -7.02 &    0.35 &   -7.44 &   88 &    1 \\
Tr14-20            & 104346.69-593254.7 & O6 V               &   31.71 &   -7.03 &    0.34 &   -7.77 &   98 &    1 \\
HD 92607           & 104012.42-594810.1 & O8 V               &   31.75 &   -7.04 &    0.19 &   -7.42 &    1 &    2 \\
Tr14-5             & 104353.63-593328.4 & O9 V               &   30.98 &   -7.05 &    0.67 &   -7.18 &   33 &    1 \\
HD 93146           & 104400.16-600509.8 & O7 V((f))          &   31.77 &   -7.07 &    0.24 &   -7.52 &   50 &    1 \\
HD 93129B          & 104357.65-593253.7 & O3.5 V((fc))       &   32.03 &   -7.08 &    0.51 &   -7.56 &   84 &    1 \\
CPD-59 2661        & 104553.71-595703.9 & O9.5 V             &   31.08 &   -7.09 &    2.40 &   -7.22 &   19 &    1 \\
Tr14-21            & 104348.70-593324.2 & O9 V               &   31.13 &   -7.10 &    0.53 &   -7.52 &   48 &    1 \\
HD 93128           & 104354.40-593257.4 & O3.5 V((fc))       &   32.11 &   -7.10 &    0.39 &   -7.52 &   26 &    3 \\
HD 93222           & 104436.23-600529.0 & O7 V((f))          &   31.94 &   -7.11 &    0.27 &   -7.59 &   11 &    3 \\
HD 93843           & 104837.74-601325.7 & O5 III(fc)         &   32.08 &   -7.14 &    0.40 &   -7.43 &   26 &    2 \\
Tr14-9             & 104355.36-593248.8 & O8 V               &   31.44 &   -7.14 &    0.19 &   -7.81 &   39 &    1 \\
HD 305438          & 104243.71-595416.6 & O8 V((f))          &   31.17 &   -7.16 &    0.38 &   -7.62 &   60 &    3 \\
HD 305536          & 104411.04-600321.8 & O9.5 V             &   31.24 &   -7.16 &    1.01 &   -7.40 &   57 &    1 \\
HD 93160           & 104407.26-593430.5 & O6 III             &   32.10 &   -7.18 &    0.25 &   -7.68 &    2 &    2 \\
Tr16-3             & 104506.70-594156.6 & O8.5 V             &   31.04 &   -7.18 &    0.36 &   -7.74 &   49 &    1 \\
HD 303316          & 104311.17-594420.8 & O6 V               &   31.53 &   -7.19 &    0.30 &   -7.77 &   89 &    1 \\
Tr16-115           & 104520.57-594251.1 & O9.5 V             &   31.01 &   -7.20 &    0.32 &   -7.65 &   18 &    1 \\
Tr14-8             & 104359.92-593225.4 & O6.5 V             &   31.48 &   -7.20 &    0.23 &   -7.83 &   89 &    1 \\
Tr16-244           & 104413.19-594310.1 & O3/4 If            &   31.80 &   -7.21 &    0.62 &   -7.33 &   34 &    1 \\
HD 93028           & 104315.33-601204.3 & O9 IV              &   31.23 &   -7.21 &    0.14 &   -7.74 &   10 &    1 \\
HD 93161B          & 104409.08-593435.3 & O6.5 V(f)          &   31.75 &   -7.26 &    0.54 &   -7.62 &   11 &    1 \\
HD 305524          & 104445.27-595441.5 & O7 V((f))          &   31.43 &   -7.28 &    0.61 &   -7.60 &   10 &    1 \\
HD 305518          & 104343.99-594817.9 & O9.5 V             &   31.27 &   -7.29 &    0.38 &   -7.96 &   61 &    1 \\
HD 93204           & 104432.34-594431.0 & O5.5 V((fc))       &   31.67 &   -7.29 &    0.30 &   -7.70 &    9 &    1 \\
HD 93027           & 104317.92-600803.1 & O9.5 IV            &   31.08 &   -7.30 &    0.27 &   -7.81 &   34 &    1 \\
HD 305539          & 104633.07-600412.9 & O7                 &   31.16 &   -7.32 &    0.48 &   -7.81 &   25 &    4 \\
HD 305532          & 104534.04-595726.7 & O6 V               &   31.39 &   -7.34 &    0.23 &   -7.88 &   83 &    1 \\
LS 1892            & 104622.48-595320.4 & O5 V(n)((fc))      &   31.53 &   -7.35 &    0.55 &   -7.64 &   47 &    2 \\
HD 93249           & 104443.88-592125.1 & O9 III             &   31.33 &   -7.36 &    0.46 &   -7.69 &   45 &    1 \\
HD 305525          & 104605.70-595049.5 & O4 V               &   31.76 &   -7.41 &    0.33 &   -7.92 &   41 &    1 \\
HD 305523          & 104429.47-595718.1 & O9 II              &   31.41 &   -7.41 &    0.30 &   -8.06 &    9 &    1 \\
Tr16-21            & 104436.73-594729.5 & O8 V               &   30.97 &   -7.47 &    0.15 &   -7.91 &   14 &    1 \\
HD 93576           & 104653.84-600441.9 & O9 IV              &   30.91 &   -7.50 &    0.34 &   -7.86 &   86 &    1 \\
LS 1809            & 104341.24-593548.1 & O7 V               &   31.03 &   -7.58 &    0.44 &   -7.93 &    1 &    2 \\
HD 93130           & 104400.38-595227.5 & O6 III             &   31.65 &   -7.61 &    0.63 &   -8.00 &   33 &    1 \\
HD 303304          & 104635.70-593700.7 & O7 V               &   31.23 &   -7.64 &    0.74 &   -7.87 &    7 &    1 \\
HD 305612          & 104716.41-600539.9 & O9 V               &   30.80 &   -7.64 &    0.54 &   -7.93 &   94 &    1 \\
HD 93632           & 104712.63-600550.8 & O5 I-IIIf          &   31.77 &   -7.67 &    0.56 &   -7.92 &   94 &    1 \\
HD 305619          & 104815.50-601556.9 & O9.7 Ib            &   31.20 &   -7.68 &    0.54 &   -8.18 &   68 &    1 \\
Tr15-19            & 104435.91-592335.7 & O9 V               & \nodata & \nodata & \nodata &   -6.85 &   10 &\nodata \\
CPD-58 2627        & 104402.44-592936.3 & O9 III             & \nodata & \nodata & \nodata &   -7.67 &    1 &\nodata \\
Coll228-67         & 104400.43-600559.8 & O9 V               & \nodata & \nodata & \nodata &   -7.80 &   29 &\nodata \\
Bo11-5             & 104715.29-600538.8 & O9 V:              & \nodata & \nodata & \nodata &   -8.04 &   11 &\nodata \\
Tr14-127           & 104400.94-593545.7 & O9 V               & \nodata & \nodata & \nodata &   -8.07 &   33 &\nodata \\
Coll228-66         & 104359.45-600513.3 & O9.5 V             & \nodata & \nodata & \nodata &   -8.10 &\nodata &\nodata \\
Tr15-2             & 104443.77-592117.2 & O9.5 III:          & \nodata & \nodata & \nodata &   -8.32 &   19 &\nodata \\
Tr14-27            & 104343.89-593346.1 & O9 V               & \nodata & \nodata & \nodata &   -8.98 &\nodata &\nodata \\
Tr15-18            & 104436.35-592420.3 & O9 I/II:(e:)       & \nodata & \nodata & \nodata & \nodata &\nodata &\nodata \\
Tr15-20            & 104435.12-592328.1 & O9 V:              & \nodata & \nodata & \nodata & \nodata &\nodata &\nodata \\
\enddata
\tablenotetext{\star}{Kolmogorov-Smirnov percent probability of constancy. Minimum of single-epoch and merged $P_{\rm KS}$ values.}
\tablenotetext{\dagger}{Number of distinct maximum likelihood blocks in merged event data (minimum 5 counts per blcok, 95\% confidence).}
\tablenotetext{\ddagger}{HD 93250 was piled up in the ACIS-I CCCP survey data: $f_{\rm X}$ not computed in ACIS EXTRACT. $L_{\rm X}$ is the mean value observed in a series of off-axis zeroth-order grating observations. See Figs. 11 and 12 and Appendix A.}
\tablenotetext{\sharp}{The ISM-corrected 0.5-8 keV $L_{\rm X}$, {\sc fitluminosity\_tc}. For the ISM-corrected 0.5-10 keV $L_{\rm X}$, see \citet{naze11}.}
\tablenotetext{\spadesuit}{$f_{\rm X}$ is the absorbed (uncorrected) 0.5-8.0~keV flux, {\sc energyflux\_t}. For the absorbed 0.5-10~keV flux, {\sc fto} see \citet{naze11}.}
\end{deluxetable}

\begin{deluxetable}{lclcccccc}
\tabletypesize{\scriptsize}
\tablecaption{O+O Binaries: XSPEC and time variability parameters}
\tablewidth{0pt}
\tablehead{
\colhead{Star} & \colhead{ACIS} & \colhead{Spectral} & 
\colhead{$P_{\rm orb}$\tablenotemark{\dagger}} &
\colhead{$\log L_{\rm X}$\tablenotemark{\sharp}} & 
\colhead{$\log \frac{L_{\rm X}}{L_{\rm bol}}$} & 
\colhead{$kT_{\rm avg}$} & 
\colhead{$P_{\rm KS}$} & \colhead{MLB}\tablenotemark{\ddagger}\\
\colhead{Name} & \colhead{Name} & \colhead{Type} & 
\colhead{(days)} &
\colhead{(erg~s$^{-1}$)} &
\colhead{} & 
\colhead{(keV)} & \colhead{(\%)\tablenotemark{\star}} & \colhead{}
}
\startdata
HD 93403           & 104544.13-592428.1 & O5 III(fc) + O7 V            & 15.093   &   33.11 &   -6.41 &    1.00 &\phn9 &    3 \\
HD 93205           & 104433.74-594415.4 & O3 V + O8 V                  & 6.0803  &   32.55 &   -6.82 &    0.30 &\phn3 &    1 \\
HD 93129A          & 104357.47-593251.3 & O2 If$\ast$\tablenotemark{\bullet} & \nodata & 32.91 & -6.85 &  0.74 &   46 &    1 \\
HD 303308          & 104505.90-594006.0 & O4.5 V((fc))\tablenotemark{\bullet} & \nodata & 32.25 & -6.89 &   0.25 &   20 &    4 \\
HD 93161A          & 104408.84-593434.4 & O8 V + O9 V                  & 8.566   &   31.94 &   -6.92 &    0.53 &   81 &    1 \\
HD 93343           & 104512.23-594500.5 & O8 V + O7-8.5 V              & 44.15    &   31.66 &   -6.98 &    3.17 &   30 &    1 \\
Tr16-9             & 104505.84-594307.7 & O9.5 V\tablenotemark{\bullet} & \nodata &  31.27 &   -7.17 &    0.70 &\phn9 &    2 \\
Tr16-34            & 104512.72-594446.2 & O8 V + O9.5 V                & 2.30000 &   31.56 &   -7.23 &    0.60 &   26 &    5 \\
Tr16-110           & 104512.88-594419.3 & O7 V + O8 V + O9 V & 3.62864, 5.034 &   31.74 &   -7.24 &    0.63 &   19 &    4 \\
QZ Car             & 104422.91-595935.9 & O9.7 I + O8 III & 20.72, 5.999 &   32.55 &   -7.26 &    1.03 &\phn5 &    1 \\
Tr16-23            & 104505.79-594519.7 & O9 III\tablenotemark{\bullet} & \nodata &  31.38 &   -7.26 &    0.32 &   83 &    1 \\
Tr16-104           & 104447.31-594353.3 & O7 V((f)) + O9.5 + B0.2 IV   & 2.1529  &   31.38 &   -7.29 &    0.53 &   49 &    1 \\
Tr16-1             & 104508.21-594049.6 & O9.5 V + B0.3 V              & 1.4693  &   30.87 &   -7.30 &    0.28 &   38 &    1 \\
Tr16-112           & 104516.52-594337.1 & O5.5-O6 V(n)((fc)) + B2 V-III          & 4.0157  &   31.78 &   -7.31 &    0.35 &   53 &    1 \\
FO 15              & 104536.33-594823.5 & O5.5 Vz + O9.5 V             & 1.1414  &   31.24 &   -7.65 &    0.50 &   41 &    5 \\
\enddata
\tablenotetext{\star}{Kolmogorov-Smirnov percent probability of constancy. Minimum of single-epoch and merged $P_{\rm KS}$ values.}
\tablenotetext{\dagger}{Orbital period(s) in days (Rauw et al. 2009, Nelan et. 2010). For triple and quadruple systems the primary period is shown in Fig.\ 13.}
\tablenotetext{\ddagger}{Number of distinct maximum likelihood blocks in merged data (minimum 5 counts, 95\% confidence).}
\tablenotetext{\bullet}{Close visual binary detected with {\it HST} FGS by Nelan et al. (2004, 2010).}
\tablenotetext{\sharp}{The ISM-corrected 0.5-8 keV $L_{\rm X}$, {\sc fitluminosity\_tc}. For the ISM-corrected 0.5-10 keV X-ray luminosity, see \citet{naze11}.}
\end{deluxetable}

\begin{deluxetable}{lclccccrc}
\tabletypesize{\scriptsize}
\tablecaption{Notable Early B stars: XSPEC and time variability parameters}
\tablewidth{0pt}
\tablehead{
\colhead{Star} & \colhead{ACIS} & \colhead{Spectral} & \colhead{$\log L_{\rm X}$\tablenotemark{\sharp}} & 
\colhead{$\log \frac{L_{\rm X}}{L_{\rm bol}}$} & 
\colhead{$kT_{\rm avg}$} & \colhead{$\log f_{\rm X}$\tablenotemark{\spadesuit}} & \colhead{$P_{\rm KS}$} & 
\colhead{MLB\tablenotemark{\dagger}} \\
\colhead{Name} & \colhead{Name} & \colhead{Type} & \colhead{(erg~s$^{-1}$)} &
\colhead{} & \colhead{(keV)} & \colhead{(erg~cm$^{-2}$~s$^{-1}$)} & \colhead{(\%)\tablenotemark{\star}} & \colhead{}
}
\startdata
SS73 24            & 104557.13-595643.1 & Be pec                       &   31.70 &   -5.65 &    3.13 &  -13.07 &   19 &    2 \\
Tr16-64            & 104504.75-594053.7 & B1.5 Vb                      &   31.37 &   -6.05 &    2.73 &  -13.48 &   30 &    2 \\
Tr16-10            & 104430.34-593726.8 & B0 V                         &   31.44 &   -6.99 &    1.68 &  -13.56 &   32 &    2 \\
Tr16-5             & 104454.06-594129.4 & B1 V                         &   31.18 &   -6.43 &    2.94 &  -13.64 &    0 &    4 \\
Tr14-28            & 104343.55-593403.4 & B2 V                         &   31.24 &   -5.98 &    2.74 &  -13.66 &    0 &    3 \\
HD 93501           & 104622.02-600118.8 & B1.5 III:                    &   31.13 &   -6.94 &  $>6.$ &  -13.69 &   30 &    1 \\
Coll228-68         & 104400.17-600607.7 & B1 Vn                        &   31.06 &   -6.46 &    2.54 &  -13.76 &    2 &    3 \\
Tr14-124           & 104405.84-593511.6 & B1 V                         &   31.13 &   -6.59 &    1.96 &  -13.77 &   12 &    2 \\
HD 93190           & 104419.63-591658.6 & B0 IV:ep                     &   31.12 &   -7.84 &    2.35 &  -13.77 &    1 &    2 \\
Tr14-18            & 104357.96-593353.4 & B1.5 V                       &   31.13 &   -6.25 &    2.13 &  -13.77 &   40 &    1 \\
LS 1813            & 104345.04-595325.0 & B2 V                         &   31.04 &   -6.34 &    1.96 &  -13.81 &    0 &    2 \\
Tr16-11            & 104422.51-593925.4 & B1.5 V                       &   31.00 &   -6.35 &    0.98 &  -13.88 &    0 &    4 \\
Tr14-19            & 104358.45-593301.5 & B1 V                         &   31.07 &   -6.54 &    2.38 &  -13.92 &   92 &    1 \\
HD 92644           & 104031.71-594643.9 & B1.5 III                     &   31.12 &   -6.82 &    0.24 &  -13.98 &   52 &    1 \\
Tr14-29            & 104405.09-593341.4 & B1.5 V                       &   30.91 &   -6.62 &    2.49 &  -14.02 &   16 &    2 \\
HD 305533          & 104513.44-595753.1 & B0.5 Vnn:\,(shell)            &\nodata &\nodata &\nodata &  -14.16 &   20 &    1 \\
HD 305599          & 104924.95-594944.0 & B0 Ib                        &   31.28 &   -6.94 &    0.23 &  -14.19 &   17 &    2 \\
HD 305439A         & 104210.35-595800.9 & B0 Ia                        &   31.11 &   -7.63 &    0.56 &  -14.20 &    7 &    2 \\
Tr16-126           & 104359.86-593524.1 & B1 V                         &\nodata &\nodata &\nodata &  -14.24 &   66 &    2 \\
HD 305538          & 104546.52-600513.5 & B0 V                         &\nodata &\nodata &\nodata &  -14.27 &   30 &    3 \\
HD 305556          & 104343.42-602027.7 & B0 Ib                        &   31.17 &   -7.36 &    0.28 &  -14.27 &    9 &    1 \\
Tr14-22            & 104348.82-593335.2 & B2 V                         &\nodata &\nodata &\nodata &  -14.33 &   64 &    1 \\
HD 92741           & 104112.33-595825.0 & B1.5 II:                     &\nodata &\nodata &\nodata &  -14.64 &    2 &    2 \\
LS 1864            & 104450.39-595545.0 & B1 V                         &\nodata &\nodata &\nodata &  -14.70 &    7 &    1 \\
HD 93026           & 104316.35-591027.2 & B1.5 V                       &\nodata &\nodata &\nodata &  -14.82 &    6 &    1 \\
HD 303225          & 104135.44-593945.6 & B1.5 V                       &\nodata &\nodata &\nodata &  -14.90 &    9 &    1 \\
HD 305606          & 104925.87-600137.3 & B2 V                         &\nodata &\nodata &\nodata &  -14.92 &    1 &    1 \\
\enddata
\tablenotetext{\star}{Kolmogorov-Smirnov percent probability of constancy. Minimum of single-epoch and merged $P_{\rm KS}$ values.}
\tablenotetext{\dagger}{Number of distinct maximum likelihood blocks in merged event data (minimum 5 counts, 95\% confidence).}
\tablenotetext{\ddagger}{HD 93501's 160-count spectrum can be fit with a power law or $kT> 6.5$~keV (90\% confidence) \citep{naze11}.}
\tablenotetext{\sharp}{The ISM-corrected 0.5-8 keV $L_{\rm X}$, {\sc fitluminosity\_tc}. For the ISM-corrected 0.5-10 keV X-ray luminosity, see \citet{naze11}.}
\tablenotetext{\spadesuit}{$f_{\rm X}$ is the absorbed (uncorrected) 0.5-8.0~keV flux, {\sc energyflux\_t}. For the absorbed 0.5-10~keV flux, {\sc fto} see \citet{naze11}.}
\end{deluxetable}

\begin{deluxetable}{lcccc}
\tabletypesize{\small}
\tablecaption{{\it Chandra} Observations of HD 93250}
\tablewidth{0in}
\tablehead{
\colhead{Date} & \colhead{MJD} & \colhead{Grating} & \colhead{OBSID} & \colhead{Exposure}
}
\startdata
2000-11-19 &  51867.12 & HETG & 632  &  89545.8\\
2003-05-02 &  52761.50 & HETG &  3745    &  94533.0\\
2004-09-21 &  53269.73 & NONE &  4495    &  56634.4\\
2005-09-05 &  53618.62 & HETG &  5400    &  33727.4\\
2005-10-22 &  53665.13 & HETG &  5399    &  39670.4\\
2006-06-19 &  53905.75 & HETG &  7341    &  53265.7\\
2006-06-22 &  53908.43 & HETG &  7342    &  49313.2\\
2006-06-23 &  53909.74 & HETG &  7189    &  17721.5\\
\enddata
\end{deluxetable}

\end{document}